\documentclass[11pt]{article}
\pdfoutput=1
\usepackage{jheppub}

\usepackage{pdflscape}
\usepackage[...]{youngtab}
\usepackage{xfrac}
\usepackage{appendix}

\usepackage{amsfonts}
\usepackage{physics}
\usepackage[usenames,dvipsnames]{xcolor}
\usepackage{tabularx,multirow,array,amsmath,mathalfa,empheq}
\usepackage{color}
\usepackage{ccaption}
\usepackage{mathtools}
\usepackage{bbold}
\usepackage{kantlipsum}
\usepackage{slashbox}
\usepackage[T1]{fontenc}
\allowdisplaybreaks

\usepackage{changepage}
\usepackage[framemethod=TikZ]{mdframed}
\usepackage{lipsum}
\usepackage{titlesec}
\mdfdefinestyle{Frame-1}{%
    linecolor=black,
    outerlinewidth=2pt,
    roundcorner=20pt,
    innerrightmargin=15pt,
    innerleftmargin=15pt,
    backgroundcolor=gray!30!white}	
\definecolor{Red}{rgb}{1,0,0}

\usepackage{graphicx}
\makeatletter

\newcommand*\bigcdot{\mathpalette\bigcdot@{.5}}
\newcommand*\bigcdot@[2]{\mathbin{\vcenter{\hbox{\scalebox{#2}{$\m@th#1\bullet$}}}}}
\makeatother
\usepackage{pgf}
\usepackage{tikz}
\usepackage[utf8]{inputenc}
\usetikzlibrary{arrows,automata}
\usetikzlibrary{positioning}
\tikzset{state/.style={rectangle, rounded corners, draw=black, very thick, minimum height=2em, inner sep=2pt, text centered,},}

\title{\boldmath Multi-boundary entanglement in Chern-Simons theory with finite gauge groups}

\author[a]{Siddharth Dwivedi}
\author[a,b]{, Andrea Addazi}
\author[c]{, Yang Zhou}
\author[d]{, Puneet Sharma}

\affiliation[a]{Center for Theoretical Physics, College of Physical Science and Technology, Sichuan University,\\
Chengdu, 610064, China}
\affiliation[b]{INFN sezione Roma Tor Vergata, I-00133 Rome, Italy}
\affiliation[c]{Department of Physics and Center for Field Theory and Particle Physics, Fudan University,\\ Shanghai 200433, China}
\affiliation[d]{Department of Mathematics, Indian Institute of Technology Jodhpur, Karwar, Jodhpur 342037, India}

\emailAdd{sdwivedi@scu.edu.cn, andrea.addazi@Ings.infn.it, yang{\_}zhou@fudan.edu.cn, puneet@iitj.ac.in}

\abstract{We study the multi-boundary entanglement structure of the states prepared in (1+1) and (2+1) dimensional Chern-Simons theory with finite discrete gauge group $G$. The states in (1+1)-$d$ are associated with Riemann surfaces of genus $g$ with multiple $S^1$ boundaries and we use replica trick to compute the entanglement entropy for such states. In (2+1)-$d$, we focus on the states associated with torus link complements which live in the tensor product of Hilbert spaces associated with multiple $T^2$. We present a quantitative analysis of the entanglement structure for both abelian and non-abelian groups. For all the states considered in this work, we find that the entanglement entropy for direct product of groups is the sum of entropy for individual groups, i.e. $\text{EE}(G_1 \times G_2) = \text{EE}(G_1)+\text{EE}(G_2)$. Moreover, the reduced density matrix obtained by tracing out a subset of the total Hilbert space has a positive semidefinite partial transpose on any bi-partition of the remaining Hilbert space.}

\usepackage[normalem]{ulem}  

\begin{document} 

	
		\vspace*{-2em}
	\maketitle
		\flushbottom

	\newpage


\section{Introduction}
\label{sec1}
\begin{figure}[h]
\centerline{\includegraphics[width=4.4in]{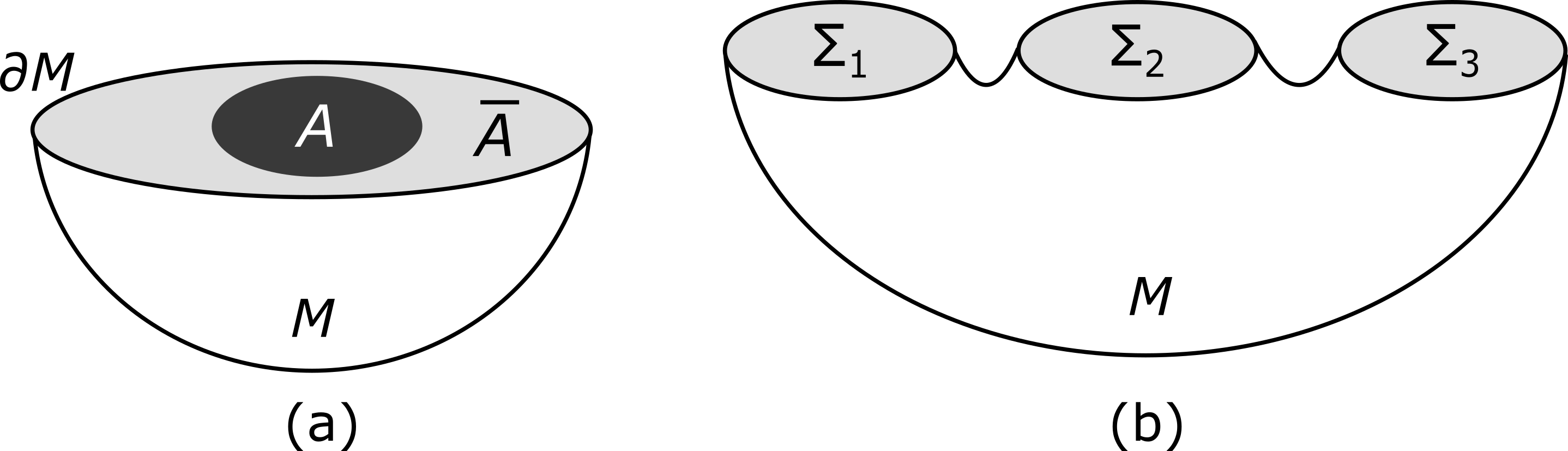}}
\caption[]{Figure (a) shows a manifold with a single boundary $\partial M$ which is partitioned into a sub-region $A$ and its complement $\bar{A}$. Figure (b) shows a manifold $M$ whose boundary has three disconnected components, i.e. $\partial M=\Sigma_1 \sqcup \Sigma_2 \sqcup \Sigma_3$.}
\label{Multi-boundary}
\end{figure}

The study of entanglement in quantum field theory (QFT) has been an active area of research and is yet far from complete. One of the important question is to understand and classify the possible patterns of entanglement that can arise in field theories. Though analyzing the entanglement structures in a generic QFT is a difficult task, a simple case where this could be done is for a class of exactly solvable QFT's called `topological quantum field theories'. In particular, there has been a lot of interest to study entanglement in (2+1) dimensional Chern-Simons theory defined on 3-manifolds $M$. The bi-partite entanglement between connected spatial sections of the boundary $\partial M$ in such theories was studied in \cite{Kitaev:2005dm,Levin:2006zz,Dong:2008ft}. A typical path integral picture has been shown in figure \ref{Multi-boundary}(a) where the boundary $\partial M$ has been bi-partitioned into connected regions $A$ and its complement $\bar{A}$. Another novel approach to study entanglement is to consider these theories on manifolds whose boundary itself consists of disconnected or disjoint components as shown in figure \ref{Multi-boundary}(b). This is usually referred as \emph{multi-boundary entanglement} and has been studied in \cite{Salton:2016qpp, Balasubramanian:2016sro,Dwivedi:2017rnj,Melnikov:2018zfn,Dwivedi:2019bzh} in the context of Chern-Simons theories whose gauge group is a compact semi-simple Lie group (SU($N$) for example). In this work, we will study the salient features of multi-boundary entanglement structures in Chern-Simons theory with finite discrete gauge groups. This suggests another pattern of entanglement in topological order apart from~\cite{Kitaev:2005dm,Levin:2006zz,Dong:2008ft} and therefore enriches our understanding of quantum information properties of physical systems with topological order.

A quantum $(d+1)$ dimensional Chern-Simons theory is defined on a compact $(d+1)$-manifold $M$ by integrating the exponentiated Chern-Simons invariant (usually referred to as the `path integral' in field theories) over the space of all gauge invariant fields. The resulting 
`partition function' is denoted as $Z(M)$~\cite{Witten:1988hf}.
The gauge fields in this case are the connections (skew-Hermitian matrices of 1-forms with vanishing trace) $A$ on the trivial SU($N$)-bundle over $M$. The partition function will therefore involve the integration over the infinite-dimensional space of connections:
\begin{equation}
Z(M) = \int e^{2 \pi i S(A)} dA ~,
\end{equation}
where $dA$ is an appropriate quantized measure defined for a connection $A$ and the integration is over all the equivalence classes (i.e. gauge invariant classes) of connections. Since the connection has been integrated out, $Z(M)$ will be a topological invariant of $M$.\footnote{To be more accurate, the $Z(M)$ constructed in this way will also depend on an additional topological structure of $M$, called as framing \cite{Witten:1988hf}. However they do not affect the entanglement structure of the states considered in this paper and hence we can ignore the framing factors in our computation.} The invariant $S(A)$, which is the Chern-Simons action, is given as,
\begin{equation}
S(A) = \frac{k}{8 \pi} \int_M \text{Tr}\left(A \wedge dA + \frac{2}{3} A \wedge A \wedge A \right) ~,
\label{action-3d}
\end{equation}
where $k$ is an integer that classifies the invariant $S(A)$. Additionally, we can also have gauge invariant operators (called Wilson loops) in the theory. Given an oriented knot $\mathcal{K}$ embedded in $M$, the Wilson loop operator is defined by taking the trace of the holonomy of $A$ around $\mathcal{K}$:
\begin{equation}
W_{R}(\mathcal{K}) = \text{Tr}_R \,P\exp\left(\oint_{\mathcal{K}} A \right) = \text{char}_R(A) ~,
\end{equation} 
where $R$ is an irreducible representation of SU($N$) with character $\text{char}_R$.
If we have a link $\mathcal{L}$ made of disjoint oriented knot components, i.e. $\mathcal{L} = \mathcal{K}_1 \sqcup \mathcal{K}_2 \sqcup \ldots \sqcup \mathcal{K}_n$, then the partition function of $M$ in the presence of link $\mathcal{L}$ can be obtained by modifying the path integral as:
\begin{equation}
Z(M; \mathcal{L}) = \int e^{2 \pi i S(A)} \prod_{j=1}^n \text{char}_{R_j}(A) \, dA ~,
\end{equation}  
where in principle we can chose different characters for different knot components. 

The partition function $Z(M)$ defined above is either a vector or a scalar depending upon whether $M$ is with or without boundary. The following two points are very crucial and are the crux of the Chern-Simons theory: 
\begin{itemize}
\item When $M$ is closed (i.e. without boundary), the quantity $Z(M)$ is simply a complex number.
\item When $M$ has a boundary $\partial M$,  the partition function $Z_{Q}(M)$ will additionally depend on the boundary field $Q$ on $\partial M$. The corresponding integral should be over the space of gauge equivalence classes of those connections on $M$ whose boundary value is $Q$. Thus $Z_{Q}(M)$ will be a function on the space $F_{\partial M}$ of gauge equivalence classes on $\partial M$.
\end{itemize}
In the quantized version of Chern-Simons theory with an appropriate measure $dA$ defined for the connections, the space $L^2(F_{\partial M}, dA) \equiv \mathcal{H}_{\partial M}$ is a unique Hilbert space associated with $\partial M$ and the partition function $Z_{Q}(M)$ will be an element of $\mathcal{H}_{\partial M}$. Henceforth, we will denote $Z_{Q}(M)$ as $\ket{\Psi}$ since it is a quantum state living in the Hilbert space $\mathcal{H}_{\partial M}$. From the topological point of view, $\ket{\Psi}$ only depends on the topology of the manifold $M$ and $\mathcal{H}_{\partial M}$ only depends on the topology of the boundary $\partial M$. Thus if we consider two topologically different manifolds $M$ and $M'$ with the same boundary $\partial M = \partial M'$, then the corresponding Chern-Simons partition functions for a particular gauge group will, in principle, give two different states $\ket{\Psi}$ and $\ket{\Psi'}$ in the same Hilbert space $\mathcal{H}_{\partial M}$. Hence, given a manifold, we can associate a quantum state to it. An important aspect of Chern-Simons theory which motivates us to study the multi-boundary entanglement is the following. If the boundary of the manifold $M$ consists of disjoint components, i.e. $\partial M = \Sigma_1 \sqcup \Sigma_2 \sqcup \ldots \sqcup \Sigma_n$ (like the one shown in Figure \ref{Multi-boundary}(b)), then the Hilbert space associated with $\partial M$ is the tensor product of Hilbert spaces associated with each component, i.e.
\begin{equation}
\mathcal{H}_{\partial M} = \mathcal{H}_{\Sigma_1} \otimes \mathcal{H}_{\Sigma_2} \otimes \ldots \otimes \mathcal{H}_{\Sigma_n} ~.
\end{equation} 
Thus the quantum state $\ket{\Psi}$ associated with $M$ lives in this tensor product of Hilbert spaces and therefore we can study its entanglement features. Various entanglement measures like von Neumann entropy and entanglement negativity etc., can be obtained by tracing out a subset of the Hilbert spaces. The entanglement of the states for a manifold with multiple $S^2$ (the genus $g=0$ Riemann surface) boundaries has been recently studied in \cite{Melnikov:2018zfn} and \cite{Dwivedi:2019bzh} for the group SU$(N)$. The states for link complement manifolds with multiple $T^2$ (the genus $g=1$ Riemann surface) boundaries was studied in \cite{Balasubramanian:2016sro, Salton:2016qpp} for the groups U(1) and SU$(2)$ and was later analyzed in \cite{Dwivedi:2017rnj} for various classical Lie groups. 

The aim of the present work is to study the entanglement structure of the states constructed in Chern-Simons theory with finite discrete gauge groups. These theories in $(2+1)$ dimension were originally introduced by Dijkgraaf and Witten \cite{dijkgraaf1990} and were later studied in detail in $(d+1)$ dimension by Freed and Quinn \cite{Freed:1991bn}. The essential algebraic and topological features of this theory are the same as that of the Chern-Simons theory with semi-simple Lie groups. However the analytical difficulties in the case of finite gauge groups are simplified because the path integral required to compute various partition functions reduces to a finite sum. In this work, we will analyze the entanglement structure of multi-boundary states in $(1+1)$ dimensional and $(2+1)$ dimensional Chern-Simons theory with finite groups. In $(1+1)$ dimension, the states are associated with Riemann surfaces with multiple $S^1$ boundaries and in $(2+1)$ dimension, we will consider the states corresponding to torus link complement manifolds with multiple disjoint $T^2$ boundaries.\footnote{If a link $\mathcal{L}$  is embedded in $S^3$, then the link complement is a three-dimensional manifold which is obtained by removing a tubular neighborhood around $\mathcal{L}$ from $S^3$, i.e $S^3 \backslash \mathcal{L} \equiv S^3 - \text{interior}(\mathcal{L}_{\text{tub}})$.}

The paper is organized as follows. In section \ref{sec2}, we discuss our set-up giving a brief review of the Chern-Simons theory with finite gauge group and the methods and tools to study the multi-boundary entanglement. In section \ref{sec3}, we study the states in $(1+1)$ dimension associated with Riemann surfaces of genus $g$. In section \ref{sec4}, we study the states in $(2+1)$ dimension associated with torus link complement. We provide quantitative analysis of the entanglement entropy for abelian as well as non-abelian groups. We conclude and discuss future questions in section \ref{sec5}.

\section{The multi-boundary entanglement set-up for finite groups}
\label{sec2}
We consider Chern-Simons theory with gauge group $G$ which is a finite discrete group. This theory is defined on a $(d+1)$ dimensional space-time $M$ where $M$ is assumed to be compact, connected $(d+1)$ dimensional manifold. The boundary of $M$ will be denoted as $\partial M$ which is a $d$-dimensional oriented and closed manifold. This theory has certain ingredients which we define and annotate in the following. 
\subsection{Chern-Simons theory with finite gauge group}
Here we briefly discuss about the theory when the gauge group is a finite discrete group. We refer the readers to \cite{Freed:1991bn} and the lecture notes by Freed \cite{Freed1993} for more details. A gauge field in this theory is a principal bundle $P$ defined over $M$ with $G$ as its structure group. In other words, $P$ is a bundle over the base space $M$ such that the group $G$ acts on $P$ from right. The action of $G$ is free, transitive and preserves the fibres of $P$. Thus by definition, the orbits of $G$-action are actually the fibres and hence topologically we have $P/G=M$. Two principal bundles $P$ and $P'$ are said to be equivalent or isomorphic if there exists a map $\varphi: P' \longrightarrow P$ which commutes with the $G$ action such that the induced map on the quotients $\varphi': M \longrightarrow M$ is an identity map. We will denote $[P]$ as the collection of all principal bundles isomorphic to $P$. Thus $[P]$ will be a `gauge invariant' field in this theory and a set of all gauge invariant fields is given as,
\begin{equation}
\phi(M) = \{[P]: \text{$P$ is a principal $G$-bundle over $M$} \} ~.
\end{equation}
If $M$ is compact, $\phi(M)$ is a finite set. Further if $M$ is connected then this set is isomorphic to:
\begin{equation}
M=\text{connected} \quad \Longrightarrow \quad \phi(M) \cong \text{Hom}(\pi_1(M), G)/G ~,
\end{equation}
where $\pi_1(M)$ is the fundamental group of $M$ and the set $\text{Hom}(\pi_1(M), G)$ is the collection of all homomorphisms from $\pi_1(M) \longrightarrow G$. The $\text{Hom}(\pi_1(M), G)/G$ denotes a quotient where $G$ acts by conjugation. Since $\phi(M)$ is a discrete set, we can assign a measure or a `mass' to each gauge invariant field which is given as,
\begin{equation}
\mu([P]) = \frac{1}{|\text{Aut}(P)|} ~,
\end{equation}
where $\text{Aut}(P)$ is the automorphism group of $P$ or the group of `gauge transformations' of $P$ and $|\text{Aut}(P)|$ is the order of this group.

Now we need to construct the action of the theory. The action is classified by the elements of the cohomology group $H^{d+1}(BG, U(1))$ where $BG$ is a connected topological space known as the classifying space of $G$.\footnote{For a discrete group $G$, the classifying space $BG$ is a connected topological space whose homotopy groups are given as: $\pi_1(BG) \cong G$ and $\pi_{n \geq 2}(BG) = 0$.} It is the base space of a principal $G$ bundle $EG$, also called universal bundle (where the action of $G$ is free). Now consider a principal $G$ bundle $P$ over $M$. Given any $P$, we can allow a bundle map $f: P \longrightarrow EG$, called as the classifying map. Since $P$ and $EG$ both are principal $G$ bundles with the quotients given as $P/G = M$ and $EG/G = BG$ respectively, the classifying map $f$ induces a quotient map $\gamma: M \longrightarrow BG$ on the base manifolds. The topology of the bundle $P$ is completely determined by the homotopy class of map $\gamma$. If $\gamma^*$ denotes the pullback, then the action is given as,
\begin{equation}
S(P) = \alpha(\gamma^*[M]) ~,
\label{Action-twisted}
\end{equation}
where $[M] \in H_{d+1}(M)$ is the fundamental class and the action is independent of the choice of map $\gamma$. The element $\alpha \in H^{d+1}(BG, U(1))$ classifies the action and plays an analogous role as that of integer $k$ which classifies the action (\ref{action-3d}) of a $(2+1)$ dimensional Chern-Simons theory with SU($N$) group.\footnote{Note that for the finite groups $H^{d+1}(BG, U(1)) \cong H^{d+2}(BG, \mathbb{Z})$. It also happens that for the connected or simply connected groups $G'$, we have $H^{4}(BG', \mathbb{Z}) \cong \mathbb{Z}$. This is precisely what happens in $(2+1)$ dimensional Chern-Simons theory with SU($N$) gauge group where the action in eq.(\ref{action-3d}) was classified by integer $k \in H^{4}(BSU(N), \mathbb{Z})$.} Thus each non-identity element $\alpha$ gives a `twisted theory'.  A huge simplification happens if we consider the class of theories given by $\alpha = 0$, the identity element of $H^{d+1}(BG, U(1))$. For this class, the action is trivial $e^{2 \pi i S(P)} = 1$ and such theories are called `untwisted theories':
\begin{equation}
\text{Untwisted theory:} \quad \alpha=0 \quad \Longrightarrow \quad e^{2 \pi i S(P)} = 1 ~.
\label{alpha=0-action}
\end{equation}

With the action and measure defined, one can carry out the path integral in the Chern-Simons theory for any compact manifold $M$ which may or may not have a boundary. If $M$ is without boundary (i.e closed), the path integral gives a complex number which is a topological invariant of $M$. This topological quantity is typically known as `partition function' of $M$ and is evaluated as,
\begin{equation}
Z(M) = \int_{P \in \phi(M)} e^{2 \pi i S(P)}\, d\mu(P) = \sum_{P \in \phi(M)} e^{2 \pi i S(P)}\, \mu(P) = \text{vol}(\phi(M))~.
\end{equation}  
On the other hand, if the manifold $M$ has an oriented boundary $\partial M$, then the path integral will be a function of $Q$ where $Q$ is the principal $G$ bundle over base $\partial M$. Now fix a $Q$ and define a set $\phi_Q(M)$ which is the collection of all principal bundles over $M$ whose restriction on $\partial M$ is Q, up to isomorphisms which are the identity on $\partial M$. In other words, define a set $\phi_Q(M)$ and $\bar{\phi}_Q(M)$ given below:
\begin{align}
\phi_Q(M) &= \{ P \longrightarrow M: \partial P \cong Q \} \nonumber \\
\bar{\phi}_Q(M) &= \text{set of isomorphic bundles in $\phi_Q(M)$}  ~.
\end{align}
Doing the path integral over all the bundles in $\bar{\phi}_Q(M)$, we will get the partition function:
\begin{equation}
Z_Q(M) = \int_{P \in \bar{\phi}_Q(M)} e^{2 \pi i S(P)}\, d\mu(P) = \text{vol}(\bar{\phi}_Q(M))~.
\end{equation} 

\begin{figure}[htbp]
	\centering
		\includegraphics[width=1.00\textwidth]{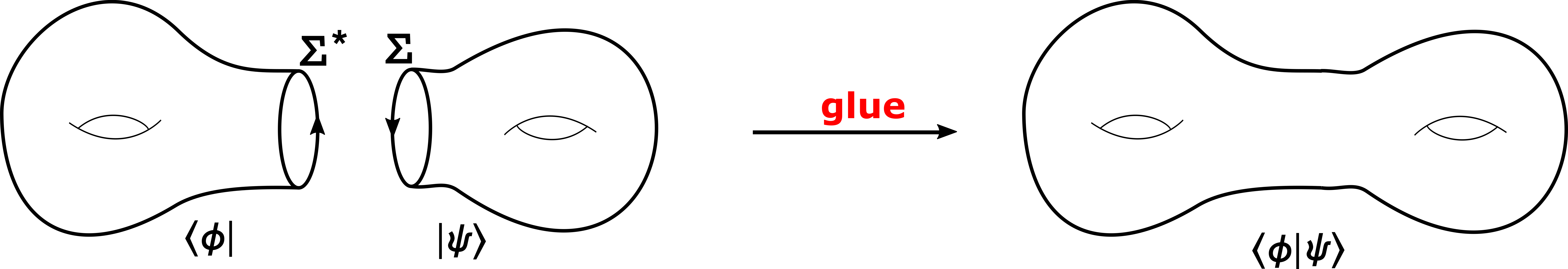}
	\caption{Two manifolds on left with same boundary but opposite orientation. The path integral on these manifolds give states $\bra{\phi} \in \mathcal{H}_{\Sigma^*}$ and $\ket{\psi} \in \mathcal{H}_{\Sigma}$. The inner product $\bra{\phi}\ket{\psi}$ will be the partition function of manifold shown in right obtained by gluing the two manifolds along the common boundary.}
	\label{gluing}
\end{figure}

Thus we see that given a manifold $M$ with boundary $\Sigma$, we can associate a quantum state $\ket{\Psi}$ which is an element of a unique Hilbert space $\mathcal{H}_{\Sigma}$ associated with the boundary. Note that $\Sigma$ has an orientation and if we reverse the orientation, we will get an oppositely oriented manifold which we shall denote as $\Sigma^*$. In fact, the Hilbert spaces associated with $\Sigma$ and $\Sigma^*$ are conjugate to each other, i.e. $\mathcal{H}_{\Sigma^*} = \mathcal{H}^*_{\Sigma}$. Thus given two states $\ket{\Psi} \in \mathcal{H}_{\Sigma}$ and $\bra{\Phi} \in \mathcal{H}_{\Sigma^*}$, there exists a natural pairing, the inner product, computed as $\langle \Phi| \Psi \rangle$ which gives a complex number. In fact, this technique can be used to compute the partition functions of complicated manifolds by gluing two disconnected pieces along common boundary whose partition functions are already known. This process is shown in figure \ref{gluing}.
We can also have a manifold $M$ whose boundary $\partial M$ has multiple disconnected components. In such a case, the Hilbert space associated with $\partial M$ will be a tensor product of Hilbert spaces associated with each of the disconnected components: 
\begin{equation}
\partial M = \Sigma_1 \sqcup \Sigma_2 \sqcup \ldots \sqcup \Sigma_n \,\, \Longrightarrow \,\, \mathcal{H}_{\partial M} = \mathcal{H}_{\Sigma_1} \otimes \mathcal{H}_{\Sigma_2} \otimes \ldots \otimes \mathcal{H}_{\Sigma_n} ~.
\end{equation} 
Thus the path integral for such a manifold will give a quantum state which will be an element of this tensor product of Hilbert spaces: 
\begin{equation}
Z_{Q_1 \sqcup Q_2 \sqcup \ldots \sqcup Q_n }(M) \equiv \ket{\Psi} \in \mathcal{H}_{\Sigma_1} \otimes \mathcal{H}_{\Sigma_2} \otimes \ldots \otimes \mathcal{H}_{\Sigma_n} ~.
\end{equation}
Since the Hilbert spaces in these cases are automatically a tensor product space, we can assign an entanglement structure to the states by tracing out a subset of the Hilbert spaces associated with some of the boundary components. In the following, we will discuss how the topological property of Chern-Simons theory enables one to compute the various entanglement measures by using the surgery (cutting and gluing) techniques.   
\subsection{Multi-boundary entanglement in Chern-Simons theory}
Consider the Chern-Simons theory defined on a $(d+1)$-dimensional manifold $M$ whose boundary consists of $n$ number of disconnected components. As mentioned earlier, to such $M$, we can associate a state $\ket{\Psi}$ which lives in a total Hilbert space ($\mathcal{H}$) that can be written as a tensor product of $n$ Hilbert spaces associated with each of the boundary components. To study the entanglement feature of this state, we bi-partition the total Hilbert space into two parts as following:
\begin{equation}
\mathcal{H} = \underbrace{\mathcal{H}_{1} \otimes \mathcal{H}_{2} \otimes \ldots \otimes \mathcal{H}_{m}}_{m} \,\,\otimes\,\, \underbrace{\mathcal{H}_{m+1} \otimes \mathcal{H}_{m+2} \otimes \ldots \otimes \mathcal{H}_{n}}_{n-m} \equiv \mathcal{H}_{A} \otimes \mathcal{H}_{B} ~.
\end{equation}
We call this bi-partitioning as $(m|n-m)$. In order to compute the entanglement measures, we must trace out $\mathcal{H}_{B}$ and obtain the reduced density matrix $\rho_A$ acting on $\mathcal{H}_{A}$. There are two ways of achieving this and depending upon the context, it is easier to use one method or the other. We will discuss both the methods in the following.

The first method is useful when the quantum states can be explicitly computed. For example, in the case of $(2+1)$-d Chern-Simons theory where $M$ is a link complement manifold whose boundary consists of multiple disconnected tori, the quantum state can be explicitly obtained (see \cite{Balasubramanian:2016sro, Dwivedi:2017rnj}) and one can continue with this method. Suppose we know the state $\ket{\Psi} \in \mathcal{H}$ which can be expanded as:
\begin{equation}
\ket{\Psi} = \sum_{e_1,\,e_2,\, \ldots,\, e_n} C_{e_1,\,e_2,\, \ldots,\, e_n} \ket{e_1,\,e_2,\, \ldots,\, e_n} ~,
\label{}
\end{equation}
where $\ket{e_i}$ is a basis of $\mathcal{H}_i$ and the coefficients $C_{e_1,\,e_2,\, \ldots,\, e_n}$ are in general complex numbers which are known. Given such a state, one associates a projection operator, denoted as $\rho_{\text{total}}$ which acts on the total Hilbert space $\mathcal{H}_A \otimes \mathcal{H}_B$. Such operator is called as density matrix operator and is given as:
\begin{equation}
\rho_{\text{total}} = \frac{\ket{\Psi} \bra{\Psi}}{\braket{\Psi}} ~,
\end{equation}
where $\braket{\Psi}$ is the normalization factor obtained by normalizing $\ket{\Psi} \bra{\Psi}$ and it ensures that the density matrix has unit trace. The dual state can be obtained as:
\begin{equation}
\bra{\Psi} = \sum_{f_1,\,f_2,\, \ldots,\, f_n} C_{f_1,\,f_2,\, \ldots,\, f_n}^{*} \bra{f_1,\,f_2,\, \ldots,\, f_n} ~,
\label{}
\end{equation} 
where $\bra{f_j}$ is a basis of the dual Hilbert space $\mathcal{H}_j^*$. The coefficients $C^*$ are the complex conjugates of coefficients $C$. Next we trace out the Hilbert space $\mathcal{H}_{B}$ which means we trace out each of the Hilbert spaces $\mathcal{H}_{m+1}, \ldots, \mathcal{H}_{n}$. This tracing out of $\mathcal{H}_{B}$ will give an operator $\rho_A$ acting on $\mathcal{H}_A$ which is called as reduced density matrix and is computed as:
\begin{equation}
\rho_A  = \text{Tr}_{\mathcal{H}_B}(\rho_{\text{total}}) = \sum_{e_{m+1},\, e_{m+2},\, \ldots,\, e_{n} } \expval{\rho_{\text{total}}}{e_{m+1},\, e_{m+2},\, \ldots,\, e_{n}} ~.
\end{equation}
Having obtained this matrix, one can calculate its spectrum $\{ \lambda_i \}$ from which various entanglement measures can be easily computed. For example, the von Neumann entropy, also called the entanglement entropy can be computed as:
\begin{equation}
\text{EE} = -\sum_{i} \lambda_i \ln \lambda_i ~.
\end{equation}
A non-zero value of the entanglement entropy tells that the quantum state $\ket{\Psi}$ is entangled on the bi-partition $\mathcal{H}_A \otimes \mathcal{H}_B$ while vanishing value of $\text{EE}$ means the state is separable (non-entangled). Typically it is a difficult problem to find whether the reduced density matrix $\rho_A$ acting on $\mathcal{H}_A$ (which now describes a mixed state) is entangled or separable. However there are certain separability criteria, such as positive partial transpose (PPT) criterion \cite{Peres:1996dw} which provides necessary but not sufficient condition for separability. For this, consider the further bi-partitioning $\mathcal{H}_A = \mathcal{H}_{A_1} \otimes \mathcal{H}_{A_2}$ and compute the partial transpose of $\rho_A$ with respect to one of the sub-system (say $A_2$). This partial transpose, denoted as $\rho_A^{\Gamma(A_2)}$ can be obtained from the density matrix $\rho_A$ as following:
\begin{equation}
\langle e_i^A, e_j^B | \rho_A^{\Gamma(A_2)} |e_k^A, e_l^B \rangle = \langle e_i^A, e_l^B | \rho_A |e_k^A, e_j^B \rangle ~.
\label{par-transpose}
\end{equation}
If $\rho_A^{\Gamma(A_2)}$ is not positive semidefinite, it necessarily implies that the density matrix $\rho_A$ is entangled. To capture this information, we define \emph{entanglement negativity} $\mathcal{N}$ as,
\begin{equation}
\mathcal{N} = \frac{||\rho_A^{\Gamma(A_2)}||-1}{2} = \sum_{i} \frac{\abs{\mu_i} - \mu_i}{2} ~,
\label{negativity}
\end{equation} 
where $\mu_i$ are the eigenvalues of $\rho_A^{\Gamma(A_2)}$. A non-zero value of $\mathcal{N}$ means that $\rho_A^{\Gamma(A_2)}$ has negative eigenvalue and is not positive semi-definite. Thus $\mathcal{N} \neq 0$ implies that $\rho_A$ is entangled.

So far we have discussed the method of computing the entanglement measures when the quantum state is known and the reduced density matrix can be explicitly obtained. The other method is generally used when the state is not known or it is difficult to compute the state. This method is called the replica trick where the tracing of the Hilbert space is achieved by means of gluing of the manifolds along corresponding boundaries. Let us briefly discuss this method for a simple case when the manifold $M$ has 3 boundary components $\Sigma_1, \Sigma_2, \Sigma_3$. There is an associated state $\ket{\Psi} \in \mathcal{H}$ and a dual state $\bra{\Psi} \in \mathcal{H}^*$  which are given below.   
\begin{align}
\ket{\Psi} = \begin{array}{c}
\includegraphics[width=0.16\linewidth]{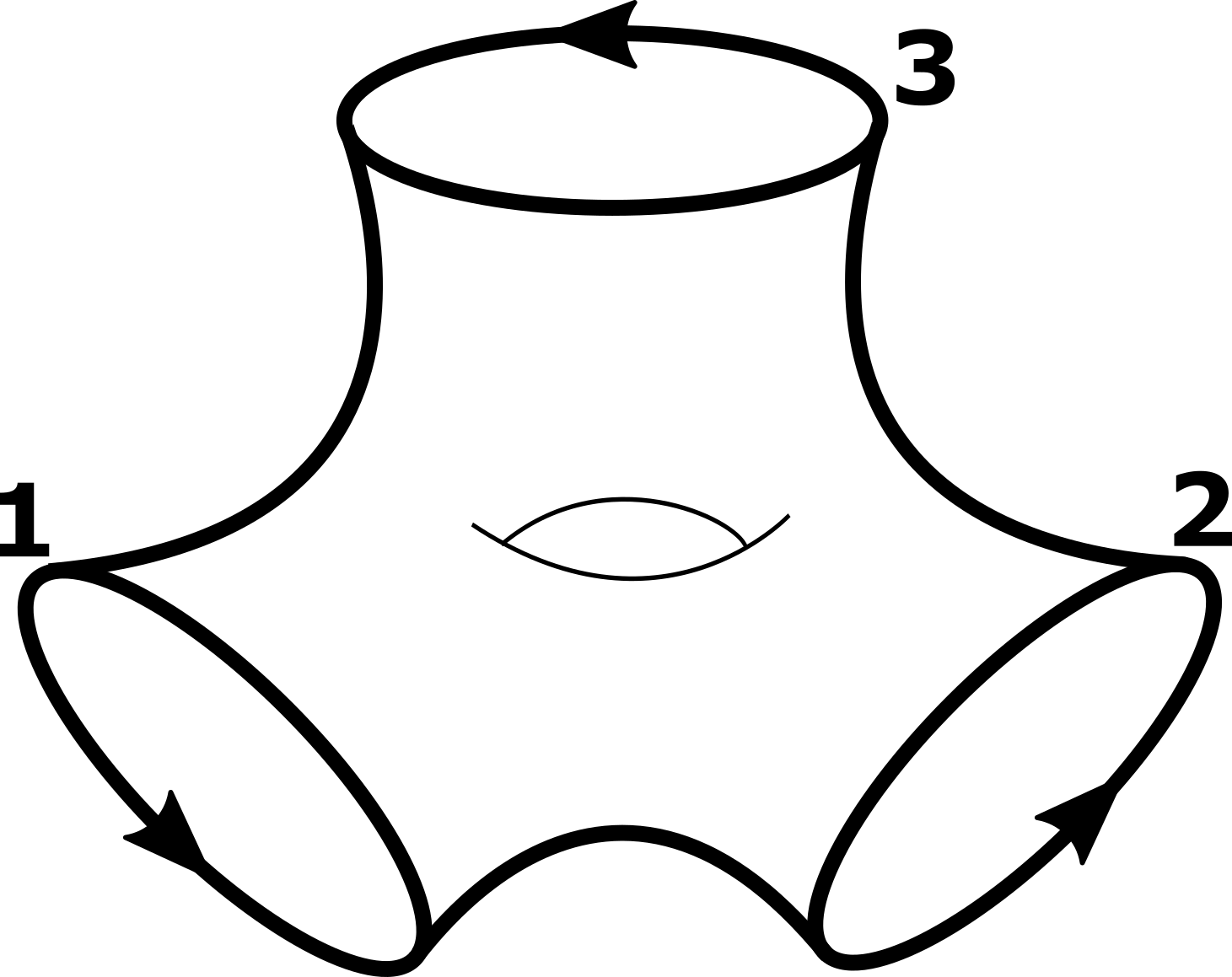}
\end{array} \quad,\quad \bra{\Psi} = \begin{array}{c}
\includegraphics[width=0.16\linewidth]{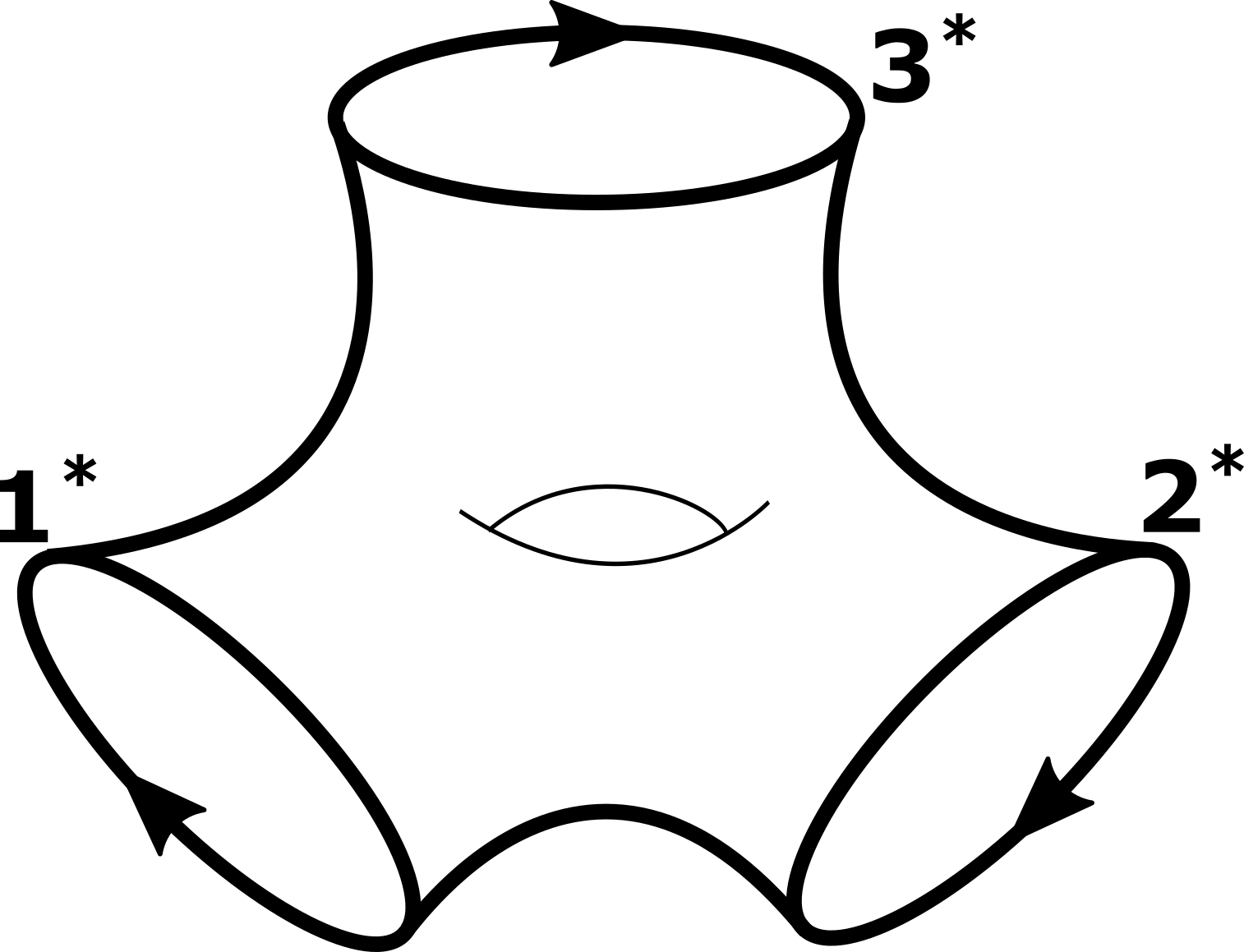}
\end{array} \quad,\quad M = \begin{array}{c}
\includegraphics[width=0.13\linewidth]{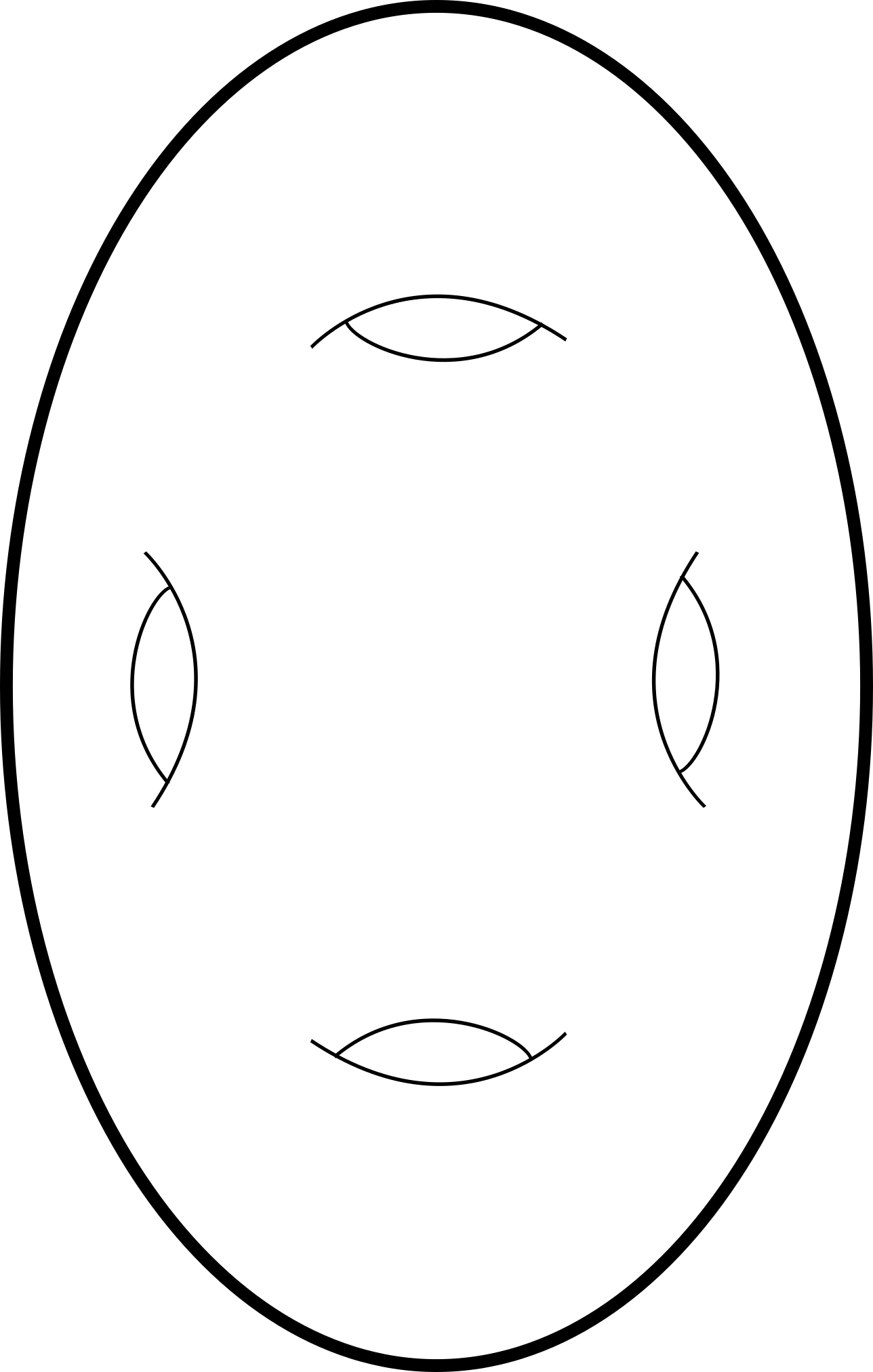}
\end{array} ~.
\label{replica-example}
\end{align}
Our first step is to compute the total density matrix
\begin{equation}
\rho_{\text{total}} = \frac{\ket{\Psi}\bra{\Psi}}{\braket{\Psi}} \quad\text{where,}\quad \braket{\Psi} = \text{Tr}_{\mathcal{H}_1} \text{Tr}_{\mathcal{H}_2} \text{Tr}_{\mathcal{H}_3} \ket{\Psi}\bra{\Psi} = Z(M) ~.
\label{}
\end{equation} 
To compute the normalization factor in the above equation, we have traced out the three Hilbert spaces. Topologically, this is achieved by considering the manifolds represented by $\ket{\Psi}$ and $\bra{\Psi}$ in eq.(\ref{replica-example}) and gluing them along the oppositely oriented boundaries (i.e. boundary $j$ glued to boundary $j^*$). This will result in a closed manifold $M$ shown in the eq.(\ref{replica-example}) and hence $\braket{\Psi}$ will be simply the partition function of $M$. Next, let us bi-partition the total Hilbert space as $(\mathcal{H}_1 \otimes \mathcal{H}_2|\mathcal{H}_3)$ and trace out $\mathcal{H}_3$ which means we glue $\rho_{\text{total}}$ along the boundaries $3$ and $3^*$. This will result in the reduced density matrix $\rho_{12}$ acting on $\mathcal{H}_1 \otimes \mathcal{H}_2$:
\begin{align}
\rho_{12} = \text{Tr}_{\mathcal{H}_3}(\rho_{\text{total}}) = \frac{1}{Z(M)} \times \begin{array}{c}
\includegraphics[width=0.16\linewidth]{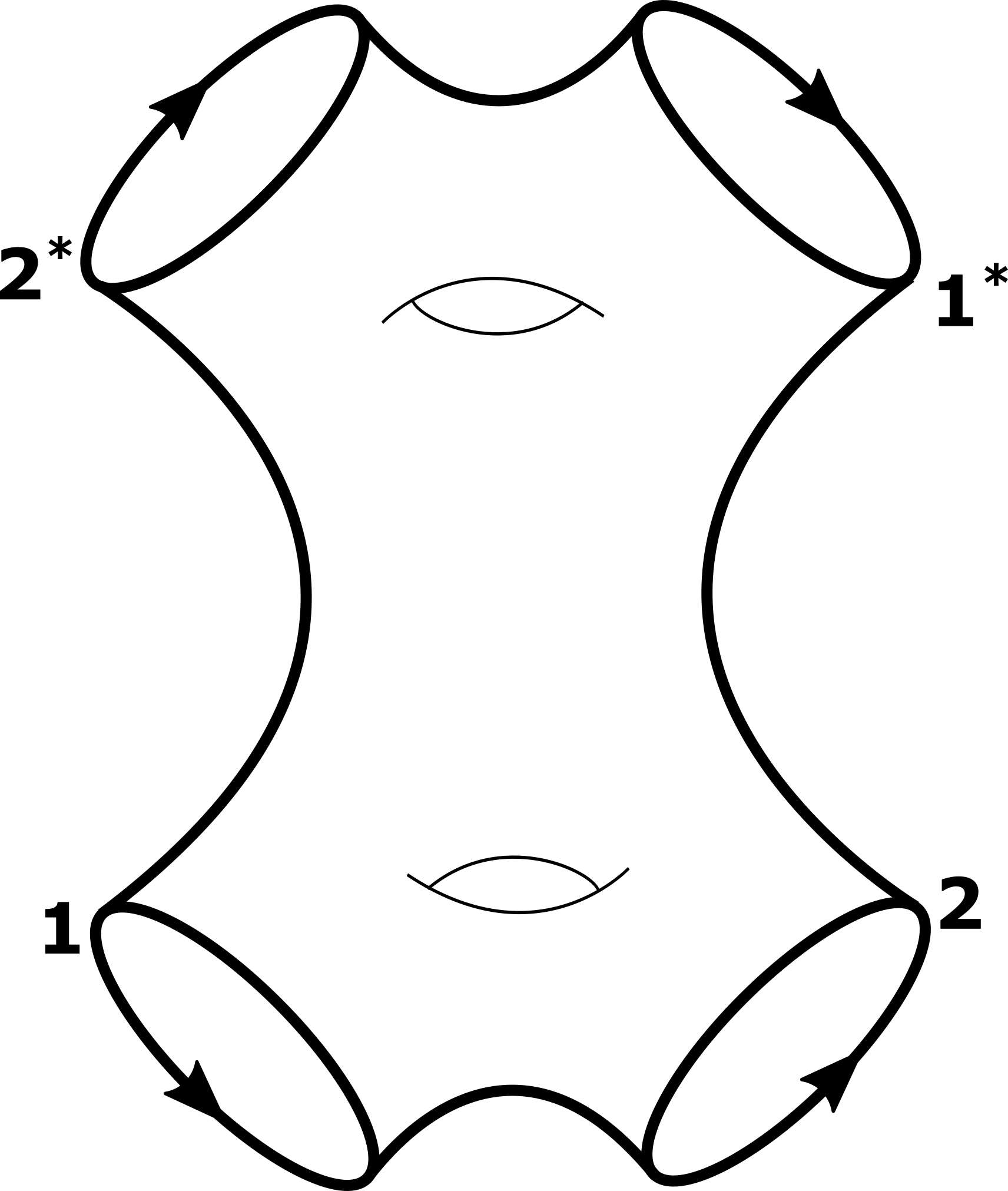}
\end{array} ~.
\label{rho12trace3}
\end{align}
Now, we know that the entanglement entropy can be computed from $\rho_{12}$ as:
\begin{equation}
\text{EE} = - \text{Tr}(\rho_{12} \ln \rho_{12}) ~.
\label{}
\end{equation}
However, we can not directly compute $\ln \rho_{12}$ from this diagrammatic approach.  In order to overcome this difficulty, we invoke replica trick. In this trick, we first compute $\rho_{12}^p$ by taking $p$ replicas of $\rho_{12}$ and glue the boundaries $1^*, 2^*$ of $i^{\text{th}}$ copy with the boundaries $1, 2$ of $(i+1)^{\text{th}}$ copy, i.e. $(1^*, 2^*)_i \longleftrightarrow(1,2)_{i+1}$ for $i=1,2,\ldots, (p-1)$. Topologically $\rho_{12}^p$ will look like this:
\begin{align}
\rho_{12}^p = \frac{1}{Z(M)^p} \times \begin{array}{c}
\includegraphics[width=0.71\linewidth]{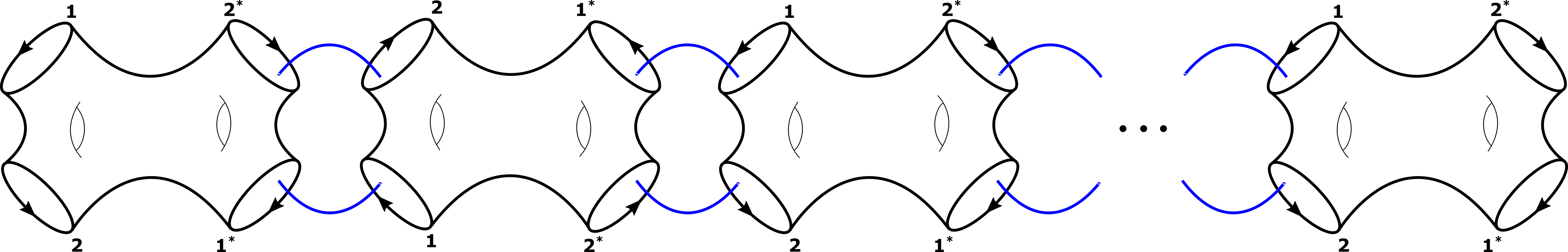}
\end{array} ~,
\label{}
\end{align}
where the blue line connecting two boundaries indicates that these boundaries are being glued. Thus resulting manifold $\rho_{12}^p$ will be a manifold with four boundaries $(1,2)_1,(1^*, 2^*)_p$. Now the trace of $\rho_{12}^p$ can be simply obtained by gluing $(1^*, 2^*)_{p} \longleftrightarrow(1,2)_{1}$ which will ultimately result in a closed manifold, say $M_p$. Thus the trace can be given as,
\begin{align}
\text{Tr}(\rho_{12}^p) = \frac{Z(M_p)}{Z(M)^p} ~.
\label{}
\end{align}
The entanglement entropy can be therefore computed as,
\begin{equation}
\text{EE}(\rho_{12}) =\lim_{p \to 1} \frac{\ln \text{Tr}(\rho_{12}^p)}{1-p}= -\lim_{p \to 1} \frac{d}{dp} \text{Tr}(\rho_{12}^p) = -\lim_{p \to 1} \frac{d}{dp} \left(\frac{Z(M_p)}{Z(M)^p}\right) ~.
\label{replica-trick-redrho}
\end{equation}
We can follow the similar steps to compute the entropy for the bi-partition $(\mathcal{H}_1 | \mathcal{H}_2 \otimes \mathcal{H}_3)$ and trace out $\mathcal{H}_2 \otimes \mathcal{H}_3$:
\begin{equation}
\text{EE}(\rho_{1}) = -\lim_{p \to 1} \frac{d}{dp} \left(\frac{Z(M_p')}{Z(M)^p}\right) ~,
\end{equation}
where the manifold $M_p'$ may be topologically different than manifold $M_p$. Another measure which checks the separability criteria of mixed states is the entanglement negativity defined in eq.(\ref{negativity}) for which we need the partial transpose of reduced density matrix with respect to one of the subsystem. We can again use the replica trick to compute the logarithmic negativity $\mathcal{N}_{\text{log}}$ which is related to negativity $\mathcal{N}$ as:  $\mathcal{N}_{\text{log}} = \ln(2\mathcal{N}+1)$. Now let us show how to compute $\mathcal{N}_{\text{log}}$ for the reduced density matrix $\rho_{12}$ of eq.(\ref{rho12trace3}). First we want the $\rho_{12}^{\Gamma_2}$ which is the partial transpose of $\rho_{12}$ obtained by swapping the boundaries $2$ and $2^*$. Next we compute its $q^{\text{th}}$ power $(\rho_{12}^{\Gamma_2})^q$ where $q$ is an even integer. It is obtained by stacking $q$ (even) copies of $\rho_{12}^{\Gamma_2}$ and gluing the boundaries $(1^*, 2)_i \longleftrightarrow(1,2^*)_{i+1}$ for $i=1,2,\ldots, (q-1)$ which will be:
\begin{align}
(\rho_{12}^{\Gamma_2})^q = \frac{1}{Z(M)^q} \times \begin{array}{c}
\includegraphics[width=0.67\linewidth]{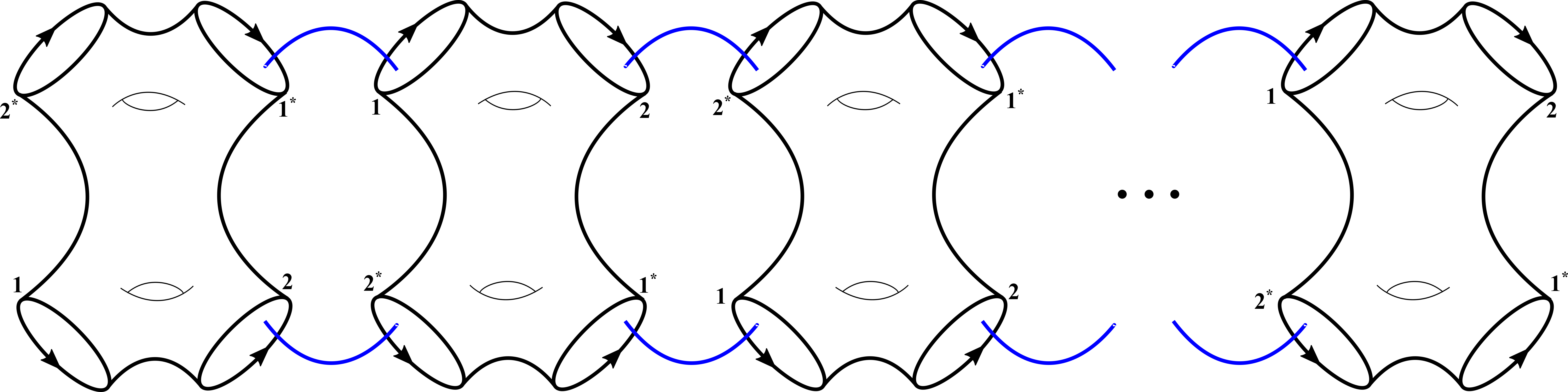}
\end{array} ~.
\label{}
\end{align}
Thus we obtain $(\rho_{12}^{\Gamma_2})^q$ with $q$ even which will be a manifold with four boundaries $(1,2^*)_1,(1^*, 2)_q$. Its trace can be obtained by gluing $(1^*, 2)_{q} \longleftrightarrow(1,2^*)_{1}$ which will result in a closed manifold, say $N_q$. The logarithmic negativity can be obtained as, 
\begin{equation}
\mathcal{N}_{\text{log}} = \lim_{q \to 1} \ln \left( \text{Tr}(\rho_{12}^{\Gamma_2})^q \right) = \lim_{q \to 1} \ln \left(\frac{Z(N_q)}{Z(M)^q}\right) ~.
\end{equation}

Having discussed the preliminaries and the basic set-up, let us move on to the computation of the entanglement structure of the states in Chern-Simons theory with finite groups. In the following section, we will consider the $(1+1)$ dimensional Chern-Simons theory where the states are associated with Riemann surfaces whose boundary consists of multiple disjoint copies of $S^1$. 
\section{Entanglement structure of states in (1+1)-d Chern-Simons theory}
\label{sec3}
Here we consider closed 2-manifolds which are the Riemann surfaces of genus $g$ without boundaries and are commonly denoted as $\Sigma_g$. In order to get compact manifolds with $n$ disjoint boundaries, imagine a Riemann surface with $n$ disks removed. This will create surfaces with $n$ number of $S^1$ boundaries. We will use the notation $\Sigma_{g,n}$ to denote the genus $g$ Riemann surfaces with $n$ boundaries. Thus, $\Sigma_g = \Sigma_{g,0}$ in our notation. We further assume that each $S^1$ boundary is oriented. The boundary of $\Sigma_{g,n}$ and the associated Hilbert space are given below:
\begin{equation}
\partial\Sigma_{g,n} = \underbrace{S^1 \sqcup S^1 \sqcup \ldots \sqcup S^1}_n \quad;\quad \mathcal{H}_{\partial\Sigma_{g,n}} = \bigotimes_{j=1}^n \mathcal{H}_{S^1} ~.
\end{equation}
To each $\Sigma_{g,n}$, we can associate a state $\ket{\Sigma_{g,n}} \in \bigotimes_{j=1}^n \mathcal{H}_{S^1}$. The Hilbert spaces associated with $S^1$ are known \cite{Freed:1991bn} and can be explicitly obtained as (see the lectures \cite{Freed1993}):
\begin{equation}
\mathcal{H}_{S^1} = \{\text{function } f: G \to \mathbb{C} \,|\, \text{$f$ is invariant under conjugation}  \} ~.
\end{equation}
Thus the dimension of each Hilbert space is given as,
\begin{equation}
\text{dim}\, \mathcal{H}_{S^1} = \text{$\#$ of irreps of group $G$} ~.
\end{equation}
The entanglement structure of the state $\ket{\Sigma_{g,n}}$ associated with $\Sigma_{g,n}$ can be studied using the replica trick. First consider $\ket{\Sigma_{g,2}}$. The state and the corresponding total density matrix are given below:
\begin{equation}
\ket{\Sigma_{g,2}} = \begin{array}{c}
\includegraphics[width=0.20\linewidth]{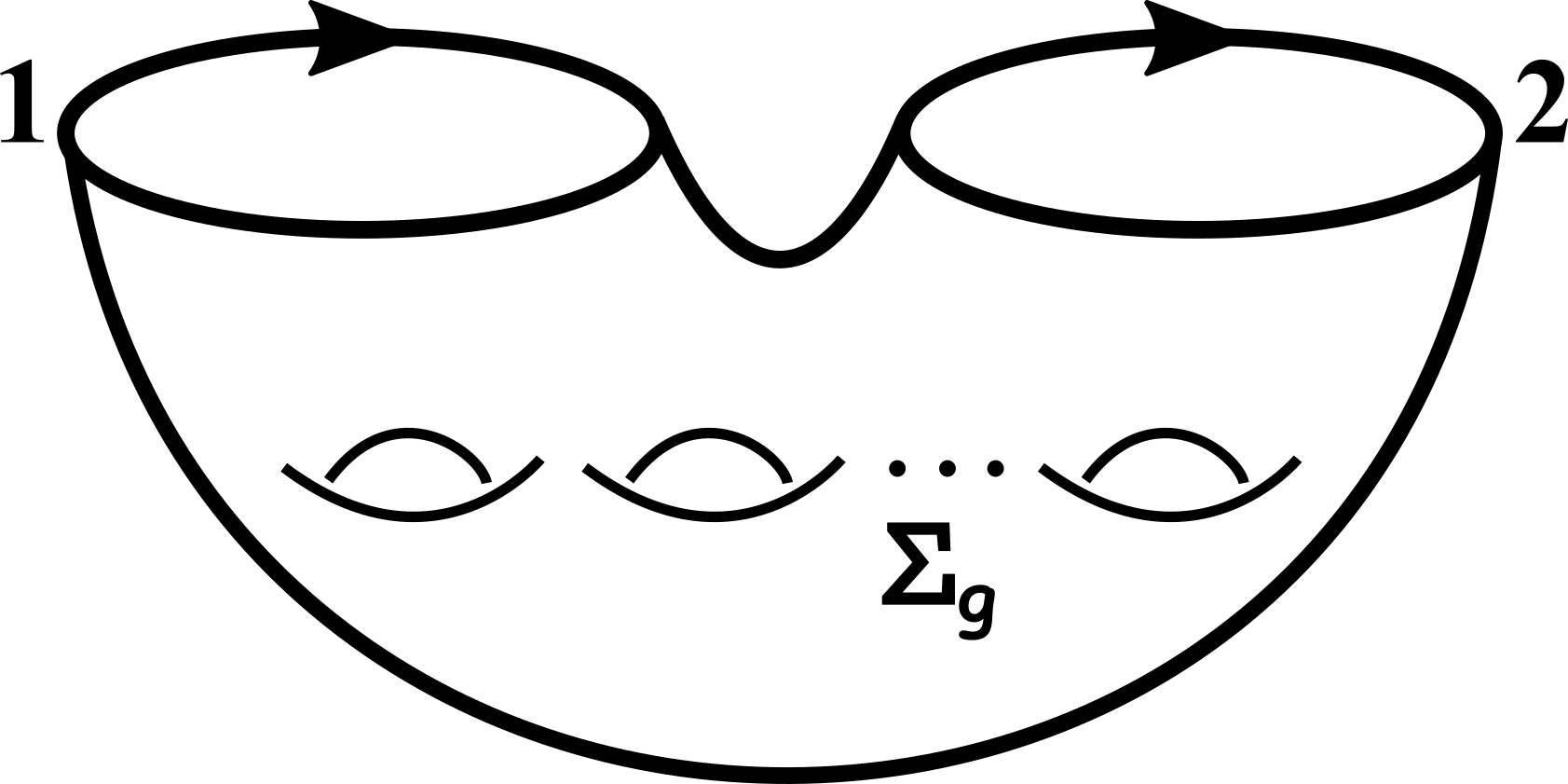}
\end{array} \quad;\quad \rho_{\text{total}} = \begin{array}{c}
\includegraphics[width=0.45\linewidth]{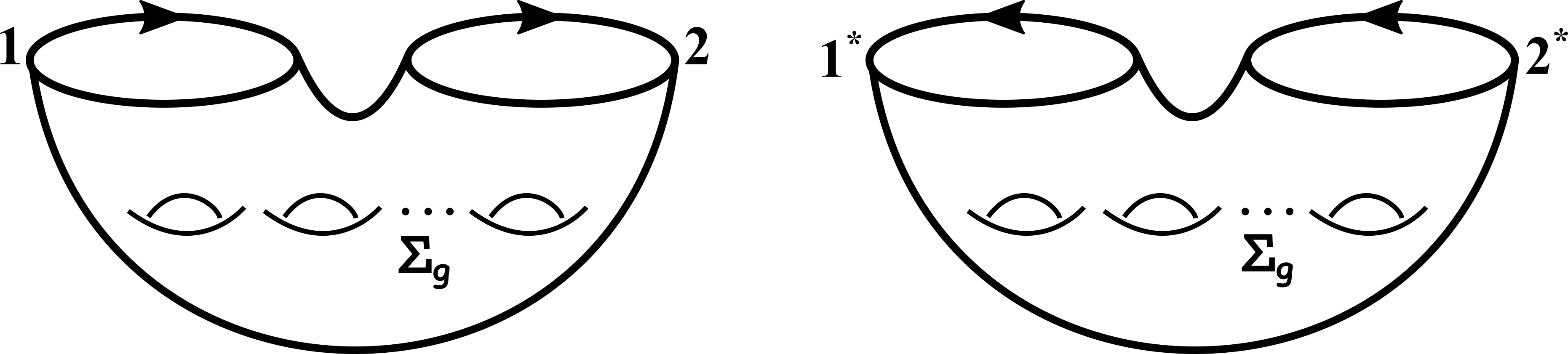}
\end{array}  ~.
\end{equation}
To obtain the entanglement structure, we trace out the Hilbert space associated with the second boundary. The reduced density matrix $\rho$ and its power $\rho^p$ can be computed as:
\begin{equation}
\rho = \frac{1}{Z(\Sigma_{2g+1})} \times \begin{array}{c}
\includegraphics[width=0.24\linewidth]{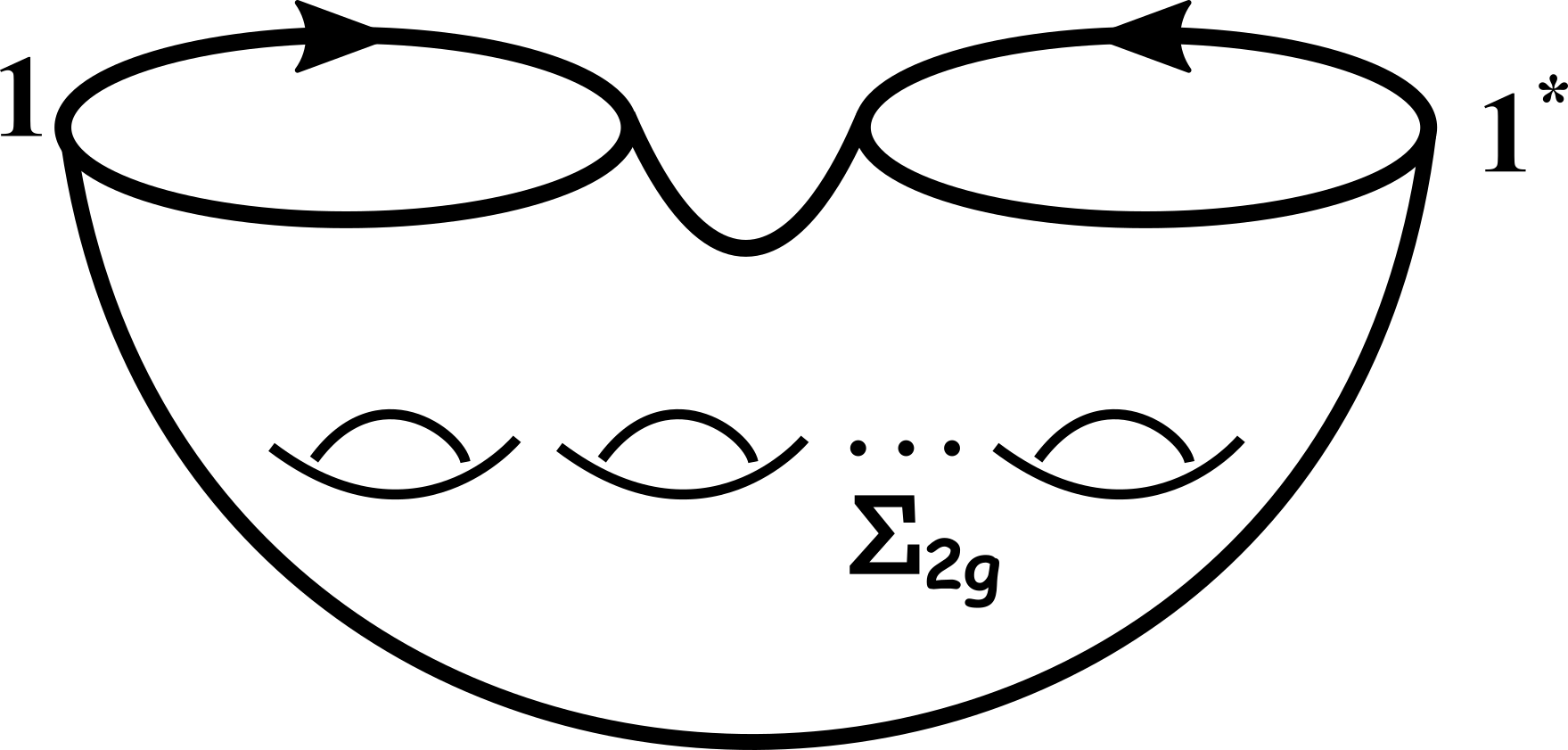}
\end{array}\, \Longrightarrow \, \rho^p = \frac{1}{Z(\Sigma_{2g+1})^p} \times \begin{array}{c}
\includegraphics[width=0.24\linewidth]{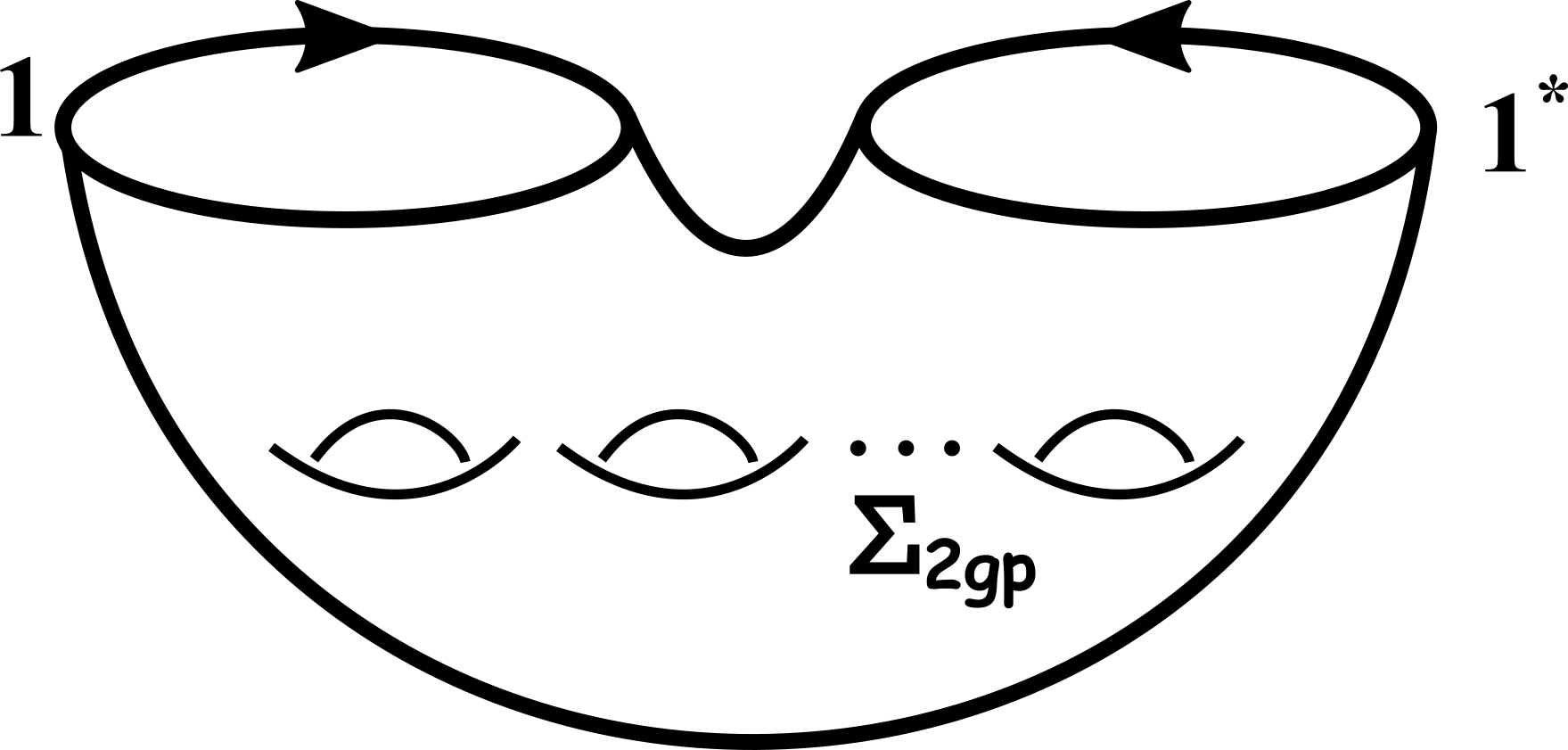}
\end{array} ~, \nonumber
\end{equation}
where the normalization factor is $\braket{\Sigma_{g,2}} = Z(\Sigma_{2g+1})$. Using the replica trick given in eq.(\ref{replica-trick-redrho}), we can compute the entanglement entropy in terms of partition functions of closed manifolds as,
\begin{equation}
\text{EE} = -\lim_{p \to 1} \frac{d}{dp} \left(\frac{Z(\Sigma_{2gp+1})}{Z(\Sigma_{2g+1})^p}\right) ~.
\end{equation}
 
Let us compute one more example for the manifold $\Sigma_{g,3}$ before giving the general result. The corresponding state and the normalization factors are,
\begin{equation}
\ket{\Sigma_{g,3}} = \begin{array}{c}
\includegraphics[width=0.25\linewidth]{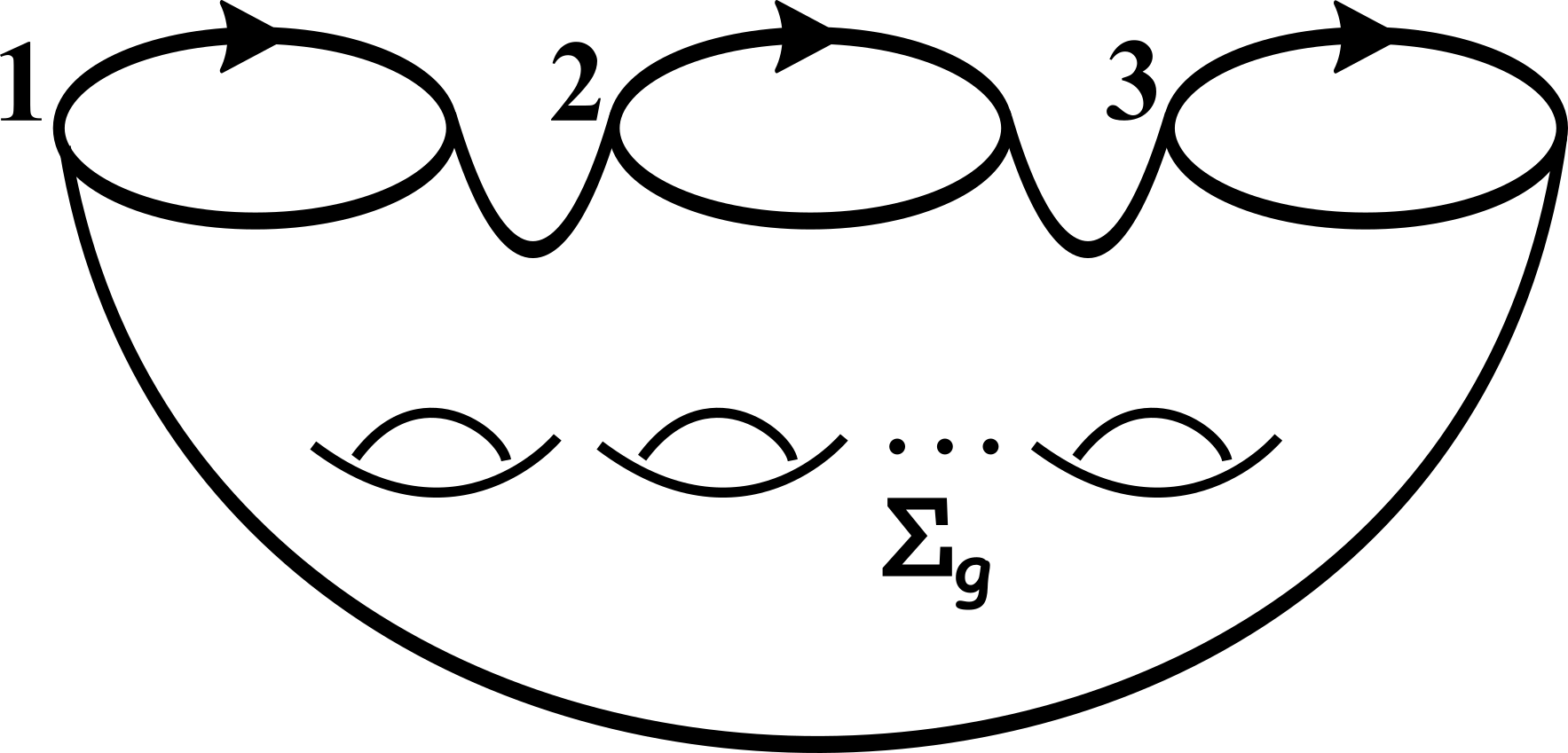}
\end{array} \quad;\quad \braket{\Sigma_{g,3}} = Z(\Sigma_{2g+2})  ~.
\end{equation}
Now let us first trace out the Hilbert space associated with boundary 3. The reduced density matrix and its power are given below:
\begin{equation}
\rho = \frac{1}{Z(\Sigma_{2g+2})} \times \begin{array}{c}
\includegraphics[width=0.24\linewidth]{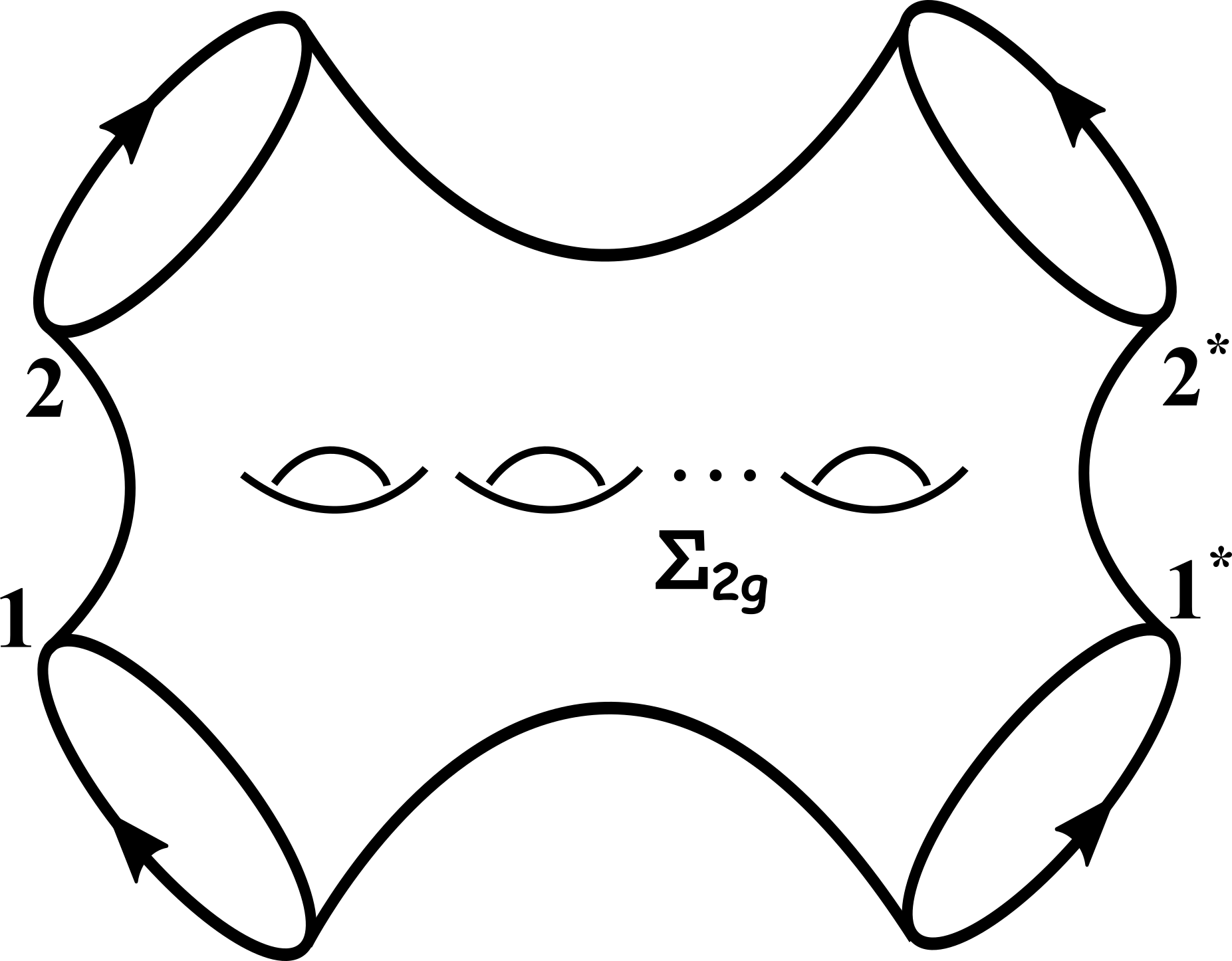}
\end{array}\, \Longrightarrow \, \rho^p = \frac{1}{Z(\Sigma_{2g+2})^p} \times \begin{array}{c}
\includegraphics[width=0.24\linewidth]{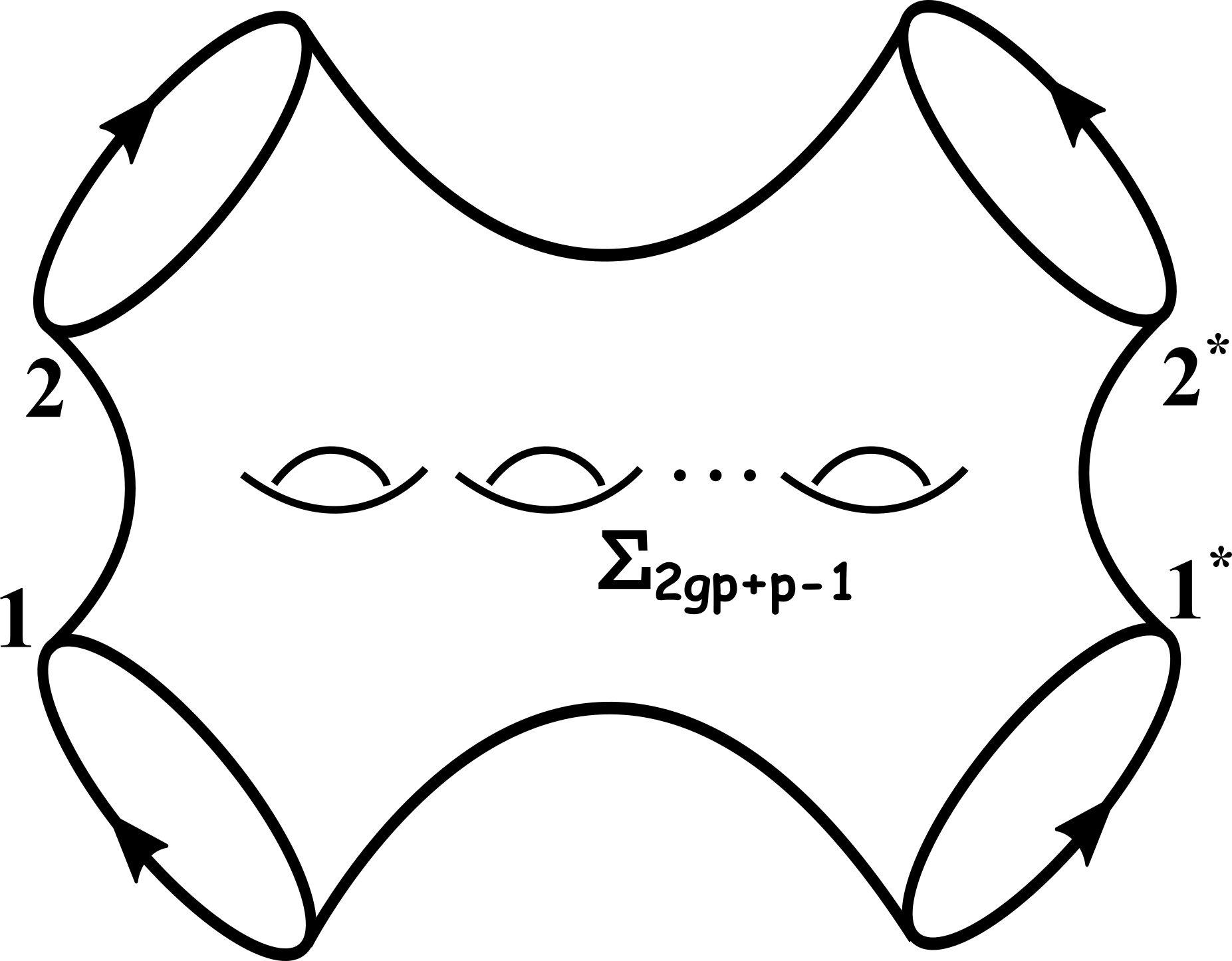}
\end{array} ~. \nonumber
\end{equation} 
The entanglement entropy can therefore be computed by obtaining the trace of $\rho^p$ and we get:
\begin{equation}
\text{EE} = -\lim_{p \to 1} \frac{d}{dp} \left(\frac{Z(\Sigma_{2gp+p+1})}{Z(\Sigma_{2g+2})^p}\right) ~.
\end{equation}  
Further it can be verified that we will get the same entropy if we trace out the Hilbert spaces associated with boundaries 2 and 3. This is true in general and we find that the entropy is independent of the choice of the bi-partition, i.e. the entropy obtained for the bi-partition $(m|n-m)$ does not depend on $m$. Further from the above reduced density matrix, we see that switching the boundaries $2$ and $2^*$ does not change the topology of the manifold. Thus the trace of the $q^{\text{th}}$ power (where $q$ is even) of the partially transposed reduced density matrix will be same as $\rho^q$. Thus the negativity associated with this reduced density matrix will vanish:
\begin{equation}
\mathcal{N}_{\text{log}} = \lim_{q \to 1} \ln \left( \text{Tr}(\rho_{12}^{\Gamma_2})^q \right) = \lim_{q \to 1} \ln \left(\frac{Z(\Sigma_{2gq+q+1})}{Z(\Sigma_{2g+2})^q}\right) = 0 ~.
\end{equation} 
This will also be true for any reduced density matrix of the general state $\ket{\Sigma_{g,n}}$ and over partial transpose of any part of the tri-partition. Thus we can now give our general result:   
\begin{equation}
\boxed{\text{EE}_{(m|n-m)}(\Sigma_{g,n}) =  -\lim_{p \to 1} \frac{d}{dp} \left(\frac{Z(\Sigma_{2gp-2p+np+1})}{Z(\Sigma_{2g+n-1})^p}\right) \quad,\quad \mathcal{N}(\rho_{\text{reduced}}) = 0 }  ~.
\label{EE-2d}
\end{equation}
This is the entanglement entropy for the states prepared in (1+1)-$d$ Chern-Simons theory. The state $\ket{\Sigma_{0,2}} \in \mathcal{H}_{S^1} \otimes \mathcal{H}_{S^1}$ is maximally entangled and hence a Bell state:
\begin{equation}
\boxed{\text{EE}(\Sigma_{0,2}) = \ln Z(\Sigma_{1}) = \ln \text{dim}\, \mathcal{H}_{S^1} \quad \Longrightarrow \quad \ket{\Sigma_{0,2}} = \text{Bell state}}~.
\end{equation} 
To quantify the entropy given in the formula in eq.(\ref{EE-2d}), we have to compute the partition functions of closed Riemann surfaces of genus $g$. In the case of untwisted Chern-Simons theory, i.e. $\alpha=0$, these partition functions can be obtained in terms of the group theoretic properties of the finite gauge group $G$ as \cite{Freed1993}:
\begin{equation}
Z(\Sigma_{g}) = |G|^{2g-2} \sum_i (\text{dim} \, R_i)^{2-2g} ~,
\label{Z-(1+1)CS}
\end{equation}
where $R_i$ denote the irreducible representations of the group $G$ and $\text{dim} \, R_i$ are their dimensions. Using this, the entropy can be explicitly computed as,
\begin{equation}
\boxed{\text{EE}(\Sigma_{g,n}) =  \ln \left( \sum_i (\text{dim} \, R_i)^{2\chi} \right) - 2\chi \, \frac{\sum_i (\text{dim} \, R_i)^{2\chi} \ln(\text{dim} \, R_i) }{\sum_i (\text{dim} \, R_i)^{2\chi}}}  ~,
\label{EE-2d-formula}
\end{equation}
where $\chi \equiv (2-2g-n)$ is the Euler characteristic of $\Sigma_{g,n}$ and is a negative integer, i.e. $\chi \leq 0$ (since we are dealing with $g\geq0$ and $n\geq2$). This entanglement entropy only depends on the coarser group theoretic details of the gauge group, the dimensions of irreducible representations. For various finite groups considered in this paper, we find that when the Euler characteristic $\chi \ll -1$ (with finite order of the group), the entropy converges and the limiting value is given as, 
\begin{equation}
\boxed{\lim_{|\chi| \to \infty} \text{EE} = \ln c(G)} ~,
\end{equation}
where $|\chi|$ denotes the absolute value of Euler characteristic and the constant $c(G)$ is an integer whose value depends on the choice of the gauge group. We give some examples in the following subsections. 
\subsection{Abelian group}
For abelian groups, all the irreducible representations have dimension 1, hence the entropy is simply given as:
\begin{equation}
\boxed{\text{EE}(\Sigma_{g,n}) = \ln |G| = \ln(\text{$\#$ of irreps of abelian group $G$}) = \ln \text{dim}\, \mathcal{H}_{S^1} } ~.
\label{abelianEE}
\end{equation}
This is the maximum possible entropy and is equal to logarithm of the dimension of the Hilbert space $\mathcal{H}(S^1)$. Further, the entropy does not capture the topological information (the Euler characteristic $\chi$) and is same for all Riemann surfaces.
In the case of non-abelian groups, the entropy depends on the topological information of the 2-manifold as well as the irreducible representations of the gauge group. Some examples of non-abelian groups are given below. 
\subsection{Non-abelian group}
\begin{figure}[htbp]
	\centering
		\includegraphics[width=1.05\textwidth]{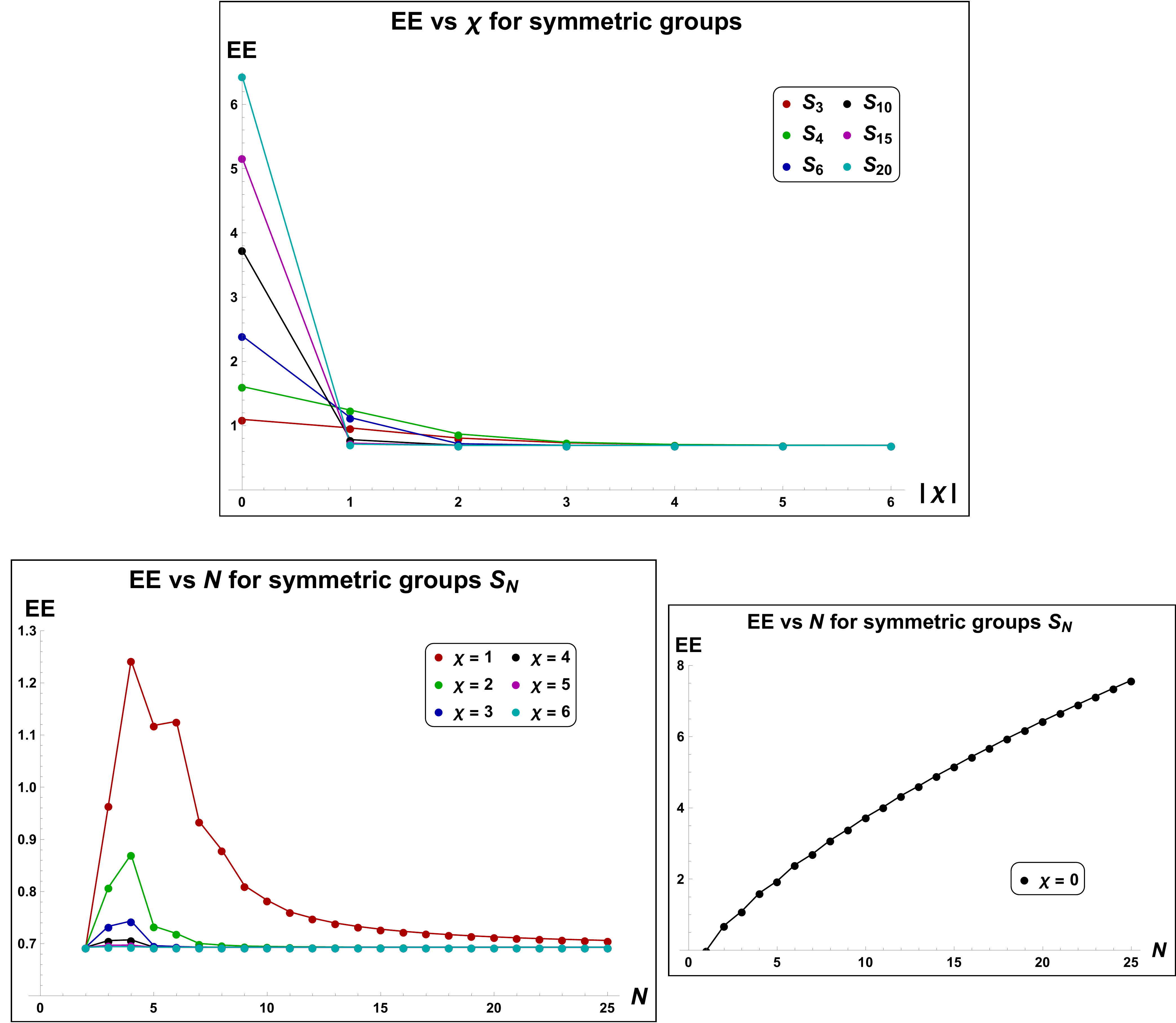}
	\caption{The variation of entropy for $S_N$ groups. For finite $N$, the entropy monotonically decreases and converge to $\ln 2$ as $|\chi| \to \infty$. The other two plots show the variation of entropy with $N$ for finite $\chi$. The $\chi=0$ corresponds to maximum entropy.}
	\label{EEplot-SN-group}
\end{figure}

\begin{figure}[htbp]
	\centering
		\includegraphics[width=1.0\textwidth]{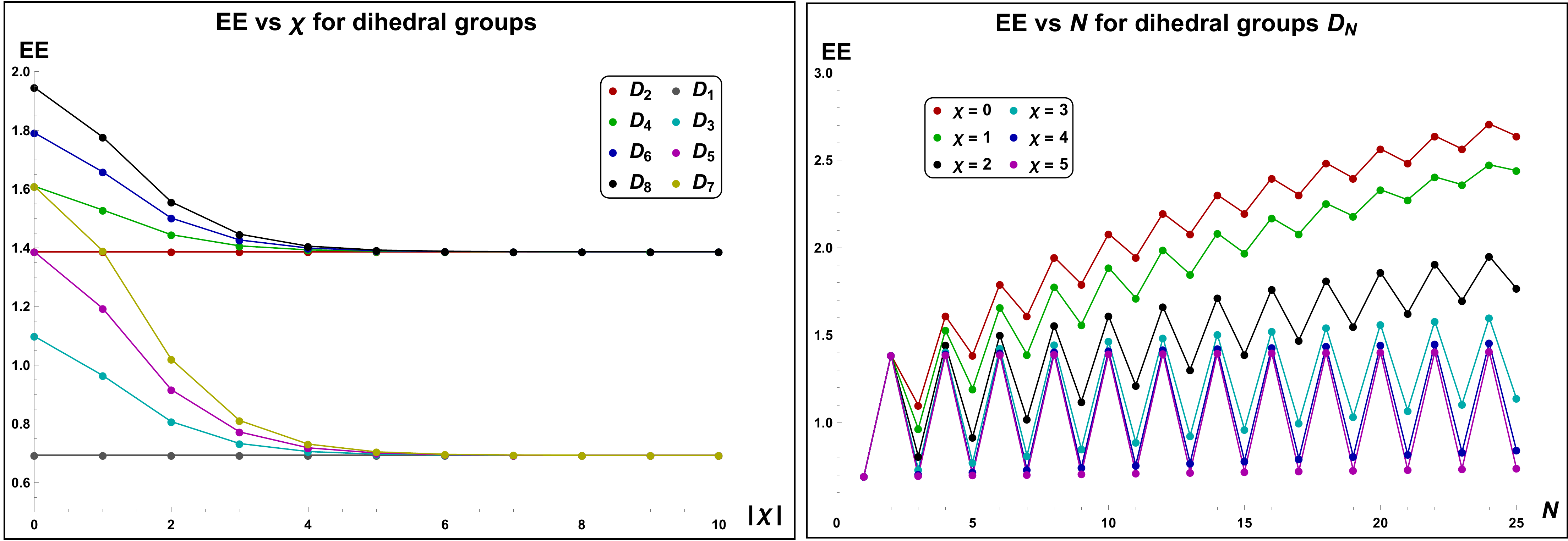}
	\caption{The variation of entropy with Euler characteristic $\chi$ and with $N$ for the dihedral group $D_N$. The asymptotic values follow the limits given in eq.(\ref{Limits-DN-group}).}
	\label{EEvsChi-DN-group}
\end{figure}

Let us first consider the symmetric group $S_N$ which is a non-abelian group for $N \geq 3$. The group theoretical details of $S_N$ can be found in the appendix. Unfortunately the close form formula for the dimensions of irreducible representation of $S_N$ are not known and hence we only present the entanglement entropy for some small values of $N$:
\begin{align}
\text{EE}(S_5) &= \ln \left(2^{1+4 \chi }+2\times 5^{2 \chi }+6^{2 \chi }+2\right)-\frac{2 \chi  \left(2^{4 \chi } \ln16 +5^{2 \chi}\ln25+6^{2 \chi } \ln6\right)}{2^{1+4 \chi }+2\times 5^{2 \chi }+6^{2 \chi }+2} \nonumber \\
\text{EE}(S_4) &= \ln \left(4^{\chi }+2\times 9^{\chi }+2\right) -\frac{2 \chi  \left(9^{-\chi } \ln2+4^{-\chi } \ln9\right)}{9^{-\chi } \left(2^{-2 \chi +1}+1\right)+2^{-2 \chi +1}}  \nonumber \\
\text{EE}(S_3) &= \ln \left(4^{\chi }+2\right)-\frac{2\chi  \ln2}{2^{-2 \chi +1}+1} ~.
\end{align}
Checking for several values of $N$, we find that:
\begin{equation}
\boxed{\lim_{|\chi| \to \infty} \text{EE}(S_N) = \ln 2} ~.
\label{largechi-SN}
\end{equation}
The variation of entanglement entropy with Euler characteristic $\chi$ and with $N$ are shown in the figure \ref{EEplot-SN-group}.
Next consider the dihedral groups $D_N$ of order $2N$ (see appendix for more details). The entropy results are given as:
\begin{equation}
\text{EE}(D_N) = \begin{cases} 
\ln \left(\frac{N}{2}+4^{-\chi +1}-1\right)+\left(\dfrac{4^{-\chi +2} \chi }{N+2^{-2 \chi +3}-2}\right)\ln2, & \text{for even $N$} \\[0.5cm]
\ln \left(N+4^{-\chi +1}-1\right)-\left(\dfrac{N+4^{-\chi +1}-2^{-2 \chi +3} \chi -1}{N+4^{-\chi +1}-1}\right)\ln2 , & \text{for odd $N$}
\end{cases} ~.
\end{equation}
From here, we can also find that,
\begin{equation}
\lim_{N \to \infty} \left(\frac{\text{EE}(D_N)}{\ln 2N}\right) = 1 \quad;\quad \lim_{|\chi| \to \infty} \text{EE}(D_N) = \begin{cases} 
\ln4, & \text{for even $N$} \\
\ln2, & \text{for odd $N$}
\end{cases} ~.
\label{Limits-DN-group}
\end{equation}
The variation of entanglement entropy with $\chi$ and with $N$ is shown in the figure \ref{EEvsChi-DN-group}.
\subsection{Entropy for direct product of groups}
If the gauge group is a direct product of two finite discrete groups, i.e. $G = G_1 \times G_2$, then the irreducible representations of $G$ are the tensor products of various irreducible representations of $G_1$ and $G_2$ respectively. In such a case, the partition functions evaluated for the group $G$ can be written as product of the partition functions computed for the individual groups. This can be directly seen from eq.(\ref{Z-(1+1)CS}):
\begin{equation}
Z_{G_1 \times G_2}(\Sigma_{g}) = Z_{G_1}(\Sigma_{g})\,\, Z_{G_2}(\Sigma_{g})  ~,
\end{equation}
where $Z_{G}(\Sigma_{g})$ denotes the partition function of closed Riemann surface of genus $g$ computed for group $G$. As a consequence, any Rényi entropy of order $p$ for the direct product of groups can be written as sum of the corresponding Rényi entropies for individual groups:
\begin{equation}
\text{RE}_p(G_1 \times G_2) = \frac{1}{(1-p)} \ln (\frac{Z_{G_1 \times G_2}(\Sigma_{2gp+1})}{Z_{G_1 \times G_2}(\Sigma_{2g+1})^p}) = \text{RE}_p(G_1) + \text{RE}_p(G_2) ~.
\end{equation}
Thus, the entanglement entropy will also follow this additive property:
\begin{equation}
\boxed{\text{EE}(G_1 \times G_2) = \text{EE}(G_1) + \text{EE}(G_2)} ~.
\end{equation}

So far we discussed the entanglement in (1+1)-$d$ Chern-Simons theory where the entanglement measures only depend on the dimensions of the irreps of the finite group. In the next section, we will consider the (2+1)-$d$ Chern-Simons theory where the states are associated with 3-manifolds whose boundaries are the Riemann surfaces. In this case, the entropy depends on the finer group theoretical details of the gauge group.  
\section{Entanglement structure of states in (2+1)-d Chern-Simons theory}
\label{sec4}
Let us now study the entanglement structure of the states in (2+1) dimensional Chern-Simons theory defined on 3-manifolds $M$ whose boundary $\partial M$ consists of multiple disjoint Riemann surfaces, i.e.
\begin{equation}
\partial M = \Sigma_{g_1} \sqcup \Sigma_{g_2} \sqcup \ldots \sqcup \Sigma_{g_n} ~,
\end{equation}
where according to our notations $\Sigma_{g}$ is a genus $g$ Riemann surface without boundary. Further we consider these surfaces to be oriented. The quantum state associated with such a manifold, which we denote as $\ket{M}$, lives in the tensor product of the Hilbert spaces associated with each boundary:
\begin{equation}
\ket{M} \in \mathcal{H}_{\Sigma_{g_1}} \otimes  \mathcal{H}_{\Sigma_{g_2}} \otimes \ldots \otimes \mathcal{H}_{\Sigma_{g_n}} ~.
\end{equation} 
Unlike the $\mathcal{H}_{S^1}$ which only depends on the irreducible representations of finite group $G$, the Hilbert spaces $\mathcal{H}_{\Sigma_{g}}$ capture more refined details of $G$, namely the conjugacy classes and the irreducible representations of centralizers of each conjugacy class. The dimensions of these Hilbert spaces are given as \cite{dijkgraaf1990,dijkgraaf1989operator}:
\begin{equation}
\text{dim} \, \mathcal{H}_{\Sigma_{g}} = \sum_{A, \, i} \left( \frac{|C_A|}{\text{dim} R_i^{A}}\right)^{2g-2} ~,
\end{equation}
where $A$ labels the conjugacy class of $G$ and $C_A$ denotes the centralizer of some representative element $a \in A$.\footnote{The conjugacy class $A$ of an element $a \in G$ is defined as: $A= \{b\in G \,|\, b = gag^{-1} \text{ for some } g\in G \}$. The centralizer $C_a$ on the other hand is a subgroup of $G$ defined as: $C_a= \{g\in G \,|\, ga =ag \}$. For any two elements $a, a' \in A$ in the same conjugacy class, the centralizers $C_a$ and $C_{a'}$ are isomorphic. Hence instead of $C_a$, we will simply denote $C_A$ to be the centralizer corresponding to any element in the conjugacy class $A$.} The irreducible representations of $C_A$ are given by $R_i^{A}$. Note that the Hilbert space associated with $S^2$ (genus 0 surface) is one dimensional. So introducing $S^2$ boundaries in the manifold will be equivalent to taking tensor products of one dimensional Hilbert spaces which will not affect the entanglement structures.\footnote{It is possible to get non-trivial Hilbert spaces if one considers $S^2$ with punctures where the punctures are connected by Wilson lines embedded in the bulk of $M$. See for example \cite{Dwivedi:2019bzh}.} The Hilbert space $\mathcal{H}_{T^2}$ associated with the torus $T^2$ however, is non trivial and in the later subsections we will study the states which live in the tensor product of multiple copies of $\mathcal{H}_{T^2}$. We leave the discussion of entanglement of states living in higher genus Hilbert spaces to future work.

The simplest state which can be prepared in this set-up is a state which can be defined on two identical boundaries related by cobordism, i.e. the manifold under consideration is $M = \Sigma_{g} \times [0,1]$. The entanglement entropy for this state can be obtained for any genus $g$ using the replica trick which we discuss in the following. 
\subsection{Cobordant boundaries as maximally entangled state}
The simplest example is to have two identical boundaries related by cobordism, i.e. we are dealing with a manifold which is of the form:
\begin{equation}
M = \Sigma_{g} \times I \quad;\quad \partial M = \Sigma_{g} \sqcup \Sigma_{g}^* ~,
\end{equation}
where $I=[0,1]$ is an interval. This manifold has two identical boundary components differing in their orientations. Topologically this 3-manifold is a cylinder whose two ends have boundaries $\Sigma_{g}$ and $\Sigma_{g}^*$. The corresponding state and the normalization factor are given below:
\begin{equation}
\ket{\Sigma_{g} \times I} = \begin{array}{c}
\includegraphics[width=0.20\linewidth]{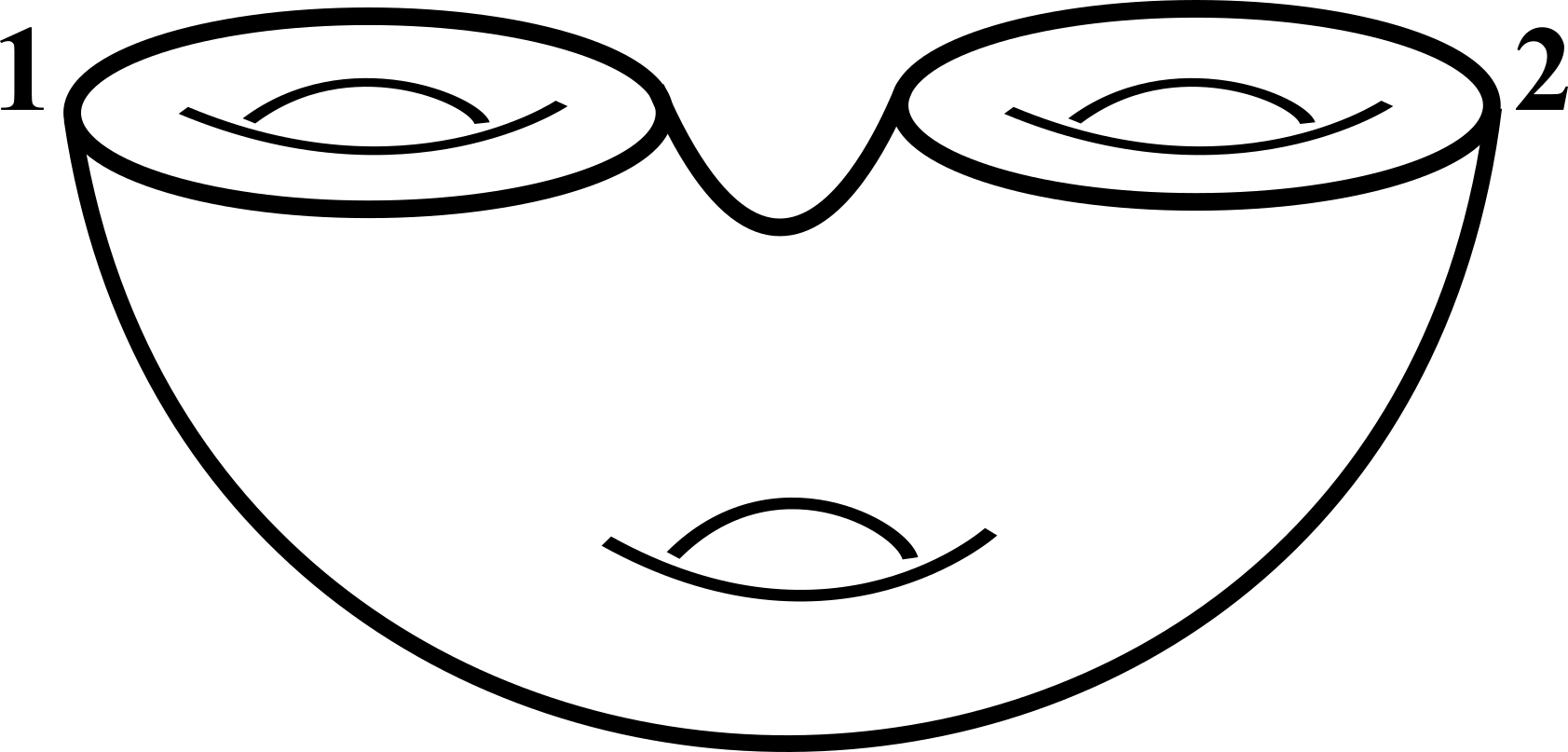}
\end{array} \,;\, \bra{\Sigma_{g} \times I}\ket{\Sigma_{g} \times I} = Z(\Sigma_{g} \times S^1) = \text{dim} \, \mathcal{H}_{\Sigma_{g}} ~.
\end{equation}
Following the earlier examples, it is a trivial exercise to trace out the Hilbert space of one of the boundaries and get the following entanglement entropy:
\begin{equation}
\text{EE} = \ln \text{dim} \, \mathcal{H}_{\Sigma_{g}} ~.
\end{equation}
For various finite groups considered in this paper, we find that when the order of the group is very large (with finite genus), the entropy goes as logarithm of the order of the group. However when genus $g$ is very large (with finite order of the group), the entropy goes linearly with $g$. The limiting values according to our observations are given as, 
\begin{equation}
\boxed{\lim_{|G| \to \infty} \left(\frac{\text{EE}}{\ln |G|}\right) = 2g \quad;\quad \lim_{g \to \infty} \left(\frac{\text{EE}}{g}\right) = 2\ln |G|} ~,
\label{EE-cobordant}
\end{equation}
For abelian groups, the entropy is,
\begin{equation}
\text{EE(abelian)} = 2g\ln|G| ~.
\end{equation}
For the dihedral groups $D_N$, the entropy can be given in a close form as,
\begin{equation}
\text{EE}(D_N) = \begin{cases} \ln\left(\dfrac{N^{2 g}+4^{g+1} N^{2 g-2}-4 N^{2 g-2}+16^g}{2}\right), &\text{for even $N$} \\[0.5cm] \ln\left(\dfrac{N^{2 g}+4^g N^{2 g-2}-N^{2 g-2}+4^g}{2}\right), &\text{for odd $N$} \end{cases} ~.
\end{equation}
For the symmetric group $S_N$, we do not have a close form for a general $N$. For lower order, the values are:
\begin{align}
\text{EE}(S_5) &= \ln\left( \tfrac{2250\times 16^g+450\times 64^g+1440\times 25^g+200\times 144^g+2800\times 36^g+18\times 400^g+16\times 900^g+25\times 576^g+14400^g}{7200} \right) \nonumber \\[0.2cm]
\text{EE}(S_4) &=  \ln\left(\tfrac{162\times 16^g+96\times 9^g+2\times 144^g+27\times 64^g+576^g}{288}\right)\quad;\quad \text{EE}(S_3) = \ln\left( \tfrac{9\times 4^g+8\times 9^g+36^g}{18}\right) ~.
\end{align}
Moreover, using the group theory, we can show that the dimension of the Hilbert space for a direct product of groups is the product of dimensions of Hilbert spaces for individual groups:
\begin{equation}
\text{dim} \, \mathcal{H}_{\Sigma_{g}}(G_1 \times G_2) = \text{dim} \, \mathcal{H}_{\Sigma_{g}}(G_1) \,\, \text{dim} \, \mathcal{H}_{\Sigma_{g}}(G_2) ~.
\end{equation}
Using this, we can again see that the entropy follows:
\begin{equation}
\boxed{\text{EE}(G_1 \times G_2) = \text{EE}(G_1) + \text{EE}(G_2)} ~.
\end{equation}

Note that the state $\ket{\Sigma_{g} \times [0,1]}$ is a maximally entangled state and is analogous to the Bell pair of quantum information theory. In the following sections, we will study the entanglement structures of more general states in which the boundaries are not related by cobordisms. We will also consider the multi-party state defined on three or more disjoint boundaries where we can study the finer entanglement features corresponding to bi-partitioning and tri-partitioning of the total Hilbert space. We will restrict to the case where each boundary is torus. Thus, let us pause for a moment and discuss briefly about the Hilbert space $\mathcal{H}_{T^2}$ associated with a torus and its basis elements which will be used later on to construct the explicit states.
\subsection{Basis of $\mathcal{H}_{T^2}$ and irreps of quantum double group}
The basis states of $\mathcal{H}_{T^2}$ for a given gauge group $G$ are given by the irreducible representations of a quantum group $Q(G)$ associated with $G$. The irreps of $Q(G)$ are in one-to-one correspondence with the primary fields of certain two-dimensional conformal field theories (CFT). For example, when $G$ is a simply connected Lie group (for example SU($N$) group), we get a 2d WZW (Wess–Zumino–Witten) model whose symmetry algebra is the affine Lie algebra $\hat{\mathfrak{g}}_k$ built from the Lie algebra $\mathfrak{g}$ associated with $G$. The quantum group is therefore $Q(G) = \hat{\mathfrak{g}}_k$ where the level $k$ of the algebra is a non-negative integer $k$. The basis of Hilbert space $\mathcal{H}_{T^2}$ in this case is given by the integrable representations of $\hat{\mathfrak{g}}_k$. For example, when $\mathfrak{g} = A_r$, the integrable representations of $\hat{\mathfrak{g}}_k$ are given by those $R=(a_1, a_2, \ldots, a_r)$ of $\mathfrak{g}$ which satisfy $a_1+a_2+\ldots+a_r \leq k$ where $a_i$ are the Dynkin labels of the highest weight of representation $R$. The integrable representations for all affine classical and exceptional Lie algebras are known (see the appendix of \cite{Dwivedi:2017rnj} for an explicit counting). Thus in this case, a basis of $\mathcal{H}_{T^2}$ will be simply $\ket{R}$ labeled by the integrable representation $R$ of $\hat{\mathfrak{g}}_k$. This basis has a topological definition and is given by the Chern-Simons partition function of a solid torus with a Wilson line transforming under $R$ placed in the bulk of solid torus along its non-contractible homology cycle. It is shown in eq.(\ref{basis-HS}).
\begin{equation}
\text{basis}(\mathcal{H}_{T^2}) = \{\ket{R}: \text{$R$ is integrable rep of $\hat{\mathfrak{g}}_k$} \} \quad;\quad \ket{R} = \begin{array}{c}
\includegraphics[width=0.13\linewidth]{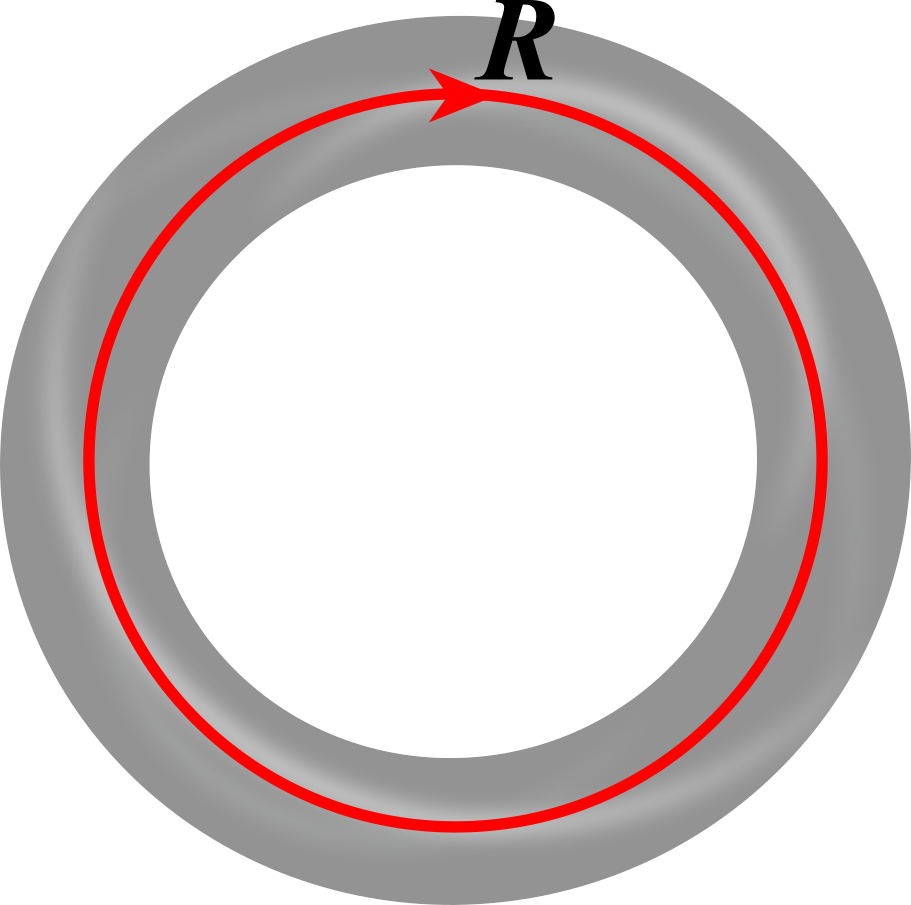}
\end{array} ~. 
\label{basis-HS}
\end{equation}
The collection of all such states for various possible integrable representations of $\hat{\mathfrak{g}}_k$ will form an orthonormal basis of the $\mathcal{H}_{T^2}$ as mentioned in eq.(\ref{basis-HS}).

Since we are dealing with a finite discrete gauge group $G$ in the present work, it will be useful to know the irreducible representations of the associated quantum group. It is known that (2+1)-$d$ Chern-Simons theory with finite gauge group corresponds to 2d rational CFT and the quantum group in this case is $D(G)$, the Drinfeld double group \cite{drinfeld1990quasi} or the quantum double group.\footnote{The group $D(G)$ here is for the untwisted (2+1)-$d$ Chern-Simons theory, i.e. for $\alpha = 0$ case as given in eq.(\ref{alpha=0-action}). For more general theories, the quantum group will be the `twisted' quantum double group denoted as $D_{\alpha}(G)$ where the cohomological twist $\alpha \in H^{3}(BG, U(1))$ is analogous to the level $k \in \mathbb{Z}$ for the quantum groups $\mathfrak{g}_k$ associated with simply connected Lie groups. For our present work, we will restrict to $D(G)$.}  The irreducible representations of $D(G)$ are labeled by pairs $(A, R_A)$ where $A$ is the conjugacy class of $G$ and $R_A$ is the irreducible representation of the centralizer $C_A$ of any representative element $a \in A$. Hence, in our set-up of finite group, a basis element of the Hilbert space $\mathcal{H}_{T^2}$ is labeled as $\ket{(A, R_A)}$. The collection of all such basis elements will form an orthonormal basis of the $\mathcal{H}_{T^2}$:
\begin{equation}
\text{basis}(\mathcal{H}_{T^2}) = \{\,\ket{(A, R_A)}: \text{$A$ is conjugacy class of $G$ and $R_A$ is an irrep of $C_A$} \}  ~. 
\label{basis-HS-finiteG}
\end{equation}
We further believe that this basis can be given a topological description similar to that given in eq.(\ref{basis-HS}) with the representation $R$ now playing the role of irrep of $D(G)$ and the topological machinery of gluing oppositely oriented boundaries can be carried out as usual. The Hilbert space $\mathcal{H}_{T^2}$ is finite dimensional with the dimension being given as:
\begin{equation}
\text{dim}(\mathcal{H}_{T^2}) = \sum_{A} \#(\sigma_A) ~,
\end{equation}
where the sum is over all the conjugacy classes of $G$ and $\#(\sigma_A)$ is the number of irreps of $C_A$. For an abelian group, it is trivial to see that $\text{dim}(\mathcal{H}_{T^2}) = |G|^2$. 
 
In order to compute the entanglement structure of the quantum states in our set-up, we also need the modular data, i.e. the generators $\mathcal{S}$ and $\mathcal{T}$ of the unitary representation of the modular group SL(2, $\mathbb{Z}$).  We briefly discuss them in the following: 
\subsubsection{Mapping class group of torus and the generators $\mathcal{S}$ and $\mathcal{T}$}
\begin{figure}[h]
\centerline{\includegraphics[width=4.4in]{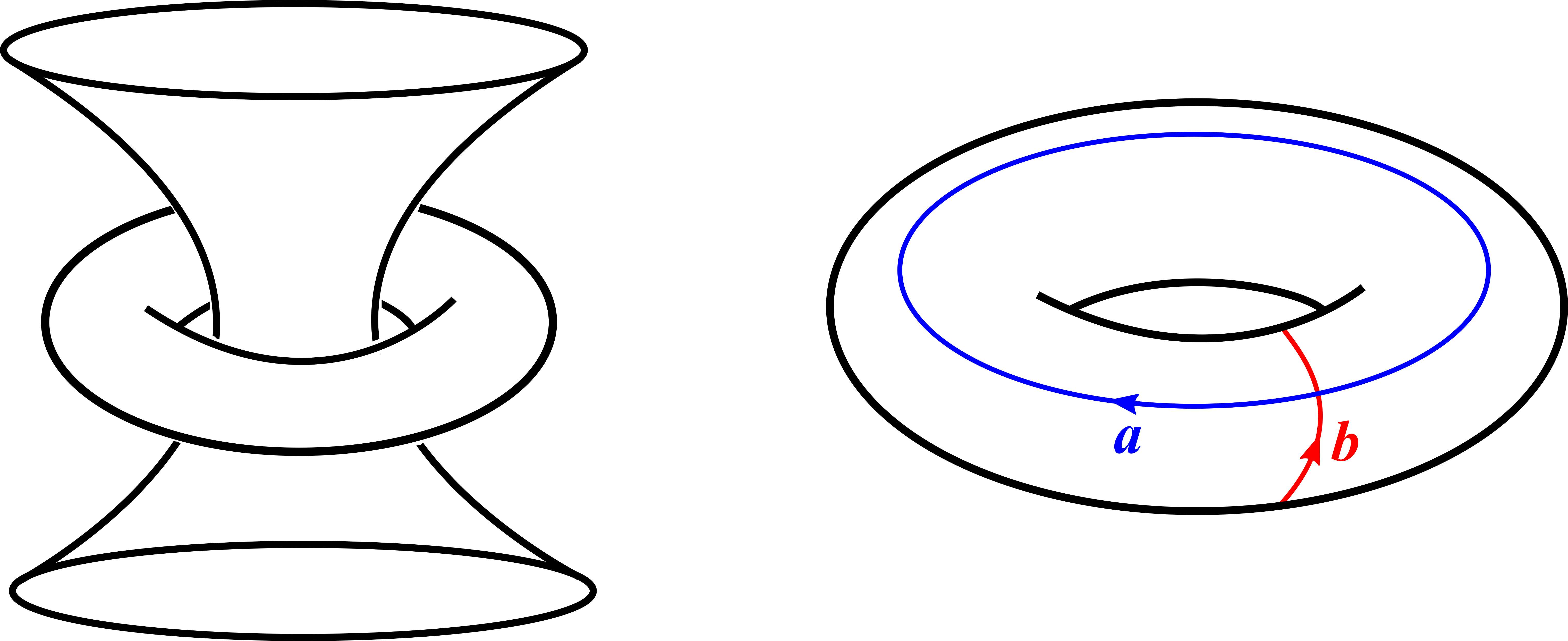}}
\caption[]{Heegaard splitting of $S^3$ into two solid tori shown on the left. The figure on the right shows the two cycles of a torus which form its homology basis.}
\label{Heegaard}
\end{figure}

It is known that the Chern-Simons partition function of any 3-manifold can be obtained from $S^3$ via a specific surgery. In such a surgery, we first perform the Heegaard splitting of $S^3$ by cutting $S^3$ along a Riemann surface $\Sigma_g$. This decomposes $S^3$ into two handle-bodies of genus $g$. For example, the splitting of $S^3$ into two solid tori is shown in the figure \ref{Heegaard}.
Next, we act on the boundary of one of the handle-body by a diffeomorphism operator 
\begin{equation}
\hat{O}: \Sigma_g \longrightarrow \Sigma_g
\end{equation} 
and then glue back the two handle-bodies along $\Sigma_g$. This will result in a closed manifold whose topology will depend on the choice of the operator $\hat{O}$. These operators are given in terms of the generators of the mapping class group MCG($\Sigma_g$) of $\Sigma_g$. In the present case, we are interested in $\Sigma_g = T^2$ for which the mapping class group is MCG($T^2$) $=$ SL(2, $\mathbb{Z}$). The unitary representation of SL(2, $\mathbb{Z}$) has two modular generators $\mathcal{S}$ and $\mathcal{T}$ which act as diffeomorphism operators  for $T^2$ by acting on its homology basis (which consists of the two cycles $a$ and $b$ as shown in the figure \ref{Heegaard}) as following:
\begin{equation}
\mathcal{S}: (a,b) \longrightarrow (b,-a) \quad;\quad  \mathcal{T}: (a,b) \longrightarrow (a,a+b) ~.
\end{equation}
These generators satisfy $\mathcal{S}^2: (a,b) \longrightarrow (-a,-b)$ and $(\mathcal{S}\mathcal{T})^3: (a,b) \longrightarrow (-a,-b)$ which simply reverses the orientations of the two cycles of the torus. Thus they satisfy the relation $S^2 = (ST)^3 = C$, where $C$ is commonly known as charge conjugation in physics literature which obeys $C^2 = \mathbb{1}$. For finite discrete groups, these generators can be explicitly written in a matrix form in the basis $\ket{(A, R_A)}$ of $\mathcal{H}_{T^2}$. Following \cite{Coste:2000tq}, their matrix elements are given as:\footnote{We would like to again remind the readers that these matrix elements are given for the untwisted Chern-Simons theory, i.e. for $\alpha = 0$. For a non-trivial twist $\alpha$, the matrix elements have to be modified as prescribed in \cite{dijkgraaf1990,Coste:2000tq}. For the present work, we only consider $\alpha = 0$ case.}
\begin{empheq}[box=\fbox]{align}
  \mathcal{S}_{(A,\,R_A) (B,\,R_B)} &= \frac{1}{|C_A||C_B|}\, \sum_{g\, \in Y_{AB}} \chi R_A(gbg^{-1})^{*} \, \chi R_B(g^{-1}ag)^{*} \nonumber \\
\mathcal{T}_{(A,\,R_A) (B,\,R_B)} &= \delta_{AB}\, \delta_{R_A R_B}\, \frac{\chi R_A(a)}{\chi R_A(e)} ~.
\label{SandT-matrices}
\end{empheq} 
Here $(A,\,R_A)$ and $(B,\,R_B)$ are the irreducible representations of $D(G)$ and label the row and column respectively of $\mathcal{S}$ and $\mathcal{T}$ matrices. The symbols $\chi R_A$ and $\chi R_B$ denote the characters of irreps $R_A$ and $R_B$ of centralizers $C_A$ and $C_B$ respectively with `$*$' being the complex conjugation. The elements $a \in A$ and $b \in B$ where $A$ and $B$ are the conjugacy classes of $G$. The notation $\chi R(x)$ simply means the character of the representation $R$ being evaluated for an element $x$. The set $Y_{AB}$ is defined as: 
\begin{equation}
Y_{AB} = \{g \in G \,|\, gbg^{-1} \in C_A \, \text{ and } \,  g^{-1}ag \in C_B  \} ~.
\end{equation}
In appendix \ref{appendix-A}, we have computed and tabulated these modular matrices for various finite groups. We will use these data to compute the states on the tensor product of $\mathcal{H}_{T^2}$ Hilbert spaces and to determine their entanglement measures. In the following sections, let us study the entanglement properties of the state associated with a 3-manifold which is the torus link complement.
\subsection{State associated with torus link complement}
\begin{figure}[h]
\centerline{\includegraphics[width=4.4in]{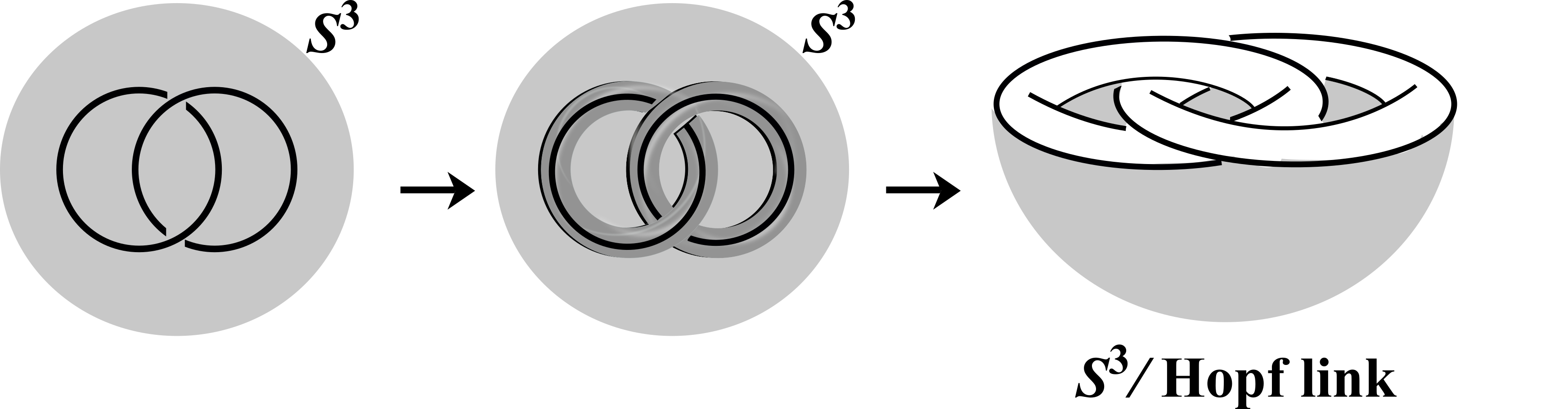}}
\caption[]{Figure showing the construction of Hopf link complement. Removing a tubular neighborhood around Hopf link embedded in $S^3$ results in Hopf link complement which is a manifold with two torus boundaries.}
\label{HLcomplement}
\end{figure}

In \cite{Balasubramanian:2016sro,Dwivedi:2017rnj,Balasubramanian:2018por}, the entanglement structure of the states were studied for the link complements. These are three dimensional manifolds obtained by removing a tubular neighborhood of a link $\mathcal{L}$ from $S^3$ and are usually denoted as $S^3/\mathcal{L}$. A typical manifold is shown in figure \ref{HLcomplement} which is the Hopf link complement.
We will denote the quantum state associated with $S^3/\mathcal{L}$ as $\ket{\mathcal{L}}$ which will depend upon the topology of the link involved. If the link $\mathcal{L}$ consists of $n$ number of knot components, then the corresponding link state lives in the following Hilbert space:   
\begin{equation}
\ket{\mathcal{L}} \in \bigotimes_{i=1}^{n} \mathcal{H}_{T^2} ~.
\end{equation}
It was further shown in \cite{Balasubramanian:2016sro} that these states can be explicitly constructed by doing surgery along $\mathcal{L}$ and are given as:  
\begin{equation}
\ket{\mathcal{L}} = \sum_{a_1,\, a_2,\,\ldots,\,a_n} Z(S^3;\,\mathcal{L}[a_1, a_2,\ldots, a_n]) \ket{a_1, a_2, \ldots, a_n} ~,
\end{equation}
where $\ket{a_i}$ is the basis of the $i^{\text{th}}$ Hilbert space and $Z(S^3;\,\mathcal{L}[a_1, a_2,\ldots, a_n])$ is the Chern-Simons partition function of $S^3$ in presence of link $\mathcal{L}$ whose $i^{\text{th}}$ knot component carries representation $a_i$. 

One special class of links is the so-called torus link for which the partition function can be determined in terms of modular generators $\mathcal{S}$ and $\mathcal{T}$. We will use the notation $T_{p,q}$ to denote a generic torus link. This link consists of $n=$ gcd($p,q$) number of linked circles and each circle winds $\frac{p}{n}$ and $\frac{q}{n}$ number of times along the two homology cycles of $T^2$. Topologically $T_{p,q}$ and $T_{q,p}$ are equivalent, hence without loss of generality we will assume $q \geq p$. The partition function of $S^3$ in presence of a torus link is known in the literature for simply connected Lie groups \cite{Stevan:2010jh,Brini:2011wi}. For example, for SU($N$) gauge group, it is given as:
\begin{equation}
Z(S^3;\,T_{p,q}[a_1, a_2,\ldots, a_n]) = \sum_{\alpha, \beta, \gamma}  \frac{\mathcal{S}_{\alpha \beta}^* \,\mathcal{S}_{0 \gamma}}{(\mathcal{S}_{0 \alpha})^{n-1}}\, (\mathcal{T}_{\gamma \gamma})^{q/p} \left(\prod_{i=1}^n \mathcal{S}_{a_i \alpha}\right)  X_{\beta \gamma}(p/n) ~,
\end{equation}
where $a_1,\ldots,a_n$ and $\alpha, \beta, \gamma$ are the integrable representations of $\hat{\text{su}}(N)_k$. The coefficients $X_{\beta \gamma}(y)$ are unique integers which can be obtained by expanding the traces of powers of holonomy operator $U$:
\begin{equation}
\text{Tr}_{\beta}(U^m) = \sum_{\gamma} X_{\beta \gamma}(m)\, \text{Tr}_{\gamma}(U) ~.
\end{equation}
For SU($N$), these coefficients can be obtained by performing the Adams operation on the characters associated with the irreps of SU($N$) group (see \cite{Stevan:2014xma} and references therein).  Performing the surgery on the link $T_{p,q}$ following \cite{Balasubramanian:2016sro}, the corresponding link state on tensor product of $n$ copies of $\mathcal{H}_{T^2}$ can be given as,
\begin{equation}
\ket{T_{p,q}} = \sum_{a_1,\ldots,a_n} Z(S^3;\,T_{p,q}[a_1, a_2,\ldots, a_n]) \ket{a_1,\ldots,a_n} ~.
\end{equation}
Since the entanglement properties of a state does not change under the unitary transformation of basis states, we can further simplify this state by changing the $i^{\text{th}}$ basis state as: $\ket{a_i} = \sum_{y_i} \mathcal{S}_{a_i y_i}^{*}\ket{y_i}$. In this basis, the state can be written as,
\begin{equation}
\ket{T_{p,q}} = \sum_{a_1,\ldots,a_n} \, \sum_{y_1,\ldots,y_n} \left(\prod_{i=1}^n \mathcal{S}_{a_i y_i}^*\right) Z(S^3;\,T_{p,q}[a_1, a_2,\ldots, a_n]) \ket{y_1,\ldots,y_n} ~.
\end{equation}
Using the symmetric and unitary property of $\mathcal{S}$ matrix, the state can be simplified as,
\begin{equation}
\ket{T_{p,q}} =  \sum_{\alpha, \beta, \gamma}  \frac{\mathcal{S}_{\alpha \beta}^* \,\mathcal{S}_{0 \gamma}}{(\mathcal{S}_{0 \alpha})^{n-1}}\, (\mathcal{T}_{\gamma \gamma})^{q/p} \,  X_{\beta \gamma}(p/n) \, \ket{\alpha,\alpha,\ldots,\alpha} ~.
\end{equation}

In the context of finite gauge group, we can write an analogous state by taking the representations $\alpha, \beta, \gamma$ to be the irreps of quantum double group $D(G)$ and replace the $\mathcal{S}$ and $\mathcal{T}$ matrices by the modular $\mathcal{S}$ and $\mathcal{T}$ matrices of finite group given in eq.(\ref{SandT-matrices}). Moreover since the basis of $\mathcal{H}_{T^2}$ is determined by the irreps of the centralizer $C_A$ associated with a conjugacy class $A$ of the finite group $G$, we can obtain the unique integer coefficients $X_{\beta \gamma}(m)$ by performing the Adams operation within the centralizer subgroup of a particular conjugacy class: 
\begin{equation}
X_{\beta \gamma}(m) = Y_{R_{i}, R_{j}}(m) \, \delta_{A_\beta A_\gamma} ~.
\label{X-coeff}
\end{equation} 
In the above equation, the representations $R_{i}$ and $R_{j}$ are the irreps of centralizer $C_{A_\beta}$ associated with the conjugacy class $A_\beta$, i.e. we have labeled the irreps $\beta$ and $\gamma$ as $\beta = (A_\beta, C_{A_\beta})$ and $\gamma = (A_\gamma, C_{A_\gamma})$. With this definition, we can now compute $Y_{R_{i}, R_{j}}(m)$ which are determined by the Adams operation in $C_{A_\beta}$ defined as:
\begin{equation}
\Psi^m \chi_i (a) = \chi_i (a^m) = \sum_j Y_{R_{i}, R_{j}}(m) \, \chi_j (a) ~,
\end{equation}   
where the Adams operation is $\Psi^m$ defined on the character $\chi_i$ of irrep $R_i$ of $C_{A_\beta}$ and is evaluated at a representative element $a \in A_{\beta}$. Using the inner product
\begin{equation}
\bra{f_1}\ket{f_2} = \frac{1}{|G|} \sum_{g \in G} f_1(g)^*\,f_2 (g) ~,
\end{equation}
defined for any two class functions $f_1$ and $f_2$ of a  group $G$ and using the fact that the irreducible characters form an orthonormal basis, the coefficients $Y_{R_{i}, R_{j}}(m)$ can be uniquely determined which turns out to be integers and are given as:
\begin{equation}
Y_{R_{i}, R_{j}}(m) = \frac{1}{|C_{A_\beta}|} \sum_{g \in C_{A_\beta}} \chi_i (g^{m})\,\chi_j (g^{-1}) ~.
\end{equation} 
With this choice, the state can be written as,
\begin{equation}
\boxed{\ket{T_{p,q}} =  \sum_{\alpha} \sum_{A} \sum_{R_i,\,R_j} \frac{\mathcal{S}_{\alpha \beta_i}^* \,\mathcal{S}_{0 \beta_j}}{(\mathcal{S}_{0 \alpha})^{n-1}}\, (\mathcal{T}_{\beta_j \beta_j})^{q/p} \,  Y_{R_i, R_j}(p/n) \, \ket{\alpha,\alpha,\ldots,\alpha}} ~,
\end{equation}
where $\alpha$ is an irrep of $D(G)$. Similarly, $\beta_i = (A,R_i)$ and $\beta_j = (A,R_j)$ are the irreps of $D(G)$ with $A$ denoting a conjugacy class and $R_i, R_j$ being the irreps of centralizer $C_A$. This is a state defined on disjoint torus boundaries and lives in the tensor product of $n$ copies of $\mathcal{H}_{T^2}$. We can write this state in a compact fashion as following. Collect all the coefficients $X_{\beta \gamma}(m)$ in eq.(\ref{X-coeff}) into a matrix $X(m)$ whose rows and columns are labeled by the irreps $\beta$ and $\gamma$. Note that in obtaining $X(m)$, we must use the same ordering of the basis which was used to define the modular $\mathcal{S}$ and $\mathcal{T}$ matrices. This will enable us to perform matrix algebra between these matrices. With our choice of coefficients, it is also clear that the matrix $X(m)$ will be block diagonal where each block will contain the Adams coefficients for various centralizer subgroups. Thus the state can be written as:
\begin{equation}
\ket{T_{p,q}} =  \sum_{\alpha}  \dfrac{\left(\mathcal{S}^*X(p/n)\mathcal{T}^{\frac{q}{p}}\mathcal{S}\right)_{\alpha 0}}{(\mathcal{S}_{0 \alpha})^{n-1}}\, \ket{\alpha,\alpha,\ldots,\alpha} ~.
\end{equation}
Using the steps prescribed in section 2, we can obtain the spectrum of the reduced density matrix and the entanglement entropy. We can clearly see that this state has a GHZ-like structure in the sense that the partial transpose of the reduced density matrix along any tri-partition is positive semi-definite and hence the reduced density matrix computed for any bi-partition is always separable. The entanglement entropy is independent of the number of Hilbert spaces being traced out and can be obtained as:
\begin{equation}
\boxed{\text{EE} = -\sum_{\alpha} \lambda_{\alpha} \ln \lambda_{\alpha} \quad;\quad \lambda_{\alpha} = \frac{1}{\text{trace}}\left| \dfrac{\left(\mathcal{S}^*X(p/n)\mathcal{T}^{\frac{q}{p}}\mathcal{S}\right)_{\alpha 0}}{(\mathcal{S}_{0 \alpha})^{n-1}} \right|^2 } ~,
\label{eigen-toruslink}
\end{equation}
where the symbol $|x|^2$ is used to denote the modulus square of $x$, i.e. $|x|^2 = x x^*$ and the factor `trace' is defined as,
\begin{equation}
\text{trace} =  \sum_{\alpha} \left| \dfrac{\left(\mathcal{S}^*X(p/n)\mathcal{T}^{\frac{q}{p}}\mathcal{S}\right)_{\alpha 0}}{(\mathcal{S}_{0 \alpha})^{n-1}} \right|^2 ~.
\end{equation}
Note that for the torus links of type $T_{p,\, pn}$, the matrix $X(1)$ will be an identity matrix and the state can be simply computed with the knowledge of $\mathcal{S}$ and $\mathcal{T}$ matrices:
\begin{equation}
\boxed{\ket{T_{p,\, pn}} =  \sum_{\alpha}  \dfrac{\left(\mathcal{S}^* \mathcal{T}^{n}\mathcal{S}\right)_{\alpha 0}}{(\mathcal{S}_{0 \alpha})^{p-1}}\, \ket{\alpha,\alpha,\ldots,\alpha}} ~.
\end{equation}
For example, the simplest non-trivial torus link is $T_{2,2}$, the Hopf link. The state associated with Hopf link can be given as,
\begin{align}
\ket{T_{2,2}} = \sum_{\alpha} \frac{\left(\mathcal{S}^* \mathcal{T} \mathcal{S}\right)_{\alpha 0}}{\mathcal{S}_{0 \alpha}} \, \ket{\alpha,\alpha} = \sum_{\alpha} \frac{\left(\mathcal{T}^* \mathcal{S}^* \mathcal{T}^*\right)_{\alpha 0}}{\mathcal{S}_{0 \alpha}} \, \ket{\alpha,\alpha} = \sum_{\alpha} \mathcal{T}_{\alpha \alpha}^* \, \ket{\alpha,\alpha} ~.
\end{align}
Since $\mathcal{T}_{\alpha \alpha}$ is just a phase, it will not affect the entanglement structure and thus we see that the Hopf link state is precisely a maximally entangled  Bell state and hence has the same entropy as that of the state $\ket{T^2 \times [0,1]}$ already discussed in earlier sections.

The entanglement structure of the torus link states shows some interesting features. For a given value of $p$, the entanglement structure of the state $\ket{T_{p,q}}$ shows a periodic behavior as we increase $q$. The fundamental period of this periodicity is determined by the positive integer $e(G)$, which is the exponent of the group.\footnote{The exponent of a discrete group is the smallest positive integer $x$ such that $g^x = e$ for all $g \in G$. In other words, it is the least common multiple of the orders of various elements of the group.} We find that the entropy satisfies: 
\begin{equation}
\text{EE}(T_{p,\,q}) = \text{EE}(T_{p,\,q+p\,e(G)}) ~.
\label{period-toruslink}
\end{equation}  
It also exhibits a nice symmetric property:
\begin{equation}
\text{EE}(T_{p,\,q}) = \text{EE}(T_{p,\,p\,e(G)-q}) ~.
\end{equation}
For a fixed $p$, the entropy fluctuates between maximum and minimum values as we vary $q$. Our numerical computations for various gauge groups show that the maxima and minima usually occur only when $q$ is a multiple of $p$, i.e for the states of the form $\ket{T_{p,\,pn}}$ where $n$ is an integer. The entropy for the state $\ket{T_{p,\,pn}}$ vanishes whenever $n$ is a multiple of $e(G)$ and obtain a maximum value whenever $n$ is coprime to $e(G)$. We observe the following:
\begin{equation}
\boxed{ \text{EE}(T_{p,\,pn}) = 
\begin{cases}
\text{EE}_{\text{min}} = 0, & \text{when $n=0$ modulo $e(G)$} \\
\text{EE}_{\text{max}} = \text{EE}(T_{p,\,p}), & \text{when gcd$\{n, e(G)\}=1$}
\end{cases}} ~.
\end{equation} 
Recalling that a torus link $T_{p,q}$ can be drawn on the surface of a torus in which each circle component winds around the two homology cycles of $T^2$ with winding numbers $\frac{p}{\text{gcd}(p,q)}$ and $\frac{q}{\text{gcd}(p,q)}$ respectively, we can interpret the above result as following. Amongst the class of torus links $T_{p,q}$ with a fixed  value of $p$, the entropy is maximum for those links in which each circle winds exactly once along the two homology cycles of the torus. These are precisely the links $T_{p,p}$. When the winding number along any of the two cycles exceeds 1, the entropy corresponding to those links will be generically smaller.\footnote{It may happen that the entropy for two different links become equal for certain gauge group. All we claim here is that the torus links in which each component has winding number 1 (along both homology cycles of torus), will have maximum value of entropy.} We have already encountered one special case of this result. For two component links, the Hopf link ($T_{2,2}$) is maximally entangled and as we increase the winding number (links of type $T_{2,2n}$), the entropy usually decreases. Further, when winding number along any of the homology cycle becomes a multiple of $e(G)$, the entropy will vanish. Thus we can rewrite the above result as following: 
\begin{empheq}[box=\fbox]{align}
  &\text{Link with both winding numbers 1} \Longrightarrow \text{EE} = \text{maximum} \nonumber \\
&\text{Link with any of the two winding number equal to multiple of $e(G)$} \Longrightarrow \text{EE} = 0 \nonumber ~.
\end{empheq}
Let us quantify the above discussion by calculating the entropy for various groups in the following sections.
\subsubsection{Entropy for abelian group}
For $\mathbb{Z}_N$ group, the centralizers associated with each of the conjugacy classes are again $\mathbb{Z}_N$. The characters along with modular $\mathcal{S}$ and $\mathcal{T}$ matrices are given in appendix-\ref{appendix-A} and the Adams operation on centralizers is performed in appendix-\ref{appendix-B}. Using these details, the spectrum of the reduced density matrix for the state $\ket{T_{p,\,q}}$ can be computed as:
\begin{equation}
\boxed{\lambda_{ab} = \frac{1}{\text{trace}}\left(\sum_{A=0}^{N-1} \sum_{x=0}^{N-1} \sin Y\right)^2+\frac{1}{\text{trace}}\left(\sum _{A=0}^{N-1} \sum _{x=0}^{N-1} \cos Y\right)^2} ~,
\end{equation}
where the eigenvalues are labeled by integers $a$ and $b$ each taking values from 0 to $(N-1)$. The factor `trace' ensures that the reduced density matrix is normalized to have unit trace and the variable $Y$ in above equation is given as,  
\begin{equation}
Y=\frac{2 \pi  \left(A q \left[\frac{p x}{\gcd (p,q)}\right]_N +a p x+A b p\right)}{N p} ~,
\end{equation}
where the symbol $[\alpha]_N \equiv \alpha \,(\text{mod } N)$. Hence the entanglement entropy can be obtained as,
\begin{equation}
\boxed{\text{EE}(T_{p,q}) = -\sum_{a=0}^{N-1}\sum_{b=0}^{N-1}\lambda_{ab} \ln \lambda_{ab}} ~.
\end{equation}
From the eigenvalues given above, the following properties are evident:
\begin{equation}
\text{EE}(T_{p,\,q+pN}) = \text{EE}(T_{p,\,q})  \quad;\quad \text{EE}(T_{p,\,q}) = \text{EE}(T_{p,\,pN-q}) ~,
\end{equation}   
where we note that being a cyclic group we have $e(\mathbb{Z}_N) = N$. The minimum and maximum values of entropy are given below:
\begin{equation}
\text{EE}(T_{p,\,pn}) = 
\begin{cases}
\text{EE}_{\text{min}} = 0, & \text{when gcd$(n, N)=N$} \\
\text{EE}_{\text{max}} = \ln N^2, & \text{when gcd$(n, N)=1$}
\end{cases} ~.
\end{equation} 
In fact, one can write a close form expression for the entropy of the states $\ket{T_{p,\,pn}}$ as following:
\begin{equation}
\boxed{\text{EE}(T_{p,\,pn}) = \ln \left(\frac{N}{\text{gcd}(n,N)}\right)^2 } 
\label{EE-torus-abelian} ~.
\end{equation}
Thus the spectrum for $\ket{T_{p,\,pn}}$ state consists of all equal eigenvalues. For general links however, the spectrum of the reduced density matrix usually breaks into several degenerate sectors. For example, for $\mathbb{Z}_2$ group, the spectrum for a generic link is given as:
\begin{equation}
\lambda(T_{p,\,q}\,;\,\mathbb{Z}_2) = \{ \lambda_1, \lambda_2, \lambda_2, \lambda_2  \} \quad;\quad \text{EE}(T_{p,\,q}\,;\,\mathbb{Z}_2) = -\lambda_1 \ln \lambda_1 -3\,\lambda_2 \ln \lambda_2 ~,
\end{equation}
where we have:
\begin{equation}
\lambda_1 = \frac{5}{8}+ \frac{3}{8}\cos\left(\frac{\pi q \left[\frac{p}{\gcd (p,q)}\right]_2}{p}\right) \quad,\quad \lambda_2 = \frac{1}{8}- \frac{1}{8}\cos\left(\frac{\pi q \left[\frac{p}{\gcd (p,q)}\right]_2}{p}\right) ~.
\end{equation}

From the above discussion, one can infer that the entropy computed for abelian groups does not capture the complete information about the topology of link complements. For example, the entropy of $\ket{T_{p,\,pn}}$ comes out to be independent of the value of $p$. So let us study the entanglement features of these states for some non-abelian groups.
\subsubsection{Entropy for dihedral group}
Here we compute the entanglement entropy for the dihedral group $D_N$ which has $2N$ elements. The characters along with modular $\mathcal{S}$ and $\mathcal{T}$ matrices are given in appendix-\ref{appendix-A} and the Adams operation on centralizers is performed in appendix-\ref{appendix-B}. The dimension of each Hilbert space, which is also the number of irreps of Drinfeld double of $D_N$ is given as:
\begin{equation}
\text{dim}(\mathcal{H}_{T^2}) = 
\begin{cases} \left(\dfrac{N^2+28}{2}\right), & \text{when $N$ is even} \\[0.5cm]
\left(\dfrac{N^2+7}{2}\right), & \text{when $N$ is odd}
\end{cases} ~.
\end{equation}
The entanglement structure has a periodic behavior as we increase $q$ for a given value of $p$. Further there is a symmetry in $q$ as given by the following:
\begin{equation}
\text{EE}(T_{p,\,q+p\,e(D_N)}) = \text{EE}(T_{p,\,q})  \quad;\quad \text{EE}(T_{p,\,p\,e(D_N)-q}) = \text{EE}(T_{p,\,q}) ~,
\end{equation}  
where the exponent for the dihedral group is $e(D_N) = \text{lcm}(2,N)$.  
Similar to the abelian case, the minimum and maximum values of entropy usually happens for those links where $q$ is a multiple of $p$:
\begin{equation}
\text{EE}(T_{p,\,pn}) = 
\begin{cases}
\text{EE}_{\text{min}} = 0, & \text{when $n=0$ modulo $e(D_N)$} \\
\text{EE}_{\text{max}} = \ln \text{dim}(\mathcal{H}_{T^2}), & \text{when gcd$(n, N)=1$}
\end{cases} ~.
\end{equation} 
For the link states $\ket{T_{p,\,pn}}$, the spectrum of the reduced density matrix can be given as:
\begin{equation}
\lambda(T_{p,\,pn}) = \begin{cases} \frac{1}{\text{trace}_1}\{\,(x+1)^2,\, (x-1)^2,\, \underbrace{1, 1, \ldots, 1}_{\alpha(x)}\,\, \} &,\,\, \text{when $n=$ even} \\ \frac{1}{\text{trace}_2}\{\, \underbrace{\tfrac{N^2}{x^2},\tfrac{N^2}{x^2},\ldots, \tfrac{N^2}{x^2}}_{\beta(x)}\,,\,\, \underbrace{1, 1, \ldots, 1}_{\gamma(x)}\,\, \} &,\,\, \text{when $n=$ odd} \end{cases} ~,
\label{EE-torus-dihedral}
\end{equation}
where $x = \frac{N}{\text{gcd}(N,n)}$ and the factors $\text{trace}_1$ and $\text{trace}_2$ are given as,
\begin{equation}
\text{trace}_1 = 2x^2 + 2 + \alpha(x) \quad,\quad \text{trace}_2 = \frac{N^2}{x^2}\beta(x) + \gamma(x)  ~.
\end{equation}
The integers $\alpha(x)$ and $\beta(x)$ are defined as following:
\begin{equation}
\alpha(x) = \begin{cases} (x^2+8)/2 &,\,\, \text{when $x=$ even} \\ (x^2-1)/2 &,\,\, \text{when $x=$ odd} \end{cases}  \quad;\quad \beta(x) = \begin{cases} (x^2+12)/2 &,\,\, \text{when $x=$ even} \\ (x^2+3)/2 &,\,\, \text{when $x=$ odd} \end{cases} \nonumber
\end{equation}
and the integer $\gamma(x)=8$ for $x=$ even and $\gamma(x)=2$ for $x=$ odd respectively.

We see that the spectrum for the state $\ket{T_{p,\,pn}}$ breaks into several degenerate parts unlike the abelian case, but the entropy is still independent of $p$. As a last example, we will study the symmetric groups where the entropy captures the topology of different $T_{p,\,pn}$ links.
\subsubsection{Entropy for symmetric group}
Unlike the abelian and dihedral groups where the entanglement structure of the $\ket{T_{p,\,pn}}$ states was the same for all values of $p$, the symmetric groups can distinguish the affect of $p$ on the entropy. Though we could not analyze these states for generic $S_N$, the computation for lower order groups show that the entropy decreases as $p$ increases for a fixed value of $n$. When $p \to \infty$ for a fixed $n$, the entropy converges to some finite value. Our calculations for the lower order $S_N$ groups also confirm that the entanglement structure has the following periodic behavior:
\begin{equation}
\text{EE}(T_{p,\,q+p\,e(S_N)}) = \text{EE}(T_{p,\,q})  \quad;\quad \text{EE}(T_{p,\,p\,e(S_N)-q}) = \text{EE}(T_{p,\,q}) ~,
\label{periodicity-SN}
\end{equation}  
where the exponent for the symmetric groups is given as:
\begin{equation}
e(S_N) = \text{lcm}(1,2,\ldots,N) ~.
\label{exponent-SN}
\end{equation}  
The minimum and maximum values of entropy for the $S_N$ group and the values at which it happens are given below:
\begin{equation}
\text{EE}(T_{p,\,pn}) = 
\begin{cases}
\text{EE}_{\text{min}} = 0, & \text{when $n=0$ modulo $e(S_N)$} \\
\text{EE}_{\text{max}} = \text{EE}(T_{p,p}), & \text{when gcd$(n, N!)=1$}
\end{cases} ~.
\label{min-max-SN}
\end{equation} 
In the following, we present results for lower order groups supporting our observations. The groups $S_2$ and $S_3$ are isomorphic to $\mathbb{Z}_2$ and $D_3$ groups respectively which we have already discussed. For $S_4$ group, we present plots in figure \ref{EE(p,pn)linkS4group} showing the variation of entropy as a function of $n$ for various $T_{p,\,pn}$ links.    
\begin{figure}[htbp]
	\centering
		\includegraphics[width=0.95\textwidth]{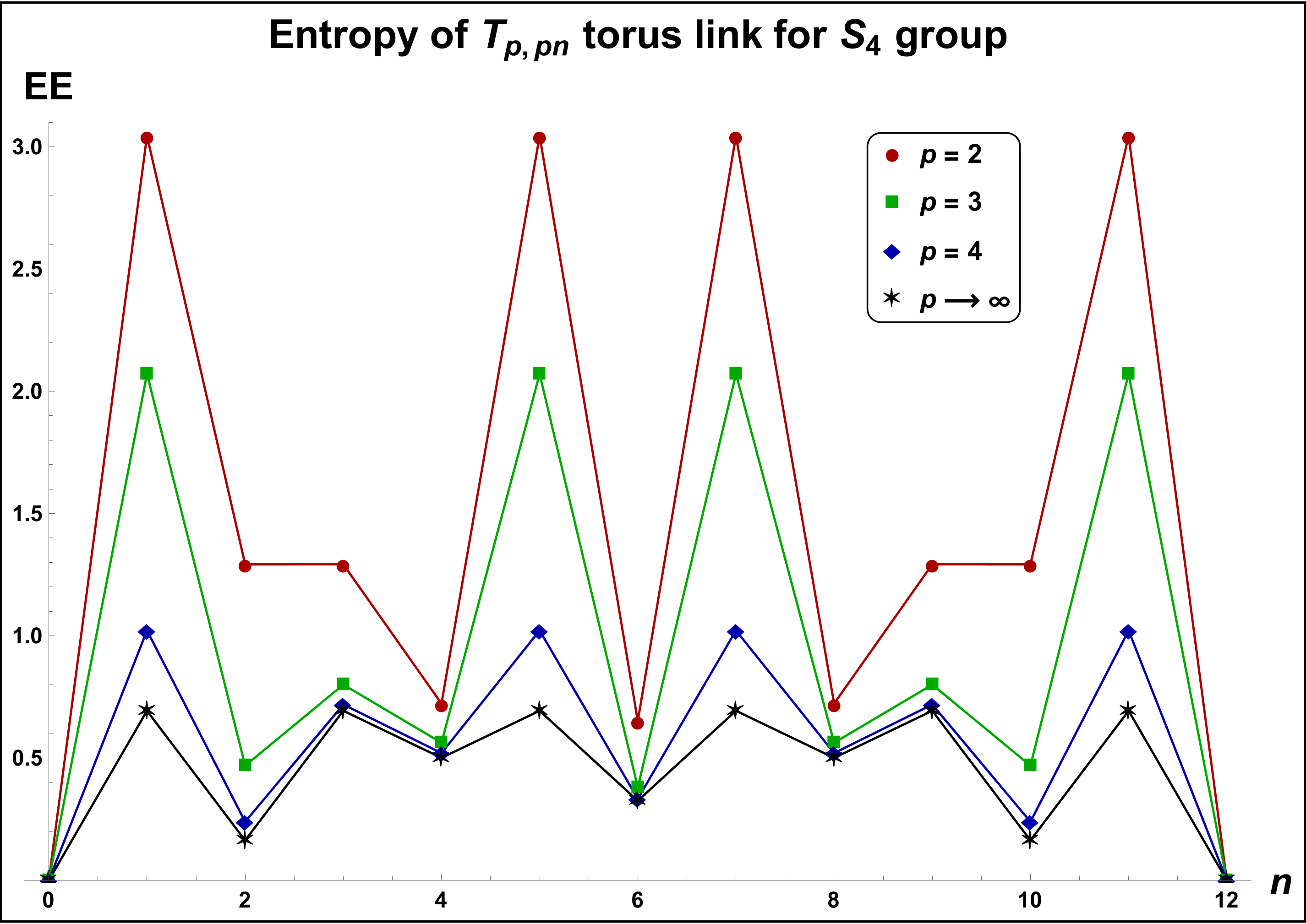}
	\caption{The variation of entropy of $\ket{T_{p,\,pn}}$ states computed for the $S_4$ group. The entropy fluctuates as $n$ increases and reaches maximum and minimum values according to eq.(\ref{min-max-SN}). We also show the entropy of $\ket{T_{p,\,pn}}$ state as $p \to \infty$.}
	\label{EE(p,pn)linkS4group}
\end{figure}
The entropy for the $T_{p,p}$ link for $S_4$ group can be obtained from the following spectrum of the reduced density matrix:
\begin{align}
\lambda(T_{p,p}) = \frac{1}{\text{trace}(p)}\left\{576^{p-2}, 576^{p-2}, 144^{p-2}, 64^{p-2}, 64^{p-2}, 64^{p-2},64^{p-2},64^{p-2},64^{p-2},  \right. \nonumber \\
\left. \mbox{} 16^{p-2},16^{p-2},16^{p-2},16^{p-2},16^{p-2},16^{p-2},16^{p-2},16^{p-2},16^{p-2},9^{p-2},9^{p-2},9^{p-2} \right\} ~,
\end{align}
where $\text{trace}(p)$ is defined as,
\begin{equation}
\text{trace}(p) = \frac{243 \times 16^p \left(4^p+24\right)+9^p \left(2^{4 p+3}+64^p+6144\right)}{165888} ~.
\end{equation}
Further we see that as $p \to \infty$ the entropy converges: 
\begin{equation}
\lim_{p \to \infty} \text{EE}(T_{p,\,p}) = \ln 2 ~.
\end{equation}
In general, the entropy of the state $\ket{T_{p,\,pn}}$ converges as $p \to \infty$ as shown in the figure \ref{EE(p,pn)linkS4group}.  

We have also given the variation of the entropy for various $\ket{T_{p,\,pn}}$ states for the group $S_5$ in the figure \ref{EE(p,pn)linkS5group}. The last plot in figure \ref{EE(p,pn)linkS5group} shows that the entropy converges as $p \to \infty$.
\begin{figure}[htbp]
	\centering
		\includegraphics[width=1.00\textwidth]{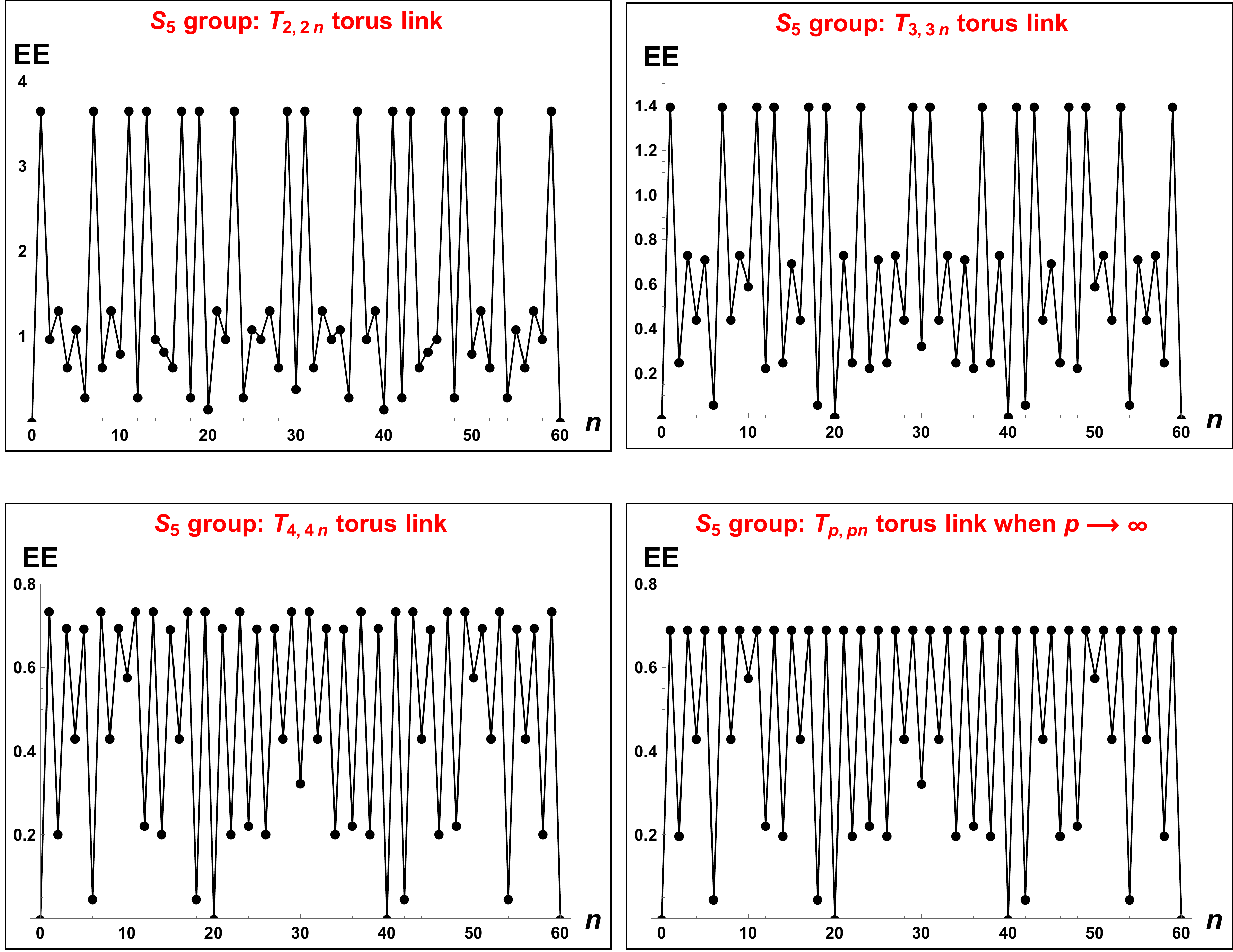}
	\caption{The variation of entropy of $\ket{T_{p,\,pn}}$ states computed for the $S_5$ group. The entropy fluctuates as $n$ increases and reaches maximum and minimum values according to eq.(\ref{min-max-SN}). We also show the entropy of $\ket{T_{p,\,pn}}$ state as $p \to \infty$ in the last plot.}
	\label{EE(p,pn)linkS5group}
\end{figure}

\subsubsection{Entropy for direct product of groups}
Consider the direct product of two groups: $G=G_1 \times G_2$. The centralizer $C_A$ corresponding to the conjugacy class $A$ of the group $G$ is obtained as:
\begin{equation}
C_A \equiv C_a = C_{(a_1,a_2)} = C_{a_1} \times C_{a_2} = C_{A_1} \times C_{A_2}  ~,
\end{equation} 
where $a_1, a_2$ and $a=(a_1,a_2)$ are the representative elements of the conjugacy classes $A_1, A_2$ and $A$ respectively of the groups $G_1,G_2$ and $G$. Thus the Adams operation on the centralizer $C_A$ would be given as:
\begin{align}
\Psi^m \text{char}_R (a) &= \text{char}_R(a^m) = \text{char}_{R_1\otimes R_2}(a_1^m, a_2^m) = \text{char}_{R_1}(a_1^m) \,\text{char}_{R_2}(a_2^m) \nonumber \\
 &= \sum_{S_1,\,S_2} Y_{R_1S_1}^{(1)}(m)\, Y_{R_2S_2}^{(2)}(m)\, \text{char}_{S_1}(a_1) \, \text{char}_{S_2}(a_2) \nonumber \\
 &= \sum_{S=S_1\otimes S_2} Y_{RS}(m)\, \text{char}_{S}(a_1, a_2) ~,
\end{align} 
where $R$ and $S$ are the irreps of the group $G$ which can be written as tensor products $R_1 \otimes R_2$ and $S_1 \otimes S_2$ of the irreps associated with $G_1$ and $G_2$. The $Y^{(1)}(m)$ and $Y^{(2)}(m)$ are the matrices containing the Adams coefficients corresponding to the centralizer subgroups $C_{A_1}$ and $C_{A_2}$ of the groups $G_1$ and $G_2$. Moreover, from the above argument, we can also see that the matrix $Y(m)$ for the centralizer $C_{A}$ of the group $G$ will be the Kronecker product of the matrices of individual groups, i.e. for any centralizer subgroup, we will have:
\begin{equation}
Y(m) = Y_1(m) \otimes Y_2(m)  ~.
\end{equation}
On the other hand, we also know that the primaries (i.e. the irreps) associated with the Drinfeld double $D(G)$ can be simply obtained from the primaries of $D(G_1)$ and $D(G_2)$ and thus the basis of the Hilbert space $\mathcal{H}_{T^2}$ can be given as:
\begin{equation}
\ket{\Phi_{ij}} = |\Phi_{i}^{(1)} \rangle \otimes |\Phi_{j}^{(2)} \rangle ~,
\end{equation}
where $\ket{\Phi_{ij}}$ is the basis of $\mathcal{H}_{T^2}$ in the context of group $G$ which can be written as a tensor product of the basis elements of $\mathcal{H}_{T^2}$ for the groups $G_1$ and $G_2$ respectively. With a proper ordering of the basis $\ket{\Phi_{ij}}$, we can also write the matrix $X(m)$ defined in eq.(\ref{X-coeff}) for the group $G$ as the Kronecker product of the matrices for the individual groups:
\begin{equation}
X(m) = X_1(m) \otimes X_2(m) ~.
\end{equation}
Further, maintaining the same ordering of the basis, the modular matrices $\mathcal{S}$ and $\mathcal{T}$ can be simply written as the Kronecker product of matrices corresponding to the two groups \cite{Coste:2000tq}:
\begin{equation}
\mathcal{S} = \mathcal{S}_1 \otimes \mathcal{S}_2 \quad;\quad \mathcal{T} = \mathcal{T}_1 \otimes \mathcal{T}_2 ~.
\end{equation} 
Using the mixed-product property of the Kronecker product and the fact that $\mathcal{T}$ is a diagonal matrix, we will have:
\begin{equation}
\mathcal{S}^*X(p/n)\mathcal{T}^{\frac{q}{p}}\mathcal{S} =  \left(\mathcal{S}_1^*X_1(p/n)\mathcal{T}_1^{\frac{q}{p}}\mathcal{S}_1\right) \otimes \left(\mathcal{S}_2^*X_2(p/n)\mathcal{T}_2^{\frac{q}{p}}\mathcal{S}_2\right) ~.
\end{equation} 
Thus it is not difficult to see from eq.(\ref{eigen-toruslink}) that the eigenvalues of the reduced density matrix of the state $\ket{T_{p,q}}$ for the group $G$ can be written as product of the eigenvalues computed for the individual groups $G_1$ and $G_2$, i.e.,
\begin{equation}
\lambda_{ij}^G = \lambda_{i}^{G_1} \, \lambda_{j}^{G_2} ~.
\end{equation} 
Hence the entanglement entropy for the direct product of two groups is simply the sum of the entropies corresponding to individual groups:
\begin{equation}
\boxed{\text{EE}(G_1 \times G_2) = \text{EE}(G_1) + \text{EE}(G_2)} ~.
\end{equation}
\section{Conclusion}
\label{sec5}
In this work, we studied the entanglement structure of multi-boundary states prepared in Chern-Simons theory with a finite discrete gauge group. The path integral of the Chern-Simons theory on  a manifold $M$ with boundary defines a state $\ket{\Psi}$ in the Hilbert space $\mathcal{H}_{\partial M}$ associated with the boundary. If the boundary consists of multiple disconnected components, then $\mathcal{H}_{\partial M}$ can be decomposed into the tensor product of the Hilbert spaces associated with each component.
Throughout this work, we have restricted to the untwisted case ($\alpha=0$ in eq.(\ref{alpha=0-action})) where the action is trivial ($e^{2 \pi i S}=1$).

In section \ref{sec3}, we considered the states associated with Riemann surfaces of genus $g$ having $n$ number of circle boundaries. The state $\ket{\Sigma_{g,n}}$ thus lives in the tensor product of $n$ copies of $\mathcal{H}_{S^1}$. We study the entanglement features of these states by using the replica trick. The entanglement entropy does not depend on the choice of the bi-partition which is evident from eq.(\ref{EE-2d}). Moreover the reduced density matrix obtained by tracing out a subset of the total Hilbert space has a positive semidefinite partial transpose over any bi-partition of the remaining Hilbert space. We verified this by computing entanglement negativity which vanishes (eq.(\ref{EE-2d})). We gave an explicit expression for the entanglement entropy in eq.(\ref{EE-2d-formula}) which depends on the dimension of the irreducible representations of the finite group $G$ and the Euler characteristic $\chi$ of the manifold $\Sigma_{g,n}$. We presented concrete results of the entropy for various abelian and non-abelian groups and find that the entropy converges to a constant value as $\chi \to -\infty$. Moreover, eq.(\ref{abelianEE}) and eq.(\ref{Limits-DN-group}) also show that the entropy has a logarithmic asymptotic behavior and goes as $\ln |G|$ as the order of the group $|G| \to \infty$ for a fixed $\chi$. However, we could not verify this for $S_N$ group due to the lack of an analytical expression for a generic $N$. We further show that the entropy is additive for a direct product of groups, i.e. $\text{EE}(G_1 \times G_2) = \text{EE}(G_1) + \text{EE}(G_2)$.

In section $\ref{sec4}$, we studied the entanglement properties of states associated with 3-manifolds. In particular, we focused on the torus link complements $S^3 / T_{p,q}$, which are obtained by cutting out a tubular neighborhood around the torus link $T_{p,q}$ from $S^3$. These manifolds have `$\text{gcd}(p,q)$' number of torus boundaries and hence the corresponding state $\ket{T_{p,q}}$ lives in the tensor product of `$\text{gcd}(p,q)$' copies of $\mathcal{H}_{T^2}$. The entanglement entropy, given in eq.(\ref{eigen-toruslink}), is independent of the choice of bi-partition and depends on finer group theoretic details of $G$ like the conjugacy classes and the centralizer subgroup associated with an element of each conjugacy class. We show that these states have a GHZ-like entanglement structure where the reduced density matrix (obtained after tracing out a subset of the total Hilbert space) is separable over any bi-partition. The state $\ket{T_{p,q}}$ has a periodic entanglement structure in the sense that the entropy shows a periodic behavior as we increase $q$ for a fixed value of $p$ as given in eq.(\ref{period-toruslink}). The fundamental period of this periodicity is controlled by the positive integer $e(G)$, which is the exponent of the group $G$. Based on our analysis for various groups, we also find that amongst the class of torus links $T_{p,q}$ for a fixed value of $p$, the entropy is maximum when $q=p$, i.e for those torus links which can be drawn on the surface of a torus such that each component winds exactly once along the two homology cycles of the torus. When the winding number along any of the homology cycle exceeds 1 (i.e. when $q>p$), the entropy is generically smaller and vanishes whenever any of the two winding numbers is a multiple of $e(G)$. This is evident from eq.(\ref{EE-torus-abelian}), eq.(\ref{EE-torus-dihedral}) and the plots shown in figure \ref{EE(p,pn)linkS4group} and figure \ref{EE(p,pn)linkS5group} respectively. We also show that the entanglement entropies corresponding to torus links evaluated for a direct product of groups satisfy $\text{EE}(G_1 \times G_2) = \text{EE}(G_1) + \text{EE}(G_2)$. It is known that the entanglement entropy associated with torus link can also be used to detect one-form anomalies as shown in \cite{Hung:2018rhg, Zhou:2019ezk} for semi-simple Lie groups. It would be interesting to carry out a similar study in the context of current set-up. Further, it will be an important exercise to extend this work to study the entanglement features of the states associated with hyperbolic link complements. We hope to report these in near future.

We also studied the states prepared on cobordant boundaries associated with manifolds of type $\Sigma_g \times [0,1]$ which are maximally entangled states. The entanglement entropy computed in this case goes as $\ln |G|$ as the order of the group $|G| \to \infty$ for a fixed value of $g$. Further it shows a linear behavior in $g$ as $g \to \infty$ for a finite order of the group. The limiting values are given in eq.(\ref{EE-cobordant}). These entropies also obey the property $\text{EE}(G_1 \times G_2) = \text{EE}(G_1) + \text{EE}(G_2)$.    

In the present work, we restricted to the case where the Chern-Simons action is trivial, i.e. $e^{2 \pi i S}=1$. It would be interesting to extend the study to the cases where the action in eq.(\ref{Action-twisted}) is classified by the elements of the cohomology group. In such a case, the partition functions and hence the entanglement structure will depend on the cohomological twists $\alpha$. This requires an elaborate study and we leave this analysis for future work.

\appendix
\section{Modular data for finite group $G$}
\label{appendix-A}
Here we tabulate the modular data, i.e. the matrix elements of generators $\mathcal{S}$ and $\mathcal{T}$ of SL($2,\mathbb{Z}$). These matrices are written in the basis of irreps of the untwisted (i.e. $\alpha=0$) quantum double group $D(G)$ associated with the finite group $G$. As discussed in section \ref{sec4}, these matrix elements are given as,
\begin{align}
  \mathcal{S}_{(A,\,R_A) (B,\,R_B)} &= \frac{1}{|C_A||C_B|}\, \sum_{g\, \in Y_{AB}} \chi R_A(gbg^{-1})^{*} \, \chi R_B(g^{-1}ag)^{*} \nonumber \\
\mathcal{T}_{(A,\,R_A) (B,\,R_B)} &= \delta_{AB}\, \delta_{R_A R_B}\, \frac{\chi R_A(a)}{\chi R_A(e)} ~,
\label{S-T-matrices}
\end{align} 
where $(A,\,R_A)$ and $(B,\,R_B)$ are the irreducible representations of $D(G)$ and label the row and column respectively of $\mathcal{S}$ and $\mathcal{T}$ matrices. The symbols $\chi R_A$ and $\chi R_B$ denote the characters of irreps $R_A$ and $R_B$ of centralizers $C_A$ and $C_B$ respectively with `$*$' being the complex conjugation. The elements $a \in A$ and $b \in B$ where $A$ and $B$ are the conjugacy classes of $G$. The notation $\chi R(x)$ simply means the character of the representation $R$ being evaluated for an element $x$. The set $Y_{AB}$ is defined as: 
\begin{equation}
Y_{AB} = \{g \in G \,|\, gbg^{-1} \in C_A \, \text{ and } \,  g^{-1}ag \in C_B  \} ~.
\end{equation}
From their structure, it is clear that $\mathcal{S}$ and $\mathcal{T}$ are block matrices, i.e. their $AB$ element labeled by conjugacy classes $A$ and $B$ is a matrix whose rows and columns are generated by various irreducible characters of $C_A$ and $C_B$ respectively. For example, if $\{A_1, A_2, A_3, \ldots\}$ are the conjugacy classes of $G$, then $\mathcal{S}$ and $\mathcal{T}$ will be of the following form:
\begin{equation}
\mathcal{S} = \left(
\begin{array}{cccc}
 M_{A_1 A_1} & M_{A_1 A_2} & M_{A_1 A_3} & \ldots \\
 M_{A_2 A_1} & M_{A_2 A_2} & M_{A_2 A_3} & \ldots \\
 M_{A_3 A_1} & M_{A_3 A_2} & M_{A_3 A_3} & \ldots \\
 \vdots & \vdots & \vdots & \vdots \\
\end{array}
\right) \quad,\quad \mathcal{T} = \left(
\begin{array}{cccc}
 N_{A_1 A_1} & 0 & 0 & \ldots \\
 0 & N_{A_2 A_2} & 0 & \ldots \\
 0 & 0 & N_{A_3 A_3} & \ldots \\
 \vdots & \vdots & \vdots & \vdots \\
\end{array}
\right) ~,
\end{equation}
where $M_{A_i A_j}$ and $N_{A_i A_j}$ are matrices. If $\{\sigma_A^{(1)}, \sigma_A^{(2)}, \sigma_A^{(3)}, \ldots\}$ denote the irreps of $C_A$ for a given class $A$, then the $(k,l)^{\text{th}}$ element of $M_{A_i A_j}$ or $N_{A_i A_j}$ are computed using $\sigma_{A_i}^{(k)}$ and $\sigma_{A_j}^{(l)}$ respectively in the eq.(\ref{S-T-matrices}). In the following, we will compute these matrices for abelian and non-abelian finite groups. 
\subsection{Modular $\mathcal{S}$ and $\mathcal{T}$ matrices for finite cyclic group: $G=\mathbb{Z}_N$}
The group $\mathbb{Z}_N$ is an abelian group. Each element is a conjugacy class in itself. Thus we can label the conjugacy class by an integer $a$ such that $0 \leq a \leq (N-1)$. The centralizer associated with each class is $C_a = \mathbb{Z}_N$ which is again abelian and has only one dimensional irreducible representations which we label by another integer $b$ which takes values $0 \leq b \leq (N-1)$. Hence we can label the irreps of quantum double group associated with $\mathbb{Z}_N$ by a pair of integers $(a, b)$. The character of the representation `$b$' evaluated for the element $a \in \mathbb{Z}_N$ can be taken as following:
\begin{equation}
\chi_b(a) = \exp(2 \pi i\frac{ a b}{N}) ~.
\end{equation}
Thus the modular elements can be easily obtained from eq.(\ref{S-T-matrices}):
\begin{empheq}[box=\fbox]{align}
  \mathcal{S}_{(a, b) (a', b')} &=  \frac{1}{N}\exp[-\frac{2 \pi i}{N}(a b' + a' b)] \nonumber \\
\mathcal{T}_{(a, b) (a', b')} &= \exp[\frac{2 \pi i}{N}(ab)] \, \delta_{a a'}\, \delta_{b b'} ~.
\label{S-T-matrices-ZN}
\end{empheq}
The values of $a$ and $a'$ label the location of each block inside $\mathcal{S}$ and $\mathcal{T}$. Each block is a matrix whose rows and columns are labeled by $b$ and $b'$ respectively. Hence the modular matrices $\mathcal{S}$ and $\mathcal{T}$ will be of the order $N^2$. We give the matrices for $\mathbb{Z}_2$ and $\mathbb{Z}_3$ in the following:
\begin{gather*}
\mathcal{S}(\mathbb{Z}_2) = {\frac{1}{2} \left(
\begin{array}{cccc}
 1 & 1 & 1 & 1 \\
 1 & 1 & -1 & -1 \\
 1 & -1 & 1 & -1 \\
 1 & -1 & -1 & 1 \\
\end{array}
\right)},\,\, \mathcal{S}(\mathbb{Z}_3) = {\frac{1}{3} \left(
\begin{array}{ccccccccc}
 1 & 1 & 1 & 1 & 1 & 1 & 1 & 1 & 1 \\
 1 & 1 & 1 & x^2 & x^2 & x^2 & x & x & x \\
 1 & 1 & 1 & x & x & x & x^2 & x^2 & x^2 \\
 1 & x^2 & x & 1 & x^2 & x & 1 & x^2 & x \\
 1 & x^2 & x & x^2 & x & 1 & x & 1 & x^2 \\
 1 & x^2 & x & x & 1 & x^2 & x^2 & x & 1 \\
 1 & x & x^2 & 1 & x & x^2 & 1 & x & x^2 \\
 1 & x & x^2 & x^2 & 1 & x & x & x^2 & 1 \\
 1 & x & x^2 & x & x^2 & 1 & x^2 & 1 & x \\
\end{array}
\right)}  \\
\mathcal{T}(\mathbb{Z}_2) = \text{diag}\left\{1,1, 1, -1 \right\},\,\, 
\mathcal{T}(\mathbb{Z}_3) = \text{diag}\left\{1, 1, 1, 1, x, x^2, 1, x^2, x \right\},
\end{gather*}
where $x = \exp(\frac{2 \pi i}{3})$.
\subsection{Modular $\mathcal{S}$ and $\mathcal{T}$ matrices for dihedral group: $G=D_N$}
This is an example of non-abelian group. The dihedral group $D_N$ has $2N$ elements which are given as,
\begin{equation}
D_N = \{e \equiv r_0, r_1, r_2, \ldots, r_{N-1}, s_0, s_1, s_2, \ldots, s_{N-1} \} ~.
\end{equation}
The composition rule is given as,
\begin{equation}
r_i r_j = r_{i+j}\,,\quad r_i s_j = s_{i+j}\,,\quad s_i r_j = s_{i-j}\,,\quad s_i s_j = r_{i-j} ~,
\end{equation}
where the addition or subtraction of the indices are performed modulo $N$. The even and odd dihedral groups corresponding to even and odd values of $N$ have different character tables. So we will have to deal them separately.
\subsubsection{Even dihedral group: $D_{2M}$}
It has $(M+3)$ number of conjugacy classes given by their representative elements as:
\begin{equation}
\text{conjugacy classes} = [r_0], [r_1], \ldots, [r_{M-1}], [r_M], [s_0], [s_{2M-1}] ~.
\end{equation}
The $D_{2M}$ group has 4 one-dimensional irreps and $(M-1)$ number of two-dimensional irreps. If we denote $\theta_i$ and $\phi_j$ as the irreducible characters of one-dimensional and two-dimensional irreps respectively, the character table for $D_{2M}$ can be given as following:
\begin{equation}
\begin{array}{|c|c|c|c|c|c|} \hline
   & [r_0] & [r_k] & [r_{M}] & [s_0] & [s_{2M-1}] \\ \hline
 \theta_1 & 1 & 1 & 1 & 1 & 1 \\
 \theta_2 & 1 & 1 & 1 & -1 & -1 \\
 \theta_3 & 1 & (-1)^k & (-1)^{M} & 1 & -1 \\
 \theta_4 & 1 & (-1)^k & (-1)^{M} & -1 & 1 \\
 \phi_j   & 2 & 2\cos(\frac{\pi j k}{M}) & 2(-1)^j & 0 & 0  \\\hline
 C_g & D_{2M} & \mathbb{Z}_{2M} & D_{2M} & \mathbb{Z}_{2} \times \mathbb{Z}_{2} & \mathbb{Z}_{2} \times \mathbb{Z}_{2} \\ \hline
\end{array} ~.
\label{table-even-D}
\end{equation}
Here each value of $k$ forms a conjugacy class where $k = 1, 2, \ldots, (M-1)$. The $\phi_j$ is the character of the $j^{\text{th}}$ two-dimensional irrep where $j=1,2,\ldots, (M-1)$. The last row of this table gives the centralizer of representative element for each conjugacy class which can be easily obtained using the composition rules. Since $r_0$ and $r_{M}$ commute with all the elements of $D_{2M}$, we immediately get $C_{r_0} = C_{r_{M}} = D_{2M}$. Further each $r_k$ commutes only with $r_i$ giving $C_{r_{k}} = \{r_0, r_1, \ldots, r_{2M-1}\} \cong \mathbb{Z}_{2M}$. The centralizers for the remaining two classes $[s_0]$ and $[s_{2M-1}]$ are isomorphic to $\mathbb{Z}_{2} \times \mathbb{Z}_{2}$ which can be checked from their elements:
\begin{equation}
C_{s_{0}} = \{r_0, r_{M}, s_0, s_{M} \} \quad,\quad C_{s_{2M-1}} = \{r_0, r_{M}, s_{M-1}, s_{2M-1} \} ~.
\end{equation}
In order to obtain the modular data, we need the irreducible characters of centralizers of each class. The characters for $D_{2M}$ are already given in table in eq.(\ref{table-even-D}). The characters for $\mathbb{Z}_{2M}$ are denoted as $\chi_x$ where $0 \leq x < 2M$. Their values for the element $r_y \in \mathbb{Z}_{2M}$ are given as: $\chi_x(r_y) = \exp(\frac{2 \pi i x y}{2M})$. Similarly we use the notations $\alpha_{ab}$ and $\beta_{ab}$ for the characters of $C_{s_{0}}$ and $C_{s_{2M-1}}$ respectively such that $0 \leq a, b \leq 1$. The values of these characters for the elements in their respective conjugacy classes are given as:
\begin{alignat}{4}
&\alpha_{ab}(r_0) = 1,\quad &\alpha_{ab}(r_M) = (-1)^b,\quad &\alpha_{ab}(s_0) = (-1)^a,\quad &\alpha_{ab}(s_M) = (-1)^{a+b} \nonumber \\
&\beta_{ab}(r_0) = 1,\quad &\beta_{ab}(r_M) = (-1)^b,\quad &\beta_{ab}(s_{M-1}) = (-1)^a,\quad &\beta_{ab}(s_{2M-1}) = (-1)^{a+b} ~.
\end{alignat}
With all the information at our hand, the modular data can be obtained. The $\mathcal{S}$ matrix in terms of blocks labeled by conjugacy classes is written as:
\begin{equation}
\mathcal{S} = \left(
 \begin{array}{c|ccccccc}
  & [r_0] & [r_1] & \ldots & [r_{M-1}] & [r_M] & [s_0] & [s_{2M-1}] \\ \hline
 \mbox{}[r_0] & A_{00} & A_{01} & \ldots & A_{0,M-1} & A_{0M} & B_{00} & B_{01} \\
 \mbox{}[r_1] & A_{10} & A_{11} & \ldots & A_{1,M-1} & A_{1M} & 0 & 0 \\
 \vdots & \vdots & \vdots & \vdots & \vdots & \vdots & \vdots & \vdots \\
 \mbox{}[r_{M-1}] & A_{M-1,0} & A_{M-1,1} & \ldots & A_{M-1,M-1} & A_{M-1,M} & 0 & 0 \\
 \mbox{}[r_M] & A_{M0} & A_{M1} & \ldots & A_{M,M-1} & A_{MM} & B_{M0} & B_{M1} \\
 \mbox{}[s_0] & B_{00}^T & 0 & \ldots & 0 & B_{M0}^T & P & Q \\
 \mbox{}[s_{2M-1}] & B_{01}^T & 0 & \ldots & 0 & B_{M1}^T & Q^T & P
\end{array}
\right) ~.
\label{S-even-dihedral}
\end{equation}
Here the blocks written as $0$ are the matrices whose all elements are 0. Further note that $A_{yx} = A_{xy}^T$ where the superscript $T$ denotes the transpose of the matrix. Next we write the elements of matrices for each block. In order to write these matrices in a compact fashion, we define a notation $[a]_{m\times n}$ which will represent a matrix of order $m \times n$ whose each entry is $a$.
\begin{equation}
A_{00} = \frac{1}{4M}\left(
\begin{array}{cc}
 \mbox{}[1]_{4 \times 4} & [2]_{4 \times (M-1)} \\
 \mbox{}[2]_{(M-1) \times 4}\quad\,\, & [4]_{(M-1) \times (M-1)}
\end{array}
\right) ,\quad A_{0x} = \frac{1}{2M}\left(
\begin{array}{c}
 \mbox{}[1]_{2 \times 2M} \\[0.1cm]
 \mbox{}[\exp(-i \pi x)]_{2 \times 2M} \\[0.1cm]
R
\end{array}
\right) ~,
\end{equation}
where $x=1,2,\ldots,(M-1)$ and $R$ is a matrix whose element is defined as $R_{pq} =2\cos(\frac{\pi p x}{M})$ such that $1 \leq p < M$ and $1 \leq q \leq 2M$. Next the matrix $A_{0M}$ will be,
\begin{equation}
A_{0M} = \frac{1}{4M}\left(
\begin{array}{cc}
 \mbox{}[1]_{2 \times 4} & [2]_{2 \times (M-1)} \\[0.1cm]
  \mbox{}[\exp(-i \pi M)]_{2 \times 4} \quad \quad & [2\exp(-i \pi M)]_{2 \times (M-1)} \\[0.1cm]
	R_1 & R_2
\end{array}
\right) ;\,\,\, (A_{xy})_{ab} = \frac{1}{M}\cos(\frac{\pi(ay+bx)}{M}) \nonumber ~,
\end{equation}
where $R_1$ is a matrix whose element is defined as $(R_1)_{pq} = 2\cos(\pi p)$ such that $1 \leq p < M$ and $1 \leq q \leq 4$. Similarly $R_2$ is a matrix whose element is defined as $(R_2)_{rs} = 4\cos(\pi r)$ such that $1 \leq r, s < M$. The element of the matrix $A_{xy}$ is given above such that $0 \leq a,b < 2M$. The matrix $A_{xM}$ for $x=1,2,\ldots,(M-1)$ will be given as:
\begin{equation}
A_{xM} = \frac{1}{2M}\left(
\begin{array}{ccccc}
 E_1 & E_1 & E_2 & E_2 & E_3
\end{array}
\right)  ~,
\end{equation}
where $E_1$ and $E_2$ are single column matrices whose elements are given as: $(E_1)_{b,1} = \exp(-i \pi b)$ and $(E_2)_{b,1} = \exp(-i \pi b) \exp(-i \pi x)$, where $0 \leq b < 2M$. The elements of matrix $E_3$ are given as $(E_3)_{pq} = 2\exp(-i \pi p) \cos(\pi q x/M)$ such that $0 \leq p < 2M$ and $1 \leq q < M$. The matrix $A_{MM}$ is given as,
\begin{equation}
A_{MM} = \frac{1}{4M}\left(
\begin{array}{ccccc}
1 & 1 & \exp(-i \pi M) & \exp(-i \pi M) & F_1 \\[0.1cm]
1 & 1 & \exp(-i \pi M) & \exp(-i \pi M) & F_1 \\[0.1cm]
\exp(-i \pi M) & \exp(-i \pi M) & 1 & 1 & F_2 \\[0.1cm]
\exp(-i \pi M) & \exp(-i \pi M) & 1 & 1 & F_2 \\[0.1cm]
F_1^T & F_1^T & F_2^T & F_2^T & F_3
\end{array}
\right)  ~,
\end{equation}
where $F_1$ and $F_2$ are single row matrices whose elements are given as $(F_1)_{1,b} = 2\cos (\pi  b)$ and $(F_2)_{1,b} = 2\cos(\pi  b) \exp(-i \pi M)$ respectively, where $1 \leq b < M$. The elements of matrix $F_3$ are given as $(F_3)_{ab} = 4\cos(\pi  a) \cos(\pi  b)$ such that $1 \leq a,b < M$.
So far we have tabulated all the blocks given by $A$ matrices.  Let us now give the results for $B$ matrices. 
\begin{equation}
B_{00} = \frac{1}{4}\left(
\begin{array}{c}
[1]_{1\times 4} \\
\mbox{}[-1]_{1\times 4} \\
\mbox{}[1]_{1\times 4} \\
\mbox{}[-1]_{1\times 4} \\
\mbox{}[0]_{(M-1)\times 4}
\end{array}
\right), 
B_{01} = \frac{1}{4}\left(
\begin{array}{c}
[1]_{1\times 4} \\
\mbox{}[-1]_{1\times 4} \\
\mbox{}[-1]_{1\times 4} \\
\mbox{}[1]_{1\times 4} \\
\mbox{}[0]_{(M-1)\times 4}
\end{array}
\right),
B_{M0} = \left(
\begin{array}{cc}
F & F \\
F & F \\
0 & 0
\end{array}
\right),
B_{M1} = \left(
\begin{array}{cc}
F & F \\
G & G \\
0 & 0
\end{array}
\right) \nonumber
~, 
\end{equation}
where each of the 0 matrix in $B_{M0}$ and $B_{M1}$ is of order $(M-1) \times 2$ and
\begin{equation}
F = \frac{1}{4}\left(
\begin{array}{cc}
 1 & -1 \\
-1 & 1
\end{array}
\right) \quad,\quad G = \frac{1}{4}\left(
\begin{array}{cc}
 -1 & 1 \\
1 & -1
\end{array}
\right) ~.
\end{equation}
Finally $P$ and $Q$ matrices are given as,
\begin{gather}
P(\text{odd } M) = \frac{1}{4}\left(
\begin{array}{cccc}
 1 & 1 & -1 & -1 \\
 1 & 1 & -1 & -1 \\
 -1 & -1 & 1 & 1 \\
 -1 & -1 & 1 & 1 \\
\end{array}
\right),\quad P(\text{even } M) = \frac{1}{2}\left(
\begin{array}{cccc}
 1 & 0 & -1 & 0 \\
 0 & 1 & 0 & -1 \\
 -1 & 0 & 1 & 0 \\
 0 & -1 & 0 & 1 \\
\end{array}
\right) \\
Q(\text{odd } M) = \frac{1}{4}\left(
\begin{array}{cccc}
 1 & -1 & -1 & 1 \\
 -1 & 1 & 1 & -1 \\
 -1 & 1 & 1 & -1 \\
 1 & -1 & -1 & 1 \\
\end{array}
\right),\quad Q(\text{even } M) = \left(
\begin{array}{cccc}
 0 & 0 & 0 & 0 \\
 0 & 0 & 0 & 0 \\
 0 & 0 & 0 & 0 \\
 0 & 0 & 0 & 0 \\
\end{array}
\right) ~.
\end{gather}
This completes the modular $\mathcal{S}$ data for the even dihedral group $D_{2M}$. The modular $\mathcal{T}$ matrix is a diagonal matrix whose diagonal elements for $D_{2M}$ group are given as,
\begin{equation}
    \begin{split}
\mathcal{T} = \text{diag}\Bigl\{ &\underbrace{1,1,\ldots,1}_{M+3} \,, \underbrace{f(1,a)}_{a=0,\ldots,2M-1}, \underbrace{f(2,a)}_{a=0,\ldots,2M-1},\ldots, \underbrace{f(M-1,a)}_{a=0,\ldots,2M-1}, 1, 1, \exp(i \pi M), \exp(i \pi M), \\
  &  \cos\pi,\cos2\pi,\ldots,\cos(M-1)\pi, 1,1,-1,-1,1,1,-1,-1   \Bigr\} ~,
\end{split}
\label{T-even-dihedral}
    \end{equation}
where $f(x,a) = \exp(i \pi ax/M)$. 
As an example to clarify these computations, we give the result for $\mathcal{S}$ and $\mathcal{T}$ matrices of $D_4$ group which are matrices of order 22 and can be written as:
\begin{gather}
\mathcal{S}(D_4) = {\left(
\begin{array}{ccccc}
 A_{00} & A_{01} & A_{02} & B_{00} & B_{01} \\
 A_{01}^T & A_{11} & A_{12} & 0 & 0 \\
 A_{02}^T & A_{12}^T & A_{22} & B_{20} & B_{21} \\
 B_{00}^T & 0 & B_{20}^T & P & Q \\
 B_{01}^T & 0 & B_{21}^T & Q^T & P
\end{array}
\right)_{22 \times 22}} \nonumber \\[0.2cm]
\mathcal{T}(D_4) = \text{diag}\left\{ 1,1,1,1,1,1,i,-1,-i,1,1,1,1,-1,1,1,-1,-1,1,1,-1,-1\right\}~.
\label{S-matrix-D4}
\end{gather} 
The various blocks in $\mathcal{S}$ are given as following:
\begin{gather}
A_{00}=\frac{1}{8} \left(
\begin{array}{ccccc}
 1 & 1 & 1 & 1 & 2 \\
 1 & 1 & 1 & 1 & 2 \\
 1 & 1 & 1 & 1 & 2 \\
 1 & 1 & 1 & 1 & 2 \\
 2 & 2 & 2 & 2 & 4 \\
\end{array}
\right), A_{01}= \frac{1}{4} \left(
\begin{array}{cccc}
 1 & 1 & 1 & 1 \\
 1 & 1 & 1 & 1 \\
 -1 & -1 & -1 & -1 \\
 -1 & -1 & -1 & -1 \\
 0 & 0 & 0 & 0 \\
\end{array}
\right), A_{02}=\frac{1}{8} \left(
\begin{array}{ccccc}
 1 & 1 & 1 & 1 & 2 \\
 1 & 1 & 1 & 1 & 2 \\
 1 & 1 & 1 & 1 & 2 \\
 1 & 1 & 1 & 1 & 2 \\
 -2 & -2 & -2 & -2 & -4 \\
\end{array}
\right) \nonumber \\
A_{11}=\frac{1}{2} \left(
\begin{array}{cccc}
 1 & 0 & -1 & 0 \\
 0 & -1 & 0 & 1 \\
 -1 & 0 & 1 & 0 \\
 0 & 1 & 0 & -1 \\
\end{array}
\right), A_{12}=\frac{1}{4} \left(
\begin{array}{ccccc}
 1 & 1 & -1 & -1 & 0 \\
 -1 & -1 & 1 & 1 & 0 \\
 1 & 1 & -1 & -1 & 0 \\
 -1 & -1 & 1 & 1 & 0 \\
\end{array}
\right),A_{22}=\frac{1}{8} \left(
\begin{array}{ccccc}
 1 & 1 & 1 & 1 & -2 \\
 1 & 1 & 1 & 1 & -2 \\
 1 & 1 & 1 & 1 & -2 \\
 1 & 1 & 1 & 1 & -2 \\
 -2 & -2 & -2 & -2 & 4 \\
\end{array}
\right) \nonumber
~.
\end{gather} 
The remaining blocks are,
\begin{gather}
B_{00}=\frac{1}{4} \left(
\begin{array}{cccc}
 1 & 1 & 1 & 1 \\
 -1 & -1 & -1 & -1 \\
 1 & 1 & 1 & 1 \\
 -1 & -1 & -1 & -1 \\
 0 & 0 & 0 & 0 \\
\end{array}
\right), B_{01}=\frac{1}{4}\left(
\begin{array}{cccc}
 1 & 1 & 1 & 1 \\
 -1 & -1 & -1 & -1 \\
 -1 & -1 & -1 & -1 \\
 1 & 1 & 1 & 1 \\
 0 & 0 & 0 & 0 \\
\end{array}
\right), B_{20}=\frac{1}{4} \left(
\begin{array}{cccc}
 1 & -1 & 1 & -1 \\
 -1 & 1 & -1 & 1 \\
 1 & -1 & 1 & -1 \\
 -1 & 1 & -1 & 1 \\
 0 & 0 & 0 & 0 \\
\end{array}
\right) \nonumber \\
B_{21}=\frac{1}{4}\left(
\begin{array}{cccc}
 1 & -1 & 1 & -1 \\
 -1 & 1 & -1 & 1 \\
 -1 & 1 & -1 & 1 \\
 1 & -1 & 1 & -1 \\
 0 & 0 & 0 & 0 \\
\end{array}
\right), P=\frac{1}{2} \left(
\begin{array}{cccc}
 1 & 0 & -1 & 0 \\
 0 & 1 & 0 & -1 \\
 -1 & 0 & 1 & 0 \\
 0 & -1 & 0 & 1 \\
\end{array}
\right), Q=\left(
\begin{array}{cccc}
 0 & 0 & 0 & 0 \\
 0 & 0 & 0 & 0 \\
 0 & 0 & 0 & 0 \\
 0 & 0 & 0 & 0 \\
\end{array}
\right) \nonumber ~.
\end{gather} 
Each entry $0$ in eq.(\ref{S-matrix-D4}) represents zero matrix of order 4. One can easily verify that $\mathcal{S}$ and $\mathcal{T}$ matrices in eq.(\ref{S-matrix-D4}) are symmetric, unitary and satisfy $\mathcal{S}^2 = (\mathcal{S}\mathcal{T})^3=\mathbb{1}$.
\subsubsection{Odd dihedral group: $D_{2M+1}$}
The odd dihedral group has $(M+2)$ number of conjugacy classes given by their representative elements as:
\begin{equation}
\text{conjugacy classes} = [r_0], [r_1], [r_2], \ldots, [r_M], [s_0] ~.
\end{equation}
This group has 2 one-dimensional irreps and $M$ number of two-dimensional irreps. If we denote $\theta_i$ and $\phi_j$ as the irreducible characters of one-dimensional and two-dimensional irreps respectively, the character table can be given as following:
\begin{equation}
\begin{array}{|c|c|c|c|} \hline
   & [r_0] & [r_x] & [s_0] \\ \hline
 \theta_1 & 1 & 1 & 1 \\
 \theta_2 & 1 & 1 & -1 \\
 \phi_j   & 2 & 2\cos(\frac{2\pi j x}{2M+1}) & 0  \\\hline
 C_g & D_{2M+1} & \mathbb{Z}_{2M+1} & \mathbb{Z}_{2} \\ \hline
\end{array} ~.
\label{table-odd-D}
\end{equation}
Here each value of $x$ forms a conjugacy class where $x = 1, 2, \ldots, M$. The $\phi_j$ is the character of the $j^{\text{th}}$ two-dimensional irrep where $j=1,2,\ldots, M$. The last row of this table gives the centralizer for each conjugacy class which can be easily obtained using the composition rules. Since each $r_x$ commutes only with $r_i$, we get $C_{r_{x}} = \{r_0, r_1, \ldots, r_{2M}\} \cong \mathbb{Z}_{2M+1}$. The centralizer for $s_0$ is given as $C_{s_{0}} = \{r_0, s_0 \} \cong \mathbb{Z}_{2}$.
In order to obtain the modular data, we need the irreducible characters of centralizers for each class. The characters for $D_{2M+1}$ are already given in table in eq.(\ref{table-odd-D}). The characters for $\mathbb{Z}_{2M+1}$ are denoted as $\chi_a$ where $0 \leq a \leq 2M$ and their values for the element $r_x$ are given as: $\chi_a(r_x) = \exp(\frac{2 \pi i a x}{2M+1})$. Similarly the character $\gamma_{b}$ for the group $C_{s_{0}}$ where $b=0,1$ are given for the elements of $C_{s_0}$ as:  $\gamma_{b}(r_0) = 1$ and $\gamma_{b}(s_0) = (-1)^b$. With all the information at our hand, the modular matrices can be obtained. The $\mathcal{S}$ matrix in terms of blocks labeled by conjugacy classes is written as:
\begin{equation}
\mathcal{S} = \left(
 \begin{array}{c|ccccc}
  & [r_0] & [r_1] & \ldots & [r_M] & [s_0] \\ \hline
 \mbox{}[r_0] & A_{00} & A_{01} & \ldots & A_{0M} & P \\
 \mbox{}[r_1] & A_{10} & A_{11} & \ldots & A_{1M} & 0 \\
 \vdots & \vdots & \vdots & \vdots & \vdots & \vdots \\
 \mbox{}[r_M] & A_{M0} & A_{M1} & \ldots & A_{MM} & 0 \\
 \mbox{}[s_0] & P^T & 0 & \ldots & 0 & Q
\end{array}
\right) ~.
\label{S-odd-dihedral}
\end{equation}
The block $A_{yx} = A_{xy}^T$ where the superscript $T$ denotes the transpose of the matrix. Next we write the elements of matrices for each block. We define the notation $[a]_{m\times n}$ which will represent a matrix of order $m \times n$ whose each entry is $a$. Using this, the elements of various blocks are given as,
\begin{equation}
A_{00} = \frac{1}{2(2M+1)}\left(
\begin{array}{cc}
 \mbox{}[1]_{2 \times 2} & [2]_{2 \times M} \\
 \mbox{}[2]_{M \times 2}\quad\,\, & [4]_{M \times M}
\end{array}
\right) ,\quad A_{0x} = \frac{1}{(2M+1)}\left(
\begin{array}{c}
 \mbox{}[1]_{2 \times (2M+1)} \\[0.1cm]
R
\end{array}
\right) ~,
\end{equation}
where $x=1,2,\ldots,M$ and $R$ is a matrix whose element is defined as $R_{pq} =2\cos(\frac{2\pi p x}{2M+1})$ such that $1 \leq p \leq M$ and $1 \leq q \leq (2M+1)$. The matrix $A_{xy}$ is a square matrix of order $(2M+1)$ whose elements are given as,
\begin{equation}
(A_{xy})_{ab} = \frac{2}{(2M+1)}\cos\left[\frac{2 \pi}{(2M+1)} \left(ay+bx\right)\right] ~,
\end{equation}
where $0 \leq a,b \leq 2M$ and the matrices $A_{xy}$ are given for $1 \leq x,y \leq M$. The matrix $P$ and $Q$ are given as,
\begin{equation}
P = \frac{1}{2}\left(
\begin{array}{c}
 \mbox{}[1]_{1 \times 2} \\
 \mbox{}[-1]_{1 \times 2} \\
 \mbox{}[0]_{M \times 2}
\end{array}
\right)\quad,\quad
Q = \frac{1}{2}\left(
\begin{array}{cc}
 1 & -1  \\
 -1 & 1
\end{array}
\right) ~.
\end{equation}
This completes the modular $\mathcal{S}$ data for the odd dihedral group $D_{2M+1}$. The modular $\mathcal{T}$ matrix is a diagonal matrix whose diagonal elements are given as,
\begin{equation}
\mathcal{T} = \text{diag}\left\{ \underbrace{1,\ldots,1}_{M+2}, \underbrace{f(1,a)}_{a=0,\ldots,2M}, \underbrace{f(2,a)}_{a=0,\ldots,2M},\ldots, \underbrace{f(M,a)}_{a=0,\ldots,2M},\, 1,\, -1 \right\} ~,
\label{T-odd-dihedral}
    \end{equation}
where $f(x,a) = \exp(\frac{2 \pi i a x}{2M+1})$. 
As an example to clarify these computations, we give the result for $\mathcal{S}$ and $\mathcal{T}$ matrices of $D_5$ group which are matrices of order 16:
\begin{align}
\mathcal{S}(D_5)&=\frac{1}{10} \left(
\begin{array}{cccccccccccccccc}
 1 & 1 & 2 & 2 & 2 & 2 & 2 & 2 & 2 & 2 & 2 & 2 & 2 & 2 & 5 & 5 \\
 1 & 1 & 2 & 2 & 2 & 2 & 2 & 2 & 2 & 2 & 2 & 2 & 2 & 2 & -5 & -5 \\
 2 & 2 & 4 & 4 & x & x & x & x & x & y & y & y & y & y & 0 & 0 \\
 2 & 2 & 4 & 4 & y & y & y & y & y & x & x & x & x & x & 0 & 0 \\
 2 & 2 & x & y & 4 & x & y & y & x & 4 & x & y & y & x & 0 & 0 \\
 2 & 2 & x & y & x & y & y & x & 4 & y & y & x & 4 & x & 0 & 0 \\
 2 & 2 & x & y & y & y & x & 4 & x & x & 4 & x & y & y & 0 & 0 \\
 2 & 2 & x & y & y & x & 4 & x & y & x & y & y & x & 4 & 0 & 0 \\
 2 & 2 & x & y & x & 4 & x & y & y & y & x & 4 & x & y & 0 & 0 \\
 2 & 2 & y & x & 4 & y & x & x & y & 4 & y & x & x & y & 0 & 0 \\
 2 & 2 & y & x & x & y & 4 & y & x & y & x & x & y & 4 & 0 & 0 \\
 2 & 2 & y & x & y & x & x & y & 4 & x & x & y & 4 & y & 0 & 0 \\
 2 & 2 & y & x & y & 4 & y & x & x & x & y & 4 & y & x & 0 & 0 \\
 2 & 2 & y & x & x & x & y & 4 & y & y & 4 & y & x & x & 0 & 0 \\
 5 & -5 & 0 & 0 & 0 & 0 & 0 & 0 & 0 & 0 & 0 & 0 & 0 & 0 & 5 & -5 \\
 5 & -5 & 0 & 0 & 0 & 0 & 0 & 0 & 0 & 0 & 0 & 0 & 0 & 0 & -5 & 5 \\
\end{array}
\right) \nonumber \\[0.2cm]
\mathcal{T}(D_5)&= \text{diag}\left\{1,1,1,1,1,z^2,z^4,-z,-z^3,1,z^4,-z^3,z^2,-z,1,-1\right\} \nonumber ~,
\end{align}
where $x=(\sqrt{5}-1)$, $y=-(\sqrt{5}+1)$, $z=e^{\frac{i \pi }{5}}$.
\subsection{Modular $\mathcal{S}$ and $\mathcal{T}$ matrices for symmetric groups: $G=S_N$}
The groups $S_2$ and $S_3$ are isomorphic to $\mathbb{Z}_2$ and $D_3$ groups respectively whose modular matrices have already been obtained. In this appendix, we give the $\mathcal{S}$ and $\mathcal{T}$ matrices for $S_4$ and $S_5$ groups. The $S_4$ group has five conjugacy classes given by the following representative elements:
\begin{equation}
a_1 = e,\, a_2 = (12),\, a_3 = (12)(34),\, a_4 = (123),\, a_5 = (1234) ~.
\end{equation}
The centralizers associated with these representative elements are given as,
\begin{align}
C_{a_1} &= S_4 \nonumber \\
C_{a_2} &= \{e,\, (12),\, (34),\, (12)(34)\} \cong \mathbb{Z}_2 \times \mathbb{Z}_2 \nonumber \\
C_{a_3} &= \{e, r, r^2, r^3, s, sr, sr^2, sr^3 \} \cong D_4 \nonumber \\
C_{a_4} &= \{e,\, (123),\, (132) \} \cong \mathbb{Z}_3 \nonumber \\ 
C_{a_5} &= \{e,\, (1234),\, (13)(24),\, (1432)   \} \cong \mathbb{Z}_4 ~,
\end{align}
where the $C_{a_3}$ is generated by $r=(1324)$ and $s=(12)$ and can be presented as $C_{a_3} = \langle r,s \, | \, r^4 = s^2 = (sr)^2 = e \rangle$ which is isomorphic to dihedral $D_4$ group. The irreps and their characters for $\mathbb{Z}_2 \times \mathbb{Z}_2$, $\mathbb{Z}_3$, $\mathbb{Z}_4$ and $D_4$ have already been given earlier in this section. So, we only need the character table for $C_{a_1} \cong S_4 $ which is given below:
\begin{equation}
\begin{array}{|c|c|c|c|c|c|} \hline
   & [a_1] & [a_2] & [a_3] & [a_4] & [a_5] \\ \hline
 R_1 & 1 & 1 & 1 & 1 & 1 \\
 R_2 & 3 & 1 & -1 & 0 & -1 \\
 R_3 & 2 & 0 & 2 & -1 & 0 \\
 R_4 & 3 & -1 & -1 & 0 & 1 \\
 R_5 & 1 & -1 & 1 & 1 & -1  \\\hline
\end{array} ~.
\label{table-S4}
\end{equation}
The modular matrices can now be computed and are given as:
\begin{align}
\mathcal{S}(S_4) &= \frac{1}{4!}\resizebox{0.8\textwidth}{!}{$\left(
\begin{array}{ccccccccccccccccccccc}
 1 & 3 & 2 & 3 & 1 & 6 & 6 & 6 & 6 & 3 & 3 & 3 & 3 & 6 & 8 & 8 & 8 & 6 & 6 & 6 & 6 \\
 3 & 9 & 6 & 9 & 3 & 6 & 6 & 6 & 6 & -3 & -3 & -3 & -3 & -6 & 0 & 0 & 0 & -6 & -6 & -6 & -6 \\
 2 & 6 & 4 & 6 & 2 & 0 & 0 & 0 & 0 & 6 & 6 & 6 & 6 & 12 & -8 & -8 & -8 & 0 & 0 & 0 & 0 \\
 3 & 9 & 6 & 9 & 3 & -6 & -6 & -6 & -6 & -3 & -3 & -3 & -3 & -6 & 0 & 0 & 0 & 6 & 6 & 6 & 6 \\
 1 & 3 & 2 & 3 & 1 & -6 & -6 & -6 & -6 & 3 & 3 & 3 & 3 & 6 & 8 & 8 & 8 & -6 & -6 & -6 & -6 \\
 6 & 6 & 0 & -6 & -6 & 12 & 0 & 0 & -12 & 6 & -6 & 6 & -6 & 0 & 0 & 0 & 0 & 0 & 0 & 0 & 0 \\
 6 & 6 & 0 & -6 & -6 & 0 & 12 & -12 & 0 & -6 & 6 & -6 & 6 & 0 & 0 & 0 & 0 & 0 & 0 & 0 & 0 \\
 6 & 6 & 0 & -6 & -6 & 0 & -12 & 12 & 0 & -6 & 6 & -6 & 6 & 0 & 0 & 0 & 0 & 0 & 0 & 0 & 0 \\
 6 & 6 & 0 & -6 & -6 & -12 & 0 & 0 & 12 & 6 & -6 & 6 & -6 & 0 & 0 & 0 & 0 & 0 & 0 & 0 & 0 \\
 3 & -3 & 6 & -3 & 3 & 6 & -6 & -6 & 6 & 9 & -3 & -3 & 9 & -6 & 0 & 0 & 0 & 6 & -6 & 6 & -6 \\
 3 & -3 & 6 & -3 & 3 & -6 & 6 & 6 & -6 & -3 & 9 & 9 & -3 & -6 & 0 & 0 & 0 & 6 & -6 & 6 & -6 \\
 3 & -3 & 6 & -3 & 3 & 6 & -6 & -6 & 6 & -3 & 9 & 9 & -3 & -6 & 0 & 0 & 0 & -6 & 6 & -6 & 6 \\
 3 & -3 & 6 & -3 & 3 & -6 & 6 & 6 & -6 & 9 & -3 & -3 & 9 & -6 & 0 & 0 & 0 & -6 & 6 & -6 & 6 \\
 6 & -6 & 12 & -6 & 6 & 0 & 0 & 0 & 0 & -6 & -6 & -6 & -6 & 12 & 0 & 0 & 0 & 0 & 0 & 0 & 0 \\
 8 & 0 & -8 & 0 & 8 & 0 & 0 & 0 & 0 & 0 & 0 & 0 & 0 & 0 & 16 & -8 & -8 & 0 & 0 & 0 & 0 \\
 8 & 0 & -8 & 0 & 8 & 0 & 0 & 0 & 0 & 0 & 0 & 0 & 0 & 0 & -8 & -8 & 16 & 0 & 0 & 0 & 0 \\
 8 & 0 & -8 & 0 & 8 & 0 & 0 & 0 & 0 & 0 & 0 & 0 & 0 & 0 & -8 & 16 & -8 & 0 & 0 & 0 & 0 \\
 6 & -6 & 0 & 6 & -6 & 0 & 0 & 0 & 0 & 6 & 6 & -6 & -6 & 0 & 0 & 0 & 0 & 12 & 0 & -12 & 0 \\
 6 & -6 & 0 & 6 & -6 & 0 & 0 & 0 & 0 & -6 & -6 & 6 & 6 & 0 & 0 & 0 & 0 & 0 & -12 & 0 & 12 \\
 6 & -6 & 0 & 6 & -6 & 0 & 0 & 0 & 0 & 6 & 6 & -6 & -6 & 0 & 0 & 0 & 0 & -12 & 0 & 12 & 0 \\
 6 & -6 & 0 & 6 & -6 & 0 & 0 & 0 & 0 & -6 & -6 & 6 & 6 & 0 & 0 & 0 & 0 & 0 & 12 & 0 & -12 \\
\end{array}
\right)$} \nonumber \\
\mathcal{T}(S_4) &= \text{diag}\left\{1,1,1,1,1,1,-1,1,-1,1,1,1,1,-1,1,(-1)^{2/3},(-1)^{4/3},1,i,-1,-i\right\} ~.
\end{align}

For the group $S_5$, there are seven conjugacy classes given by the following representative elements:
\begin{equation}
b_1 = e,\, b_2 = (12),\, b_3 = (12)(34),\, b_4 = (123),\, b_5 = (123)(45),\, b_6 = (1234),\, b_7 = (12345) \nonumber ~.
\end{equation}
The centralizers associated with these representative elements are given as,
\begin{align}
C_{b_1} &= S_5 \nonumber \\
C_{b_2} &= \langle r=(12)(345),\,s=(34) \,\, | \,\, r^6 = s^2 = (sr)^2 = e \rangle \cong D_6 \nonumber \\
C_{b_3} &= \langle r=(1324),\,s=(12) \,\, | \,\, r^4 = s^2 = (sr)^2 = e \rangle \cong D_4 \nonumber \\
C_{b_4} &= \langle r=(123)(45) \,\, | \,\, r^6 = e \rangle \cong \mathbb{Z}_6 \nonumber \\ 
C_{b_5} &= \langle r=(123)(45) \,\, | \,\, r^6 = e \rangle \cong \mathbb{Z}_6 \nonumber \\ 
C_{b_6} &= \langle r=(1234) \,\, | \,\, r^4 = e \rangle \cong \mathbb{Z}_4 \nonumber \\
C_{b_7} &= \langle r=(12345) \,\, | \,\, r^5 = e \rangle \cong \mathbb{Z}_5 ~.
\end{align}
The irreducible characters for all these centralizers have been already tabulated before, except for $S_5$ group, which is given below:
\begin{equation}
\begin{array}{|c|c|c|c|c|c|c|c|} \hline
   & [b_1] & [b_2] & [b_3] & [b_4] & [b_5] & [b_6] & [b_7] \\ \hline
 R_1 & 1 & 1 & 1 & 1 & 1 & 1 & 1\\
 R_2 & 4 & 2 & 0 & 1 & -1 & 0 & -1 \\
 R_3 & 5 & 1 & 1 & -1 & 1 & -1 & 0 \\
 R_4 & 6 & 0 & -2 & 0 & 0 & 0 & 1 \\
 R_5 & 5 & -1 & 1 & -1 & -1 & 1 & 0  \\
 R_6 & 4 & -2 & 0 & 1 & 1 & 0 & -1  \\
 R_7 & 1 & -1 & 1 & 1 & -1 & -1 & 1  \\ \hline
\end{array} ~.
\label{table-S5}
\end{equation}
Thus the modular matrices can be computed for $S_5$ group and are given in eq.(\ref{ST-for-S5}).
\begin{landscape}
\begin{align}
\mathcal{S}(S_5) &= \frac{1}{5!}\resizebox{1.40\textwidth}{!}{$ \left(
\begin{array}{ccccccccccccccccccccccccccccccccccccccc}
 1 & 4 & 5 & 6 & 5 & 4 & 1 & 10 & 10 & 10 & 10 & 20 & 20 & 15 & 15 & 15 & 15 & 30 & 20 & 20 & 20 & 20 & 20 & 20 & 20 & 20 & 20 & 20 & 20 & 20 & 30 & 30 & 30 & 30 & 24 & 24 & 24 & 24 & 24 \\
 4 & 16 & 20 & 24 & 20 & 16 & 4 & 20 & 20 & 20 & 20 & 40 & 40 & 0 & 0 & 0 & 0 & 0 & 20 & 20 & 20 & 20 & 20 & 20 & -20 & -20 & -20 & -20 & -20 & -20 & 0 & 0 & 0 & 0 & -24 & -24 & -24 & -24 & -24 \\
 5 & 20 & 25 & 30 & 25 & 20 & 5 & 10 & 10 & 10 & 10 & 20 & 20 & 15 & 15 & 15 & 15 & 30 & -20 & -20 & -20 & -20 & -20 & -20 & 20 & 20 & 20 & 20 & 20 & 20 & -30 & -30 & -30 & -30 & 0 & 0 & 0 & 0 & 0 \\
 6 & 24 & 30 & 36 & 30 & 24 & 6 & 0 & 0 & 0 & 0 & 0 & 0 & -30 & -30 & -30 & -30 & -60 & 0 & 0 & 0 & 0 & 0 & 0 & 0 & 0 & 0 & 0 & 0 & 0 & 0 & 0 & 0 & 0 & 24 & 24 & 24 & 24 & 24 \\
 5 & 20 & 25 & 30 & 25 & 20 & 5 & -10 & -10 & -10 & -10 & -20 & -20 & 15 & 15 & 15 & 15 & 30 & -20 & -20 & -20 & -20 & -20 & -20 & -20 & -20 & -20 & -20 & -20 & -20 & 30 & 30 & 30 & 30 & 0 & 0 & 0 & 0 & 0 \\
 4 & 16 & 20 & 24 & 20 & 16 & 4 & -20 & -20 & -20 & -20 & -40 & -40 & 0 & 0 & 0 & 0 & 0 & 20 & 20 & 20 & 20 & 20 & 20 & 20 & 20 & 20 & 20 & 20 & 20 & 0 & 0 & 0 & 0 & -24 & -24 & -24 & -24 & -24 \\
 1 & 4 & 5 & 6 & 5 & 4 & 1 & -10 & -10 & -10 & -10 & -20 & -20 & 15 & 15 & 15 & 15 & 30 & 20 & 20 & 20 & 20 & 20 & 20 & -20 & -20 & -20 & -20 & -20 & -20 & -30 & -30 & -30 & -30 & 24 & 24 & 24 & 24 & 24 \\
 10 & 20 & 10 & 0 & -10 & -20 & -10 & 40 & -20 & 20 & -40 & -20 & 20 & 30 & -30 & 30 & -30 & 0 & 20 & -20 & 20 & -20 & 20 & -20 & 20 & -20 & 20 & -20 & 20 & -20 & 0 & 0 & 0 & 0 & 0 & 0 & 0 & 0 & 0 \\
 10 & 20 & 10 & 0 & -10 & -20 & -10 & -20 & 40 & -40 & 20 & -20 & 20 & -30 & 30 & -30 & 30 & 0 & 20 & -20 & 20 & -20 & 20 & -20 & 20 & -20 & 20 & -20 & 20 & -20 & 0 & 0 & 0 & 0 & 0 & 0 & 0 & 0 & 0 \\
 10 & 20 & 10 & 0 & -10 & -20 & -10 & 20 & -40 & 40 & -20 & 20 & -20 & -30 & 30 & -30 & 30 & 0 & 20 & -20 & 20 & -20 & 20 & -20 & -20 & 20 & -20 & 20 & -20 & 20 & 0 & 0 & 0 & 0 & 0 & 0 & 0 & 0 & 0 \\
 10 & 20 & 10 & 0 & -10 & -20 & -10 & -40 & 20 & -20 & 40 & 20 & -20 & 30 & -30 & 30 & -30 & 0 & 20 & -20 & 20 & -20 & 20 & -20 & -20 & 20 & -20 & 20 & -20 & 20 & 0 & 0 & 0 & 0 & 0 & 0 & 0 & 0 & 0 \\
 20 & 40 & 20 & 0 & -20 & -40 & -20 & -20 & -20 & 20 & 20 & 40 & -40 & 0 & 0 & 0 & 0 & 0 & -20 & 20 & -20 & 20 & -20 & 20 & 20 & -20 & 20 & -20 & 20 & -20 & 0 & 0 & 0 & 0 & 0 & 0 & 0 & 0 & 0 \\
 20 & 40 & 20 & 0 & -20 & -40 & -20 & 20 & 20 & -20 & -20 & -40 & 40 & 0 & 0 & 0 & 0 & 0 & -20 & 20 & -20 & 20 & -20 & 20 & -20 & 20 & -20 & 20 & -20 & 20 & 0 & 0 & 0 & 0 & 0 & 0 & 0 & 0 & 0 \\
 15 & 0 & 15 & -30 & 15 & 0 & 15 & 30 & -30 & -30 & 30 & 0 & 0 & 45 & -15 & -15 & 45 & -30 & 0 & 0 & 0 & 0 & 0 & 0 & 0 & 0 & 0 & 0 & 0 & 0 & 30 & -30 & 30 & -30 & 0 & 0 & 0 & 0 & 0 \\
 15 & 0 & 15 & -30 & 15 & 0 & 15 & -30 & 30 & 30 & -30 & 0 & 0 & -15 & 45 & 45 & -15 & -30 & 0 & 0 & 0 & 0 & 0 & 0 & 0 & 0 & 0 & 0 & 0 & 0 & 30 & -30 & 30 & -30 & 0 & 0 & 0 & 0 & 0 \\
 15 & 0 & 15 & -30 & 15 & 0 & 15 & 30 & -30 & -30 & 30 & 0 & 0 & -15 & 45 & 45 & -15 & -30 & 0 & 0 & 0 & 0 & 0 & 0 & 0 & 0 & 0 & 0 & 0 & 0 & -30 & 30 & -30 & 30 & 0 & 0 & 0 & 0 & 0 \\
 15 & 0 & 15 & -30 & 15 & 0 & 15 & -30 & 30 & 30 & -30 & 0 & 0 & 45 & -15 & -15 & 45 & -30 & 0 & 0 & 0 & 0 & 0 & 0 & 0 & 0 & 0 & 0 & 0 & 0 & -30 & 30 & -30 & 30 & 0 & 0 & 0 & 0 & 0 \\
 30 & 0 & 30 & -60 & 30 & 0 & 30 & 0 & 0 & 0 & 0 & 0 & 0 & -30 & -30 & -30 & -30 & 60 & 0 & 0 & 0 & 0 & 0 & 0 & 0 & 0 & 0 & 0 & 0 & 0 & 0 & 0 & 0 & 0 & 0 & 0 & 0 & 0 & 0 \\
 20 & 20 & -20 & 0 & -20 & 20 & 20 & 20 & 20 & 20 & 20 & -20 & -20 & 0 & 0 & 0 & 0 & 0 & 40 & -20 & -20 & 40 & -20 & -20 & 40 & -20 & -20 & 40 & -20 & -20 & 0 & 0 & 0 & 0 & 0 & 0 & 0 & 0 & 0 \\
 20 & 20 & -20 & 0 & -20 & 20 & 20 & -20 & -20 & -20 & -20 & 20 & 20 & 0 & 0 & 0 & 0 & 0 & -20 & -20 & 40 & -20 & -20 & 40 & 20 & 20 & -40 & 20 & 20 & -40 & 0 & 0 & 0 & 0 & 0 & 0 & 0 & 0 & 0 \\
 20 & 20 & -20 & 0 & -20 & 20 & 20 & 20 & 20 & 20 & 20 & -20 & -20 & 0 & 0 & 0 & 0 & 0 & -20 & 40 & -20 & -20 & 40 & -20 & -20 & 40 & -20 & -20 & 40 & -20 & 0 & 0 & 0 & 0 & 0 & 0 & 0 & 0 & 0 \\
 20 & 20 & -20 & 0 & -20 & 20 & 20 & -20 & -20 & -20 & -20 & 20 & 20 & 0 & 0 & 0 & 0 & 0 & 40 & -20 & -20 & 40 & -20 & -20 & -40 & 20 & 20 & -40 & 20 & 20 & 0 & 0 & 0 & 0 & 0 & 0 & 0 & 0 & 0 \\
 20 & 20 & -20 & 0 & -20 & 20 & 20 & 20 & 20 & 20 & 20 & -20 & -20 & 0 & 0 & 0 & 0 & 0 & -20 & -20 & 40 & -20 & -20 & 40 & -20 & -20 & 40 & -20 & -20 & 40 & 0 & 0 & 0 & 0 & 0 & 0 & 0 & 0 & 0 \\
 20 & 20 & -20 & 0 & -20 & 20 & 20 & -20 & -20 & -20 & -20 & 20 & 20 & 0 & 0 & 0 & 0 & 0 & -20 & 40 & -20 & -20 & 40 & -20 & 20 & -40 & 20 & 20 & -40 & 20 & 0 & 0 & 0 & 0 & 0 & 0 & 0 & 0 & 0 \\
 20 & -20 & 20 & 0 & -20 & 20 & -20 & 20 & 20 & -20 & -20 & 20 & -20 & 0 & 0 & 0 & 0 & 0 & 40 & 20 & -20 & -40 & -20 & 20 & 40 & 20 & -20 & -40 & -20 & 20 & 0 & 0 & 0 & 0 & 0 & 0 & 0 & 0 & 0 \\
 20 & -20 & 20 & 0 & -20 & 20 & -20 & -20 & -20 & 20 & 20 & -20 & 20 & 0 & 0 & 0 & 0 & 0 & -20 & 20 & 40 & 20 & -20 & -40 & 20 & -20 & -40 & -20 & 20 & 40 & 0 & 0 & 0 & 0 & 0 & 0 & 0 & 0 & 0 \\
 20 & -20 & 20 & 0 & -20 & 20 & -20 & 20 & 20 & -20 & -20 & 20 & -20 & 0 & 0 & 0 & 0 & 0 & -20 & -40 & -20 & 20 & 40 & 20 & -20 & -40 & -20 & 20 & 40 & 20 & 0 & 0 & 0 & 0 & 0 & 0 & 0 & 0 & 0 \\
 20 & -20 & 20 & 0 & -20 & 20 & -20 & -20 & -20 & 20 & 20 & -20 & 20 & 0 & 0 & 0 & 0 & 0 & 40 & 20 & -20 & -40 & -20 & 20 & -40 & -20 & 20 & 40 & 20 & -20 & 0 & 0 & 0 & 0 & 0 & 0 & 0 & 0 & 0 \\
 20 & -20 & 20 & 0 & -20 & 20 & -20 & 20 & 20 & -20 & -20 & 20 & -20 & 0 & 0 & 0 & 0 & 0 & -20 & 20 & 40 & 20 & -20 & -40 & -20 & 20 & 40 & 20 & -20 & -40 & 0 & 0 & 0 & 0 & 0 & 0 & 0 & 0 & 0 \\
 20 & -20 & 20 & 0 & -20 & 20 & -20 & -20 & -20 & 20 & 20 & -20 & 20 & 0 & 0 & 0 & 0 & 0 & -20 & -40 & -20 & 20 & 40 & 20 & 20 & 40 & 20 & -20 & -40 & -20 & 0 & 0 & 0 & 0 & 0 & 0 & 0 & 0 & 0 \\
 30 & 0 & -30 & 0 & 30 & 0 & -30 & 0 & 0 & 0 & 0 & 0 & 0 & 30 & 30 & -30 & -30 & 0 & 0 & 0 & 0 & 0 & 0 & 0 & 0 & 0 & 0 & 0 & 0 & 0 & 60 & 0 & -60 & 0 & 0 & 0 & 0 & 0 & 0 \\
 30 & 0 & -30 & 0 & 30 & 0 & -30 & 0 & 0 & 0 & 0 & 0 & 0 & -30 & -30 & 30 & 30 & 0 & 0 & 0 & 0 & 0 & 0 & 0 & 0 & 0 & 0 & 0 & 0 & 0 & 0 & -60 & 0 & 60 & 0 & 0 & 0 & 0 & 0 \\
 30 & 0 & -30 & 0 & 30 & 0 & -30 & 0 & 0 & 0 & 0 & 0 & 0 & 30 & 30 & -30 & -30 & 0 & 0 & 0 & 0 & 0 & 0 & 0 & 0 & 0 & 0 & 0 & 0 & 0 & -60 & 0 & 60 & 0 & 0 & 0 & 0 & 0 & 0 \\
 30 & 0 & -30 & 0 & 30 & 0 & -30 & 0 & 0 & 0 & 0 & 0 & 0 & -30 & -30 & 30 & 30 & 0 & 0 & 0 & 0 & 0 & 0 & 0 & 0 & 0 & 0 & 0 & 0 & 0 & 0 & 60 & 0 & -60 & 0 & 0 & 0 & 0 & 0 \\
 24 & -24 & 0 & 24 & 0 & -24 & 24 & 0 & 0 & 0 & 0 & 0 & 0 & 0 & 0 & 0 & 0 & 0 & 0 & 0 & 0 & 0 & 0 & 0 & 0 & 0 & 0 & 0 & 0 & 0 & 0 & 0 & 0 & 0 & 96 & -24 & -24 & -24 & -24 \\
 24 & -24 & 0 & 24 & 0 & -24 & 24 & 0 & 0 & 0 & 0 & 0 & 0 & 0 & 0 & 0 & 0 & 0 & 0 & 0 & 0 & 0 & 0 & 0 & 0 & 0 & 0 & 0 & 0 & 0 & 0 & 0 & 0 & 0 & -24 & -12 \left(\sqrt{5}-3\right) & -24 \left(\sqrt{5}+1\right) & 24 \left(\sqrt{5}-1\right) & 12 \left(\sqrt{5}+3\right) \\
 24 & -24 & 0 & 24 & 0 & -24 & 24 & 0 & 0 & 0 & 0 & 0 & 0 & 0 & 0 & 0 & 0 & 0 & 0 & 0 & 0 & 0 & 0 & 0 & 0 & 0 & 0 & 0 & 0 & 0 & 0 & 0 & 0 & 0 & -24 & -24 \left(\sqrt{5}+1\right) & 12 \left(\sqrt{5}+3\right) & -12 \left(\sqrt{5}-3\right) & 24 \left(\sqrt{5}-1\right) \\
 24 & -24 & 0 & 24 & 0 & -24 & 24 & 0 & 0 & 0 & 0 & 0 & 0 & 0 & 0 & 0 & 0 & 0 & 0 & 0 & 0 & 0 & 0 & 0 & 0 & 0 & 0 & 0 & 0 & 0 & 0 & 0 & 0 & 0 & -24 & 24 \left(\sqrt{5}-1\right) & -12 \left(\sqrt{5}-3\right) & 12 \left(\sqrt{5}+3\right) & -24 \left(\sqrt{5}+1\right) \\
 24 & -24 & 0 & 24 & 0 & -24 & 24 & 0 & 0 & 0 & 0 & 0 & 0 & 0 & 0 & 0 & 0 & 0 & 0 & 0 & 0 & 0 & 0 & 0 & 0 & 0 & 0 & 0 & 0 & 0 & 0 & 0 & 0 & 0 & -24 & 12 \left(\sqrt{5}+3\right) & 24 \left(\sqrt{5}-1\right) & -24 \left(\sqrt{5}+1\right) & -12 \left(\sqrt{5}-3\right) \\
\end{array}
\right) $} \nonumber \\
\mathcal{T}(S_5) &= \text{diag}\left\{1,\,1,\,1,\,1,\,1,\,1,\,1,\,1,\,1,\,-1,\,-1,\,-1,\,1,\,1,\,1,\,1,\,1,\,-1,\,1,\,(-1)^{4/3},\,(-1)^{2/3},\,1,\,(-1)^{4/3},\,(-1)^{2/3}, \right. \nonumber \\
& \left. 1,\,(-1)^{1/3},\,(-1)^{2/3},\,-1,\,(-1)^{4/3},\,(-1)^{5/3},\,1,\,i,\,-1,\,-i,\,1,\,(-1)^{2/5},\,(-1)^{4/5},\,(-1)^{6/5},\,(-1)^{8/5}\right\}  ~.
\label{ST-for-S5}
\end{align}
\end{landscape}
\section{Adams operation on centralizers of conjugacy classes of $G$}
\label{appendix-B}
Consider the group $G$ whose conjugacy classes are labeled as $A$ with $a \in A$ being some representative element of $A$. The centralizer corresponding to $a$ is denoted as $C_A$. We wish to perform Adams operation on $C_A$ as following:
\begin{equation}
\Psi^m \chi_{\alpha}(a) = \chi_{\alpha}(a^m) = \sum_{\beta} Y_{\alpha \beta}(m) \, \chi_{\beta}(a) ~,
\end{equation}  
where $\chi_{\alpha}$ denotes the irreducible characters of $C_A$. We know that the set of class functions of the group $C_A$ form a vector space with the inner product defined as following:
\begin{equation}
\bra{f_1}\ket{f_2} = \frac{1}{|C_A|} \sum_{g \in C_A} f_1(g^{-1})\,f_2 (g) ~.
\end{equation}
Using the fact that the irreducible characters form an orthonormal basis, the coefficients $Y_{\alpha \beta}$ can be uniquely determined as following:
\begin{equation}
\boxed{Y_{\alpha \beta}(m) = \bra{\chi_{\beta}}\ket{\Psi^m \chi_{\alpha}} = \frac{1}{|C_A|} \sum_{g \in C_{A}} \chi_{\alpha} (g^{m})\,\chi_{\beta} (g^{-1})} ~.
\end{equation} 
For our computation of entropy, we require coefficients $X_{ab}(m)$ where $a=(A, R_A^i)$ and $b=(B, R_B^j)$ are the irreps of the quantum double group. Here $A$ and $B$ label the conjugacy classes of the group $G$. The notations $R_A^i$ and $R_B^j$ represent the $i^{\text{th}}$ and $j^{\text{th}}$ irreps of the centralizers $C_A$ and $C_B$ respectively. We define these coefficients in terms of coefficients $Y_{\alpha \beta}(m)$ as:
\begin{equation}
X_{ab}(m) = Y_{\alpha \beta}(m) \, \delta_{AB} ~,
\end{equation}
where $\alpha$ and $\beta$ are now irreps of centralizer $C_A$. We can collect all the coefficients $X_{ab}(m)$ into a matrix $X(m)$ written in the basis of the irreps of quantum double group. While defining this matrix, we must take the same ordering of the basis as we did while obtaining the modular $\mathcal{S}$ and $\mathcal{T}$ matrices. This will ensure that we can perform the matrix algebra between these matrices. From the definition, it is clear that the matrix $X(m)$ will be block diagonal where each block will consist of matrices $Y(m)$ defined for each centralizer. The rows and columns of $Y(m)$ are labeled by irreps of that particular centralizer and its elements are precisely $Y_{\alpha \beta}(m)$. In the following, we will compute the coefficients $Y_{\alpha \beta}(m)$ for centralizers of various finite groups.  
\subsection{For $\mathbb{Z}_N$ group}
The centralizer corresponding to each conjugacy class is again $\mathbb{Z}_N$, i.e. $C_A = \mathbb{Z}_N$. We label the elements of $\mathbb{Z}_N$ by $g$ and the irreducible representation by $\alpha$ with both the labels taking values from 0 to $(N-1)$. The character can be given as:
\begin{equation}
\chi_{\alpha}(g) = \exp(\frac{2 \pi i \alpha g}{N}) ~.
\end{equation}
The coefficients $Y_{\alpha \beta}$ for each centralizer $C_A$ can therefore be obtained as,
\begin{equation}
\boxed{Y_{\alpha \beta}(m) = \delta_{\beta,\,m \alpha(\text{mod } N)}} ~.
\label{Adams-ZN}
\end{equation}
We can now collect all these coefficients in the matrix $Y(m)$ whose rows and columns are indexed by $\alpha$ and $\beta$ respectively taking values from 0 to $(N-1)$. Since this matrix will be the same for all the centralizers, the required matrix $X(m)$ will be a block diagonal matrix made up of $N$ blocks and each block is given by $Y(m)$. 
\subsection{For $D_N$ group}
We have already discussed the structure of dihedral group in previous appendix. Its order is $2N$ and its elements are given as,
\begin{equation}
D_N = \{r_0, r_1, r_2, \ldots, r_{N-1}, s_0, s_1, s_2, \ldots, s_{N-1} \} ~.
\end{equation}
For even dihedral group $D_{2M}$, there are $(M+3)$ conjugacy classes and hence $(M+3)$ centralizers. All the centralizers are isomorphic to three groups: $D_{2M}, \mathbb{Z}_{2M}$ and $\mathbb{Z}_{2} \times \mathbb{Z}_{2}$. For the group $D_{2M}$ itself, the characters of the one dimensional irreps evaluated at powers of various elements can be written in a  compact fashion as:
\begin{alignat}{4}
\theta_1(r_x^m) &= 1, \quad && \theta_2(r_x^m) = 1, \quad && \theta_3(r_x^m) = (-1)^{xm}, \quad && \theta_4(r_x^m) = (-1)^{xm} \nonumber \\
\theta_1(s_x^m) &= 1, \quad && \theta_2(s_x^m) = (-1)^m, \quad && \theta_3(s_x^m) = (-1)^{xm}, \quad && \theta_4(s_x^m) = (-1)^{xm+m} ~,
\end{alignat} 
while the characters of two dimensional irreps are:
\begin{equation}
\phi_j(r_x^m) = 2 \cos \left(\frac{\pi jxm}{M}\right), \quad \phi_j(s_x^m) = 1+(-1)^m ~,
\end{equation} 
where $j$ takes values from 1 to $(M-1)$ and $m$ is any integer. Using these, the matrix $Y(m)$ can be written as a block matrix as: 
\begin{equation}
Y_{D_{2M}} = \left(
\begin{array}{ccccc}
 1 & 0 & 0 & 0 & [0]_{1\times (M-1)} \\
 x_{+} & x_{-} & 0 & 0 & [0]_{1\times (M-1)} \\
 x_{+} & 0 & x_{-} & 0 & [0]_{1\times (M-1)} \\
 x_{+} & 0 & 0 & x_{-} & [0]_{1\times (M-1)} \\
 \mbox{}[x_{+}]_{(M-1)\times 1}\,\, & [-x_{+}]_{(M-1)\times 1}\,\, & [0]_{(M-1)\times 1}\,\, & [0]_{(M-1)\times 1} & F(m) \\
\end{array}
\right) ~,
\end{equation}  
where the notation $[x]_{a \times b}$ represent a matrix of order $a \times b$ whose every element is $x$, where we have defined:
\begin{equation}
x_{\pm} = \left(\frac{1\pm(-1)^m}{2}\right)
\end{equation}
and $F(m)$ is a square matrix of order $(M-1)$ whose elements are given as:
\begin{equation}
F_{ab} =\frac{1}{2M}\left( \frac{\cos\left(2 \pi  a m-\frac{\pi am}{M}\right)-\cos\left(2\pi b-\frac{\pi  b}{M}\right)+\cos\left(\frac{\pi b}{M}\right)-\cos\left(\frac{\pi am}{M}\right)}{\cos \left(\frac{\pi  b}{M}\right)-\cos \left(\frac{\pi  a m}{M}\right)}\right) ~,
\end{equation}
where $a$ and $b$ label the rows and columns and take values from 1 to $(M-1)$. The other two non-isomorphic centralizers are $\mathbb{Z}_{2M}$ and $\mathbb{Z}_{2} \times \mathbb{Z}_{2}$. We have already given the matrix $Y(m)$ for the $\mathbb{Z}_N$ group in eq.(\ref{Adams-ZN}). Thus we can now construct the matrix $X(m)$ which is a block diagonal matrix given as:
\begin{equation}
X_{D_{2M}} =\text{diag}\left\{ Y_{D_{2M}},\, Y_{Z_{2M}},\,\ldots,\,Y_{Z_{2M}},\, Y_{D_{2M}},\, Y_{Z_2} \otimes Y_{Z_2}  \right\} ~,
\end{equation}
where there are $(M-1)$ copies of $Y_{Z_{2M}}$ and $\otimes$ is the Kronecker product. 

Similarly we can also obtain these matrices for the odd dihedral group $D_{2M+1}$. This group has $(M+2)$ conjugacy classes and the corresponding centralizers are isomorphic to the groups: $D_{2M+1}, \mathbb{Z}_{2M+1}$ and $\mathbb{Z}_{2}$. Thus we only need to compute $Y_{D_{2M+1}}$. For the group $D_{2M+1}$, the characters of powers of various elements can be compactly written as:
\begin{alignat}{3}
\theta_1(r_x^m) &= 1, \quad && \theta_2(r_x^m) = 1, \quad && \phi_j(r_x^m) = 2 \cos \left(\frac{2\pi jxm}{2M+1}\right) \nonumber \\
\theta_1(s_x^m) &= 1, \quad && \theta_2(s_x^m) = (-1)^m, \quad && \phi_j(s_x^m) = 1+(-1)^m ~,
\end{alignat} 
where $m$ is any integer and $j$ takes values from 1 to $M$. Thus we can now obtain:
\begin{equation}
Y_{D_{2M+1}} = \left(
\begin{array}{ccc}
 1 & 0 & [0]_{1\times M} \\
 x_{+} & x_{-} & [0]_{1\times M} \\
 F(m)\,\, & G(m)\,\, & H(m)
\end{array}
\right) ~.
\end{equation}
Here $F(m)$ and $G(m)$ are single column matrices whose rows labeled by $\alpha$ taking values from 1 to $M$ are given as:
\begin{equation}
F_{\alpha} = x_{\alpha} + x_{+} \quad;\quad G_{\alpha} = x_{\alpha} - x_{+} ~,
\end{equation}
where we have defined
\begin{equation}
x_{\alpha} = \frac{\sin\left(\frac{\pi  \alpha m (4 M+1)}{2 M+1}\right) \csc \left(\frac{\pi  \alpha m}{2 M+1}\right)+1}{4 M+2} \quad;\quad x_{\pm} = \left(\frac{1\pm(-1)^m}{2}\right) ~.
\end{equation}
Similarly the matrix $H(m)$ is a square matrix of order $M$ whose elements are given as,
\begin{equation}
H_{ab} = \frac{\sin \left(\frac{\pi(4 M+1)(a m+b)}{2M+1}\right) \csc \left(\frac{\pi(am+b)}{2M+1}\right)+\sin\left(\frac{\pi(4 M+1) (am-b)}{2M+1}\right) \csc\left(\frac{\pi(am-b)}{2M+1}\right)+2}{4M+2}  ~,
\end{equation}
where $a$ and $b$ label the rows and columns respectively with each taking values from 1 to $M$.
The matrix $X(m)$ therefore can now be written as:
\begin{equation}
X_{D_{2M+1}} =\text{diag}\left\{ Y_{D_{2M+1}},\, Y_{Z_{2M+1}},\,\ldots,\,Y_{Z_{2M+1}}\, Y_{Z_2}  \right\} ~,
\end{equation}
where there are $M$ copies of $Y_{Z_{2M+1}}$. As a consistency check, one can compute $X(m)$ for $m=1$ for any dihedral group, which will turn out to be the identity matrix as expected. 
\subsection{For $S_N$ group}
The group $S_1$ is a trivial group. The groups $S_2$ and $S_3$ are isomorphic to $\mathbb{Z}_2$ and $D_3$ respectively for which we have already studied the Adams operation. For the $S_4$ group, there are five conjugacy classes denoted as $\text{cc}_1,\text{cc}_2,\text{cc}_3,\text{cc}_4,\text{cc}_5$ with the same ordering as given in its character table. The associated centralizers with these conjugacy classes are respectively $S_4, \mathbb{Z}_2 \times \mathbb{Z}_2, D_4, \mathbb{Z}_3, \mathbb{Z}_4$. We already know the coefficients $Y_{\alpha \beta}(m)$ for all the groups except $S_4$, so let us compute them for $S_4$ group. First we will need the various irreducible characters for the $m^{\text{th}}$ power of various elements of $S_4$. We find that for a fixed value of $m$, the $m^{\text{th}}$ power of all the elements of a particular conjugacy class will belong to a single conjugacy class (maybe different from the original class). Thus the characters $\chi(g^m)$ of $m^{\text{th}}$ power of elements belonging to a particular conjugacy class ($g \in \text{cc}_i$) will be the same. Thus we can write a character table for the characters of $m^{\text{th}}$ power of various elements of $S_4$ which is given in table (\ref{table-char-S4-power-m}).
\begin{equation}
\begin{array}{|c|c|c|c|c|c|} \hline
& g \in \text{cc}_1 & g \in \text{cc}_2 & g \in \text{cc}_3 & g \in \text{cc}_4 & g \in \text{cc}_5 \\
& \chi(g^m) & \chi(g^m) & \chi(g^m) & \chi(g^m) & \chi(g^m) \\ \hline
\chi_1 & 1 & 1 & 1 & 1 & 1 \\
\chi_2 & 3 & 2h_{02}+1 & 4h_{02}-1 & 3h_{03} & 4h_{04}-1 \\
\chi_3 & 2 & 2h_{02} & 2 & 3h_{03}-1 & 2h_{02} \\
\chi_4 & 3 & 4h_{02}-1 & 4h_{02}-1 & 3h_{03} & 3h_{04}+h_{14}-h_{24}+h_{34} \\
\chi_5 & 1 & 2h_{02}-1 & 1 & 1 & 2h_{02}-1 \\ \hline
\end{array}
\label{table-char-S4-power-m}
\end{equation}
In this table, $m$ can be any integer and we have defined $h_{ab} = \delta_{a,[m]_b}$ with $[m]_b$ denoting the value of $m$ modulo $b$. The Adams coefficients are thus given as,
\begin{equation}
Y_{\alpha \beta}(m) =  \frac{1}{24} \left( \sum_{i=1}^5 |\text{cc}_i| \, \chi_{\alpha} (g^{m})\,\chi_{\beta} (g^{-1}) \right) ~,
\end{equation} 
where the sum is over various conjugacy classes. Since the characters of $m^{\text{th}}$ powers of all elements belonging to a class are same, the element $g$ can be taken as any element belonging to that particular class in the above summation and we put the counting factor $|\text{cc}_i|$ which is the number of elements in the $i^{\text{th}}$ class. These coefficients can be arranged in a matrix with $\alpha$ and $\beta$ labeling its rows and columns:
 \begin{equation}
Y_{S_4} = \left(
\begin{array}{ccccc}
 1 & 0 & 0 & 0 & 0 \\
 h_{02}+h_{03}+h_{04} & 1-h_{04} & h_{02}-h_{03} & h_{04}-h_{02} & h_{03}-h_{04} \\
 h_{02}+h_{03} & 0 & 1-h_{03} & 0 & h_{03}-h_{02} \\
 h_{02}+h_{03}-f_1 & f_2 & h_{02}-h_{03} & 1-h_{02}-f_2 & f_1+h_{03} \\
 h_{02} & 0 & 0 & 0 & 1-h_{02} \\
\end{array}
\right)~,
\end{equation}
where we have defined:
\begin{equation}
f_1 = \frac{1-2h_{02}-3h_{04}-h_{14}+h_{24}-h_{34}}{4},\quad f_2 = \frac{1+2h_{02}-3h_{04}-h_{14}+h_{24}-h_{34}}{4}~.
\end{equation}
Thus the block diagonal matrix $X(m)$ containing the Adams coefficients for various centralizers will be:
\begin{equation}
X_{S_4} =\text{diag}\left\{ Y_{S_4},\, Y_{\mathbb{Z}_2 \times \mathbb{Z}_2},\, Y_{D_{4}},\, Y_{\mathbb{Z}_3},\, Y_{\mathbb{Z}_4}  \right\} ~.
\end{equation}

Next, we go to the $S_5$ group. There are seven conjugacy classes which we shall denote as $\text{cc}_1,\text{cc}_2,\text{cc}_3,\text{cc}_4,\text{cc}_5,\text{cc}_6,\text{cc}_7$ with the same ordering as given in its character table. The associated centralizers with these conjugacy classes are respectively $S_5, D_6, D_4, \mathbb{Z}_6, \mathbb{Z}_6, \mathbb{Z}_4, \mathbb{Z}_5$. We only need to compute coefficients $Y_{\alpha \beta}(m)$ for $S_5$ group. The characters for $m^{\text{th}}$ power of various elements of $S_5$ is given in table (\ref{table-char-S5-power-m}).
\begin{equation}
\begin{array}{|c|c|c|c|c|c|c|c|} \hline
& g \in \text{cc}_1 & g \in \text{cc}_2 & g \in \text{cc}_3 & g \in \text{cc}_4 & g \in \text{cc}_5 & g \in \text{cc}_6 & g \in \text{cc}_7 \\
& \chi(g^m) & \chi(g^m) & \chi(g^m) & \chi(g^m) & \chi(g^m) & \chi(g^m) & \chi(g^m) \\ \hline
\chi_1 & 1 & 1 & 1 & 1 & 1 & 1 & 1 \\
\chi_2 & 4 & 2h_{02}+2 & 4h_{02} & 3 h_{03}+1 & f_3 & 4 h_{04} & 5 h_{05}-1 \\
\chi_3 & 5 & 4 h_{02}+1 & 4 h_{02}+1 & 6 h_{03}-1 & f_4 & 6 h_{04}+2 h_{24}-1 & 5 h_{05} \\
\chi_4 & 6 & 6 h_{02} & 8 h_{02}-2 & 6 h_{03} & 6 h_{06} & 6 h_{04}-2 h_{24} & 5 h_{05}+1 \\
\chi_5 & 5 & 6 h_{02}-1 & 4 h_{02}+1 & 6 h_{03}-1 & 6 h_{06}-1 & 4 h_{04}+1 & 5 h_{05} \\
\chi_6 & 4 & 6 h_{02}-2 & 4 h_{02} & 3 h_{03}+1 & 3 h_{06}-3 h_{36}+1 & 4 h_{04} & 5 h_{05}-1 \\
\chi_7 & 1 & 2 h_{02}-1 & 1 & 1 & 2 h_{02}-1 & 2 h_{04}+2 h_{24}-1 & 1 \\ \hline
\end{array}
\label{table-char-S5-power-m}
\end{equation}
where we have defined,
\begin{equation}
f_3 = 4 h_{06}-h_{16}+h_{26}+2 h_{36}+h_{46}-h_{56}\,\,,\quad f_4 = 5 h_{06}+h_{16}-h_{26}+h_{36}-h_{46}+h_{56}~.
\end{equation}
The Adams coefficients are thus given as,
\begin{equation}
Y_{\alpha \beta}(m) =  \frac{1}{120} \left( \sum_{i=1}^7 |\text{cc}_i| \, \chi_{\alpha} (g^{m})\,\chi_{\beta} (g^{-1}) \right) ~.
\end{equation} 
These coefficients can be arranged in a order 7 matrix $Y_{S_5}$ with $\alpha$ and $\beta$ labeling its rows and columns, which we will not present here.
Thus the block diagonal matrix $X(m)$ containing the Adams coefficients for various centralizers will be:
\begin{equation}
X_{S_5} =\text{diag}\left\{ Y_{S_5},\, Y_{D_6},\, Y_{D_{4}},\, Y_{\mathbb{Z}_6},\, Y_{\mathbb{Z}_6},\, Y_{\mathbb{Z}_4},\, Y_{\mathbb{Z}_5}  \right\} ~.
\end{equation}

\vspace{0.5cm}
\textbf{Acknowledgements} 
SD would like to thank P. Ramadevi and Vivek Kumar Singh for useful discussions. YZ would like to thank Ling-Yan Hung and Yidun Wan for discussions. The work of SD is supported by the NSFC grant 11975158. YZ is supported by NSFC grant 11905033.
\bibliographystyle{JHEP}
\bibliography{EEDiscrete} 

\end{document}